\documentclass[prl,aps,twocolumn,amsmath,amsmath,amssymb,superscriptaddress]{revtex4-1}

\usepackage{graphicx}
\usepackage{dcolumn}
\usepackage{bm}
\usepackage[colorlinks,linkcolor=blue,citecolor=blue,urlcolor=blue]{hyperref}
\usepackage{float}
\usepackage[dvipsname]{xcolor}
\usepackage[utf8]{inputenc} 
\usepackage[T1]{fontenc}
\usepackage[normalem]{ulem}
\usepackage{amsmath}

\begin{document}

\title{Ultrafast Nonequilibrium Enhancement of Electron-Phonon Interaction in 2H-MoTe$_2$}

\author{Nina Girotto Erhardt}
\thanks{These two authors contributed equally}
\affiliation{European Theoretical Spectroscopy Facility, Institute of Condensed Matter and Nanosciences, Université catholique de Louvain, Louvain-la-Neuve, Belgium}
\affiliation{Centre for Advanced Laser Techniques, Institute of Physics, 10000 Zagreb, Croatia}

\author{Sotirios Fragkos}
\thanks{These two authors contributed equally}
\affiliation{Universit\'e de Bordeaux - CNRS - CEA, CELIA, UMR5107, F33405 Talence, France}

\author{Dominique Descamps}
\affiliation{Universit\'e de Bordeaux - CNRS - CEA, CELIA, UMR5107, F33405 Talence, France}

\author{Stéphane Petit}
\affiliation{Universit\'e de Bordeaux - CNRS - CEA, CELIA, UMR5107, F33405 Talence, France}

\author{Michael Schüler}
\affiliation{PSI Center for Scientific Computing, Theory and Data, 5232 Villigen PSI, Switzerland}
\affiliation{Department of Physics, University of Fribourg, CH-1700 Fribourg, Switzerland}

\author{Dino Novko}
\email{dino.novko@gmail.com}
\affiliation{Centre for Advanced Laser Techniques, Institute of Physics, 10000 Zagreb, Croatia}

\author{Samuel Beaulieu}
\email{samuel.beaulieu@u-bordeaux.fr}
\affiliation{Universit\'e de Bordeaux - CNRS - CEA, CELIA, UMR5107, F33405 Talence, France}

\begin{abstract}
Understanding nonequilibrium electron–phonon interactions at the microscopic level and on ultrafast timescales is a central goal of modern condensed matter physics. Combining time- and angle-resolved extreme ultraviolet photoemission spectroscopy with constrained density functional perturbation theory, we demonstrate that photoexcited carrier density can serve as a tuning knob to enhance electron–phonon interactions in nonequilibrium conditions. Specifically, nonequilibrium band structure mapping and valley-resolved ultrafast population dynamics in semiconducting transition-metal dichalcogenide 2H-MoTe$_2$ reveal band-gap renormalizations and reduced population lifetimes as photoexcited carrier densities increase. Through theoretical analysis of photoinduced electron and phonon energy and linewidth renormalizations, we attribute these transient features to nonequilibrium modifications of electron–phonon coupling matrix elements. The present study advances our understanding of microscopic coupling mechanisms enabling control over relaxation pathways in driven solids.
\end{abstract}

\maketitle

Electron–phonon coupling (EPC) plays a fundamental role in shaping various material properties, including charge transport, thermal conductivity, optical responses, and phase transitions~\cite{Giustino2017}. In transition-metal dichalcogenides (TMDCs), EPC is a key driver of charge-ordered states and superconductivity~\cite{zhu2015,otto21,BinSubhan2021,kang2018,Jung2024,Costanzo2015}, where its influence is primarily dictated by the strength of the EPC matrix elements and their screening~\cite{zhu2015,weber2011,varma1977,weber2018}. Notably, in multivalley semiconducting TMDCs, EPC exhibits an unconventional enhancement with increasing static equilibrium doping, attributed to the inefficient screening of strongly coupled out-of-plane phonon modes~\cite{sohier2019}. This behavior has significant implications for superconducting properties~\cite{Jung2024,Costanzo2015}.

In nonequilibrium settings, ultrafast spectroscopic techniques are well suited to investigate EPC-related phenomena. While time-resolved optical spectroscopies allow for investigating energy and carrier relaxation processes~\cite{Chi2019, Sayers2023}, ultrafast electron diffraction tracks energy flow into and within the lattice degree-of-freedom in a momentum-resolved fashion~\cite{Chatelin14, Seiler2021, Fukuda2023}. Moreover, time- and angle-resolved photoemission spectroscopy (trARPES) provides an energy- and momentum-resolved view on nonequilibrium phenomena driven by EPC such as transient band renormalizations, nonequilibrium phase transitions, coherent phonons, and electronic population lifetimes~\cite{Boschini2024}. In photoexcited solids, EPC is responsible for electron relaxation on multiple different timescales~\cite{sidiropoulos21, kampfrath05, cappelluti2022}, anisotropic non-thermal phonon relaxation~\cite{seiler21}, and thermalization bottlenecks~\cite{Perfetti2007, Johannsen2013, novko2020}. However, the microscopic mechanisms underlying photoinduced modifications of EPC remain poorly understood. For instance, experimental time-resolved studies of graphene demonstrated an apparent enhancement\,\cite{pomarico17} and a reduction\,\cite{ishioka08} of EPC strength, depending on the excitation conditions. However, theoretical consensus on the origin of these modifications has not been reached\,\cite{hu22,Girotto2023}. A similar debate exists for photoexcited TMDCs. Indeed, a recent time-dependent density-functional-theory study showed that photoexcitation of MoS$_2$ enhances EPC strength of optical phonons due to reduced electronic screening\,\cite{liu22}. In contrast, femtosecond electron diffuse scattering and corresponding \textit{ab initio} simulations of time-dependent Boltzmann equations (TDBE) with additional carrier screening found that photoexcitation reduces EPC of optical phonons, suppressing intravalley phonon-related scattering\,\cite{pan2025}. It is thus crucial to combine state-of-the-art experimental and theoretical techniques sensitive to both electron and lattice degrees of freedom to resolve the debate on microscopic mechanisms underlying photoinduced EPC modifications. This is particularly relevant for debated ultrafast light-induced EPC-related phenomena, such as photoinduced or quenched superconductivity\,\cite{fausti11,zhang14} and structural order\,\cite{rohwer11,stojchevska14}.

In this work, we present a joint experimental and theoretical study of nonequilibrium EPC in 2H-MoTe$_2$. Using time- and angle-resolved extreme ultraviolet (XUV) photoemission spectroscopy with a momentum microscope detection scheme, we track the evolution of the electronic structure and excited-state population dynamics in a valley-resolved manner, as a function of the photoexcited carrier density. By correlating these measurements with constrained density functional perturbation theory (cDFPT), providing non-thermal momentum-resolved renormalizations of phonons~\cite{Murray2007,liu2022,Girotto2023}, we directly link the experimentally observed band gap renormalizations and carrier lifetime reductions to photoinduced modifications of EPC matrix elements. This study highlights the critical role of EPC and the corresponding screening effects arising from photoinduced carrier distribution in TMDCs -- an essential feature that is often overlooked in conventional thermalization theories, including effective temperature models and TDBE\,\cite{Waldecker16,caruso2022}.

\begin{figure}[t]
\begin{center}
\includegraphics[width=8.6cm]{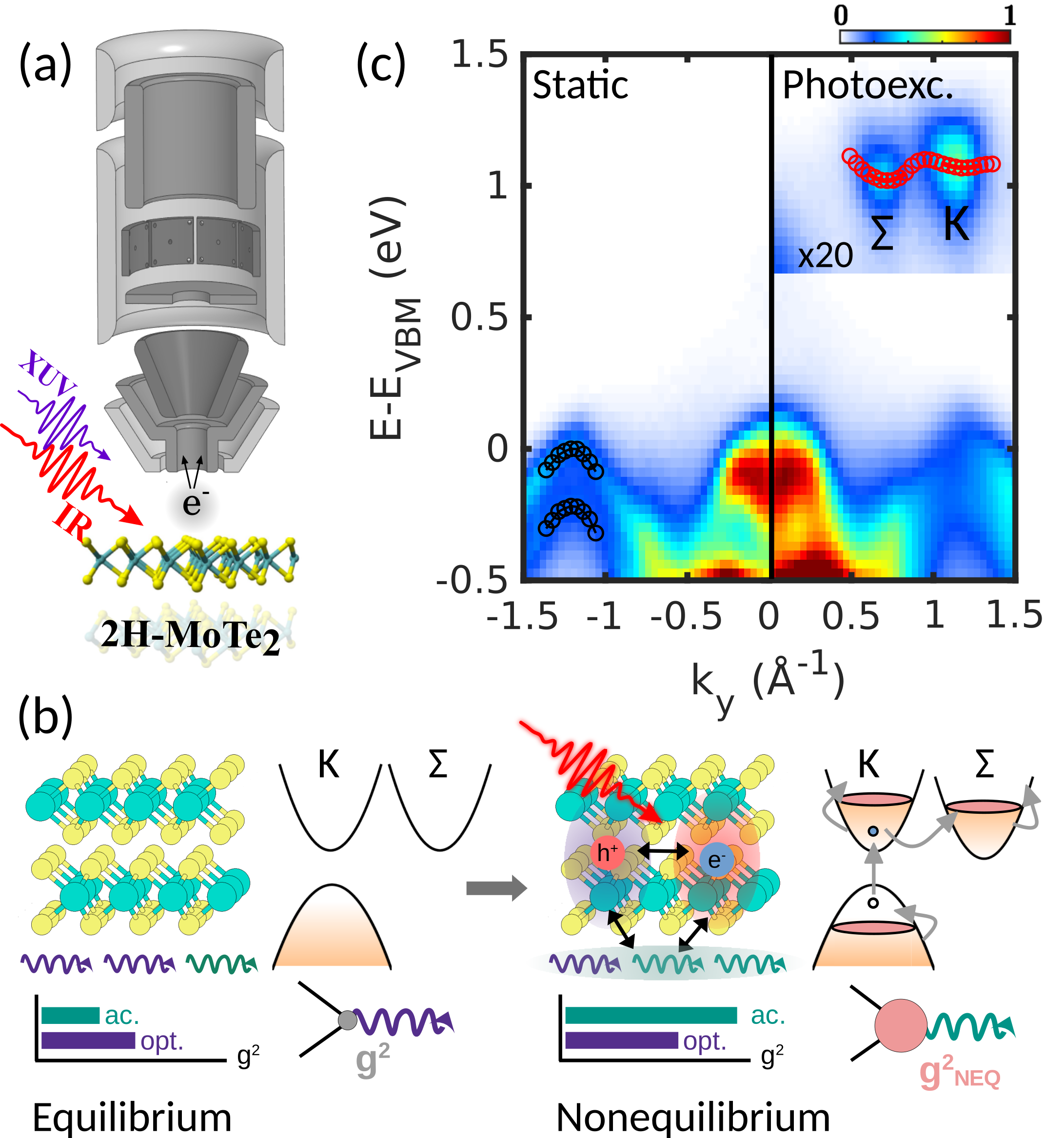}
\caption{\textbf{Nonequilibrium Band Structure Mapping and EPC in 2H-MoTe$_2$}. \textbf{(a)} Infrared pump (1.2~eV, 135~fs) and XUV probe (21.6~eV) pulses are focused on a bulk 2H-MoTe$_2$ sample in the interaction chamber of a time-of-flight momentum microscope, at an incidence angle of 65$^{\circ}$ and with the light incidence plane along the crystal mirror plane ($\Gamma$-M direction).
\textbf{(b)} Schematic of the equilibrium (left) and nonequilibrium (right) EPC, showing the relative coupling strength ($g^2$) with acoustic and optical phonon branches in both scenarios. \textbf{(c)} Static (left) and photoexcited (at pump-probe overlap, right) energy-momentum cut along $\Gamma$-$\Sigma$-K high symmetry direction, recorded using s-polarized IR pump and p-polarized XUV probe pulses.}
\label{fig1}
\end{center}
\end{figure}

The experimental setup [Fig.~\ref{fig1}(a)] is articulated around a polarization-tunable ultrafast XUV beamline~\cite{Comby22} coupled with a momentum microscope apparatus~\cite{tkach24, tkach24-2}. More details can be found in Sec.\,S1 of the Supplemental Material (SM)~\cite{SM} as well as in Ref.~\cite{Fragkos2025}. Figure~\ref{fig1}(b) presents a schematic illustrating the equilibrium (left) and photoinduced screened nonequilibrium (right) EPC, emphasizing the relative coupling strength ($g^2$) with acoustic and optical phonon branches. Figure~\ref{fig1}(c), we show the static and nonequilibrium band structure mapping of 2H-MoTe$_2$ along K-$\Sigma$-$\Gamma$ high-symmetry direction, obtained with XUV pulses only (left subpanel) and at the temporal overlap between IR and XUV pulses (right subpanel). Section\,S2 of SM contains an extended discussion about these band position fitting procedures~\cite{SM}. 

\begin{figure}[t]
\begin{center}
\includegraphics[width=8.6cm]{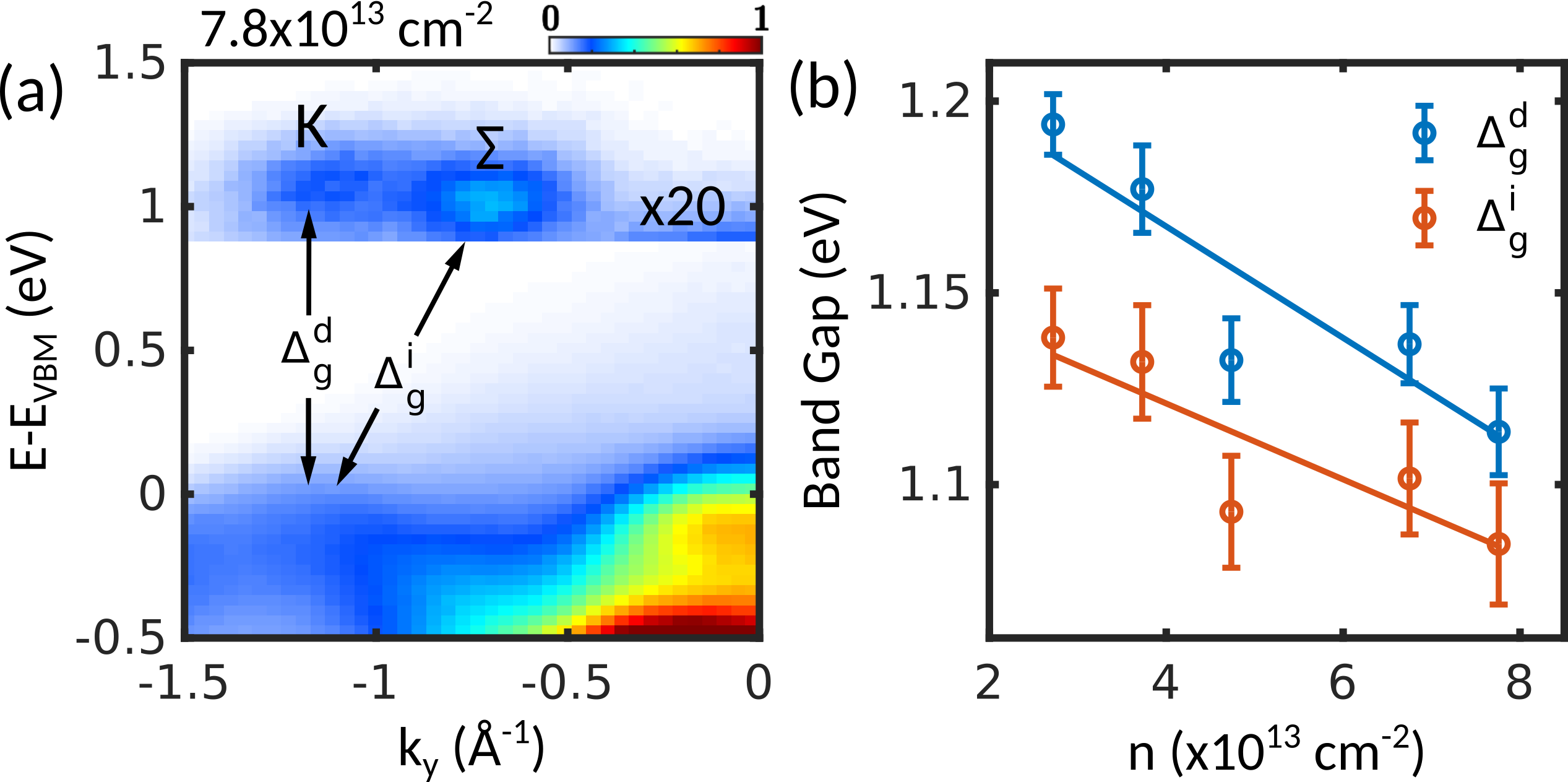}
\caption{\textbf{Photoexcited Carrier Density-Dependent Direct and Indirect Band Gap Renormalizations} \textbf{(a)} Energy-momentum cut along K-$\Sigma$-$\Gamma$ high symmetry direction, at pump-probe temporal overlap, using s-polarized IR pump and p-polarized XUV probe pulses, with arrows showing the direct ($\Delta_g^d$) and indirect ($\Delta_g^i$) band gaps. \textbf{(b)} Photoexcited carrier density-dependent direct and indirect band gap values.}
\label{fig2}
\end{center}
\end{figure}

In TMDCs, where Coulomb potential is poorly screened and where strong many-body interactions are present, electron-hole pairs from optical excitation enhance Coulomb screening and modify EPC, leading to band gap renormalization in nonequilibrium settings~\cite{Cunningham2017, Liu19, Meckbach20}. The investigation of photoexcited carrier density-dependent band gap renormalizations thus provides valuable information about many-body effects determining the optical and electronic properties of TMDCs. Data presented in Fig.~\ref{fig2}(b) reveals a direct band gap of $\Delta_{g}^{d}$ = 1.19 ± 0.01~eV at K and an indirect band gap of $\Delta_{g}^{i}$ = 1.14 ± 0.01~eV at $\Sigma$ valley (for the lowest investigated photoexcited carrier density -- 2.7 $\times$10$^{13}$~cm$^{-2}$). When photoexcited carrier density is increased from 2.7 $\times$10$^{13}$~cm$^{-2}$ (corresponding to a fluence of 0.9~mJ/cm$^2$) to 7.8 $\times$10$^{13}$~cm$^{-2}$ (corresponding to a fluence of 2.7~mJ/cm$^2$), both $\Delta_g^d$ and $\Delta_g^i$ monotonically decrease by a few tens of meV. 

\begin{figure}[t]
\begin{center}
\includegraphics[width=8.6cm]{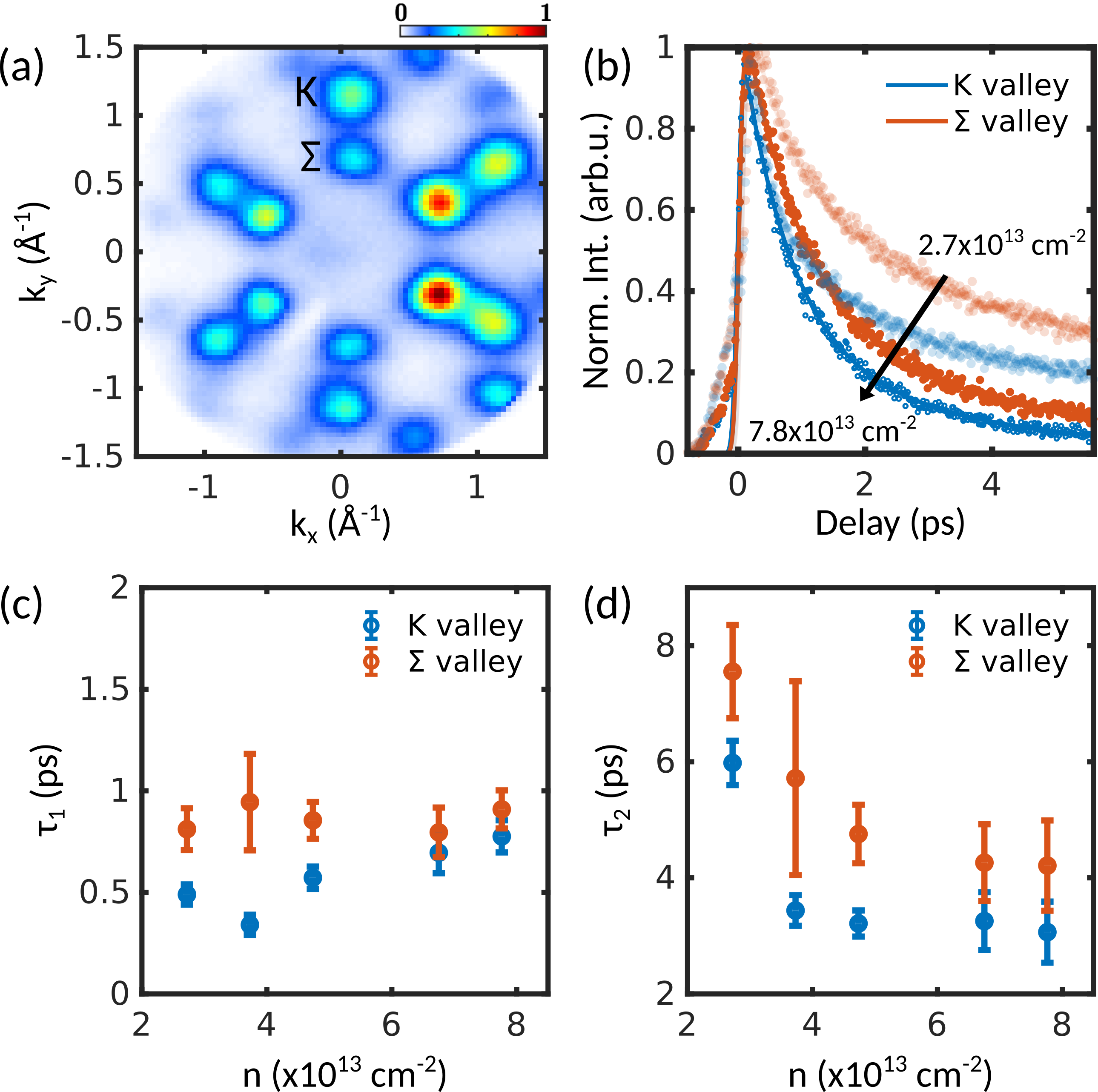}
\caption{\textbf{Photoexcited Carrier Density-Dependent Valley-Resolved Population Lifetimes} \textbf{(a)} Constant energy contour (CEC) of excited states near the conduction band minima at K and $\Sigma$ valleys. The forward-backward asymmetry in photoemission intensity arises from the photoemission transition dipole matrix elements. \textbf{(b)} Time-resolved population dynamics in K (blue) and $\Sigma$ (red) conduction band valleys, for two photoexcited carrier densities (2.7 $\times$10$^{13}$~cm$^{-2}$ and 7.8 $\times$10$^{13}$~cm$^{-2}$). \textbf{(c)-(d)} Valley-resolved photoexcited carrier density-dependent time constants ($\tau_1$ and $\tau_2$, respectively). }
\label{fig3}
\end{center}
\end{figure}

Beyond band gap renormalization, the density of photoexcited electron-hole pairs can significantly influence the ultrafast relaxation mechanisms that govern the nonequilibrium behavior of materials. We thus turn our attention to the investigation of ultrafast valley-resolved electron dynamics in photoexcited 2H-MoTe$_2$ as a function of photoexcited carrier density. Integrating photoemission intensities in the conduction band for six K and $\Sigma$ valleys [Fig.~\ref{fig3}(a)] as a function of pump-probe delay yields the valley-resolved population dynamics shown in Fig.~\ref{fig3}(b), here exemplified for the lowest (2.7 $\times$10$^{13}$~cm$^{-2}$) and highest (7.8 $\times$10$^{13}$~cm$^{-2}$) investigated photoexcited carrier densities. At our pump photon energy (1.2 eV), the only accessible bright optical transition is located around the BZ boundary (K valleys). Following photoexcitation at K valleys, carriers undergo ultrafast intervalley scattering to the $\Sigma$ global conduction band minima, as reported in similar semiconducting 2H-TMDCs~\cite{Bertoni16, Hein16, Wallauer16, Dong21}. The population dynamics at K and $\Sigma$ valleys can be fitted by a double-exponential decay function multiplied by a sigmoid function that accounts for initial photoexcitation using an IR pump with a finite duration (see Sec.\,S3 in SM\,\cite{SM}). In Figs.~\ref{fig3}(c)-(d), we report the extracted valley-resolved and photoexcited carrier density-dependent fast ($\tau_1$) and slow ($\tau_2$) time constants associated with the fitted double-exponential decays. 

Figure~\ref{fig3}(c) reveals that the fast time constant ($\tau_1$) in the $\Sigma$ valleys appears to be largely independent of photoexcited carrier density, remaining slightly below 1 ps. However, the dynamics in the K valleys exhibit a different trend. Indeed, we observe a global increase in lifetime with carrier densities (with a small deviation from this trend at low density). This behavior at K valley can be reasonably attributed to the hot phonon bottleneck effect\,\cite{Perfetti2007,shi2019,cappelluti2022,chi2020} (see Secs.\,S4 and S5~\cite{SM}). Strikingly, the slower time constant ($\tau_2$) exhibits a strong and opposite photoexcited carrier density-dependence [Fig.~\ref{fig3}(d)]. Indeed, population lifetimes at both K and $\Sigma$ valleys become significantly shorter as photoexcited carrier density increases. While these results demonstrate the possibility of tailoring ultrafast population lifetimes using different excitation conditions, which have also been discussed in previous studies~\cite{liu22, Girotto2023, pan2025}, the microscopic origin behind these renormalized nonequilibrium scattering processes has been subject to debate. Indeed, such phenomena are often discussed in the context of modified scattering phase space versus photoinduced alteration of EPC matrix elements\,\cite{otto21,ishioka08,hu22,liu22,Girotto2023}. In the case of TMDCs, a recent ultrafast electron diffraction study suggested that modifications to the EPC matrix elements do not play a decisive role for TiSe$_2$~\cite{otto21}, while a few experimental and theoretical surveys reported significant modifications (both increase and decrease) in screening of EPC in photoexcited MoS$_2$\,\cite{liu22,pan2025}. To resolve the debate about the microscopic origin of this nonequilibrium tuning knob of EPC, we performed state-of-the-art cDFPT calculations\,\cite{Girotto2023}. 

\begin{figure}[t]
\begin{center}
\includegraphics[width=8.6cm]{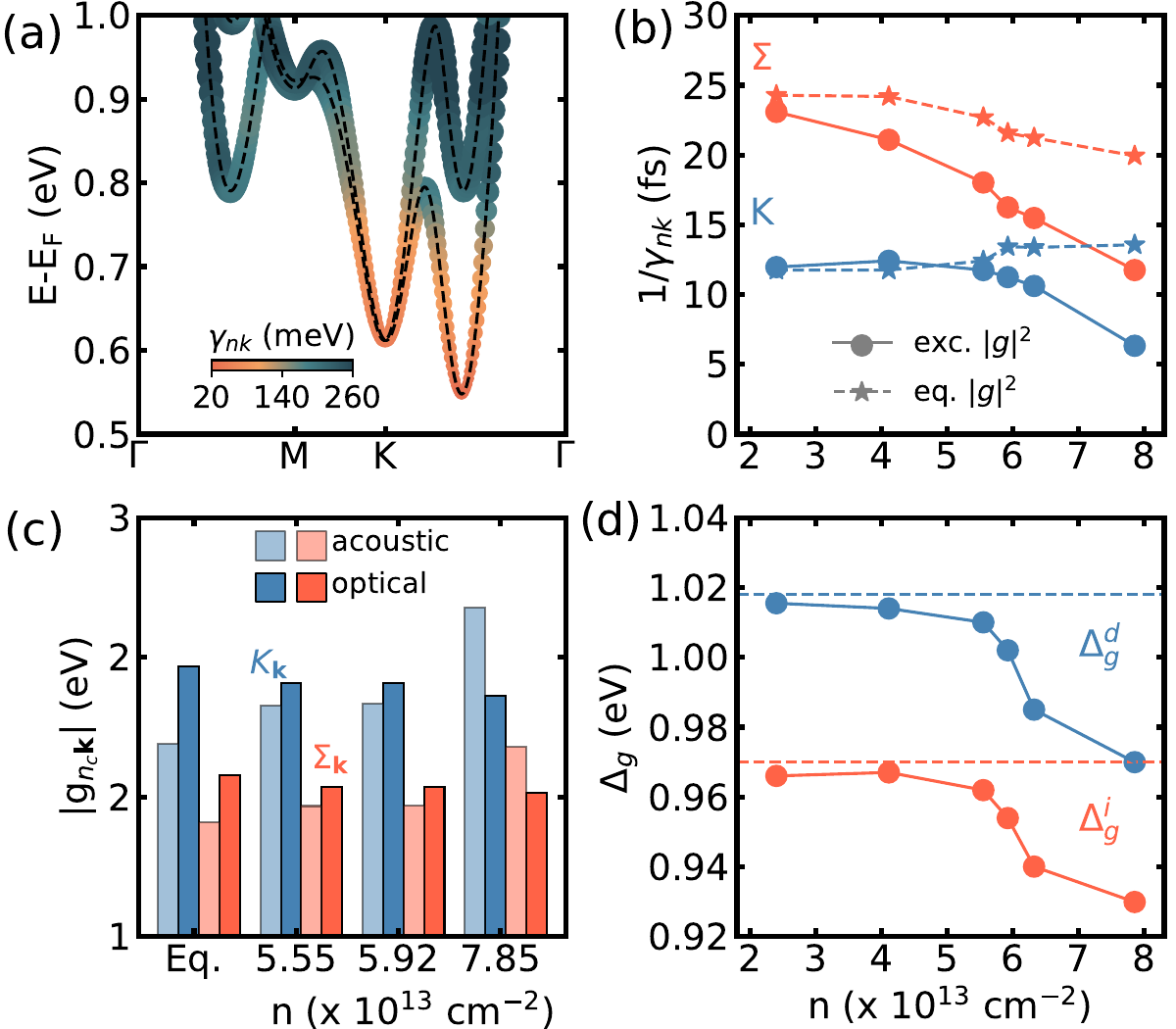}
\caption{\textbf{Equilibrium and Nonequilibrium Electronic Properties of Bilayer 2H-MoTe$_2$.} \textbf{(a)} Band-projected electron linewidth ($\gamma_{nk}$) for conduction bands of bilayer 2H-MoTe$_2$ along $\Gamma$-M-K-$\Gamma$ path. \textbf{(b)} Single-particle electron lifetimes ($1/\gamma_{nk}$) for various photoexcited carriers densities. Asterisk-shaped points and dashed lines represent results obtained using equilibrium EPC matrix elements (eq. $ \left | g \right |^2$), while round-shaped points and full lines show results of cDFPT calculations (including screening of EPC matrix elements by photoexcited carriers, exc. $ \left | g \right |^2$). \textbf{(c)} Equilibrium and nonequilibrium EPC matrix elements ($ \left | g_{q,\nu}\right |$) at the bottom of K (blue) and $\Sigma$ (red) valleys, separated in their acoustic (faint colors) and optical (bold colors) phonon contributions. \textbf{(d)} Direct ($\Delta^d_g$, blue) and indirect ($\Delta^i_g$, red) band gaps as a function of photoexcited carrier density. Dashed horizontal lines represent the associated band gap values for which no photoexcited carrier screening of $ \left | g\right |^2$ is considered.}
\label{fig:electrons}
\end{center}
\end{figure}

By performing first-principles density-functional-theory calculations on 2H-MoTe$_2$ bilayer, the conduction band minima are located at $\Sigma$ high symmetry points, in agreement with experimental measurements [Fig.~\ref{fig:electrons}(a)]. To analyze the linewidths, scattering rates, and carrier lifetimes, we computed the electron spectral function, which incorporates the Fan-Migdal electron self-energy arising from EPC~\cite{Giustino2017}. The self-energy is computed from many-body perturbation theory (MBPT) using either unscreened EPC matrix elements from DFPT~\cite{Baroni2001} or screened EPC matrix elements from cDFPT. Therefore, combining MBPT with cDFPT captures both the effect of modified scattering phase space and renormalized EPC matrix element in nonequilibrium conditions, while a combination with DFPT includes only the former.

In Fig.~\ref{fig:electrons}(a), we show the band-projected conduction electron linewidth ($\gamma_{n\mathbf{k}}$) due to electron-phonon scattering using the equilibrium DFPT approach (see SM for computational details\,\cite{SM}), which exhibit a monotonic increase of linewidth with energy. This behavior is quite general~\cite {valla1999,valla2000,bernardi2014} and persists in nonequilibrium settings~\cite{Hein16,duvel2022}. The inverse of electron linewidth ($1/\gamma_{n\mathbf{k}}$) corresponds to single-particle lifetimes. In Fig.~\ref{fig:electrons}(b), we show how single-particle lifetime decreases as a photoexcited carrier density increases, in both K and $\Sigma$ valleys. This effect is entirely due to a change in EPC matrix elements, as shown in Fig.~\ref{fig:electrons}(c) (see also Fig.\,S3\,\cite{SM}). The linewidth calculation with equilibrium EPC matrix elements [dashed line in Fig.~\ref{fig:electrons}(d)] misses out on the decreasing trend of the electron lifetime. Visible consequences of the photo-induced change in the EPC matrix elements occur in the larger carrier concentration regime. For the lower concentrations, the experimentally observed decrease of $\tau_2$ might be attributed to electron-electron scattering\,\cite{bauer2015}. In the EPC regime, we evaluate $ \left | g_{n_c\mathbf{k}}\right | = \sum_{\mathbf{q}\nu m}\left | g_{\nu n_c m}(\mathbf{k},\mathbf{q})\right |$ for a fixed $\mathbf{k}$ and the first conduction band ($n_c$) and observe that for both valleys ($\mathbf{k} = \mathrm{K}, \Sigma$), coupling of the first conduction band to acoustic phonons strengthens upon photoexcitation, while coupling to the optical modes weakens. In addition, consistent with experimental observations, EPC is stronger (shorter lifetime) at K valley (compared to $\Sigma$). However, single-particle lifetimes are two orders of magnitude smaller than the experimentally measured population lifetimes. This behavior is expected considering that the latter reflects the collective two-particle scattering process\,\cite{yang2015}, where the scattering rate is additionally reduced by electron-hole interaction (vertex correction). Nevertheless, the observed trend in photoexcited carrier density-dependent lifetimes is not affected by this additional process\,\cite{yang2015}. Also, it was shown that Boltzmann equations overestimate (underestimate) the scattering rate (lifetime) for stronger interactions and far from equilibrium\,\cite{picano21}. Moreover, as seen experimentally, we observe photoinduced direct ($\Delta^d_g$) and indirect ($\Delta^i_g$) band gap shrinkage with increasing photoexcited carrier densities, when properly accounting for renormalization of EPC matrix elements in nonequilibrium conditions. The impact of dynamical modification of EPC on band gap shrinkage is evident when comparing results obtained using screened (round markers and solid lines, cDFPT) and unscreened (dashed lines, DFPT) EPC matrix elements in Fig.~\ref{fig:electrons}(d). The weak band gap renormalization for the lower carrier excitation ($<4\times 10^{13}$\,cm$^{-2}$) might be enhanced by the inclusion of the electron-electron scattering effects~\cite{Champagne2023}. Similarly, the overall underestimation of the absolute band gap (by around 0.15\,eV) could be improved by accounting for the non-local electron correlations as in the GW approach\,\cite{Champagne2023}.

\begin{figure}[t]
\begin{center}
\includegraphics[width=8.6cm]{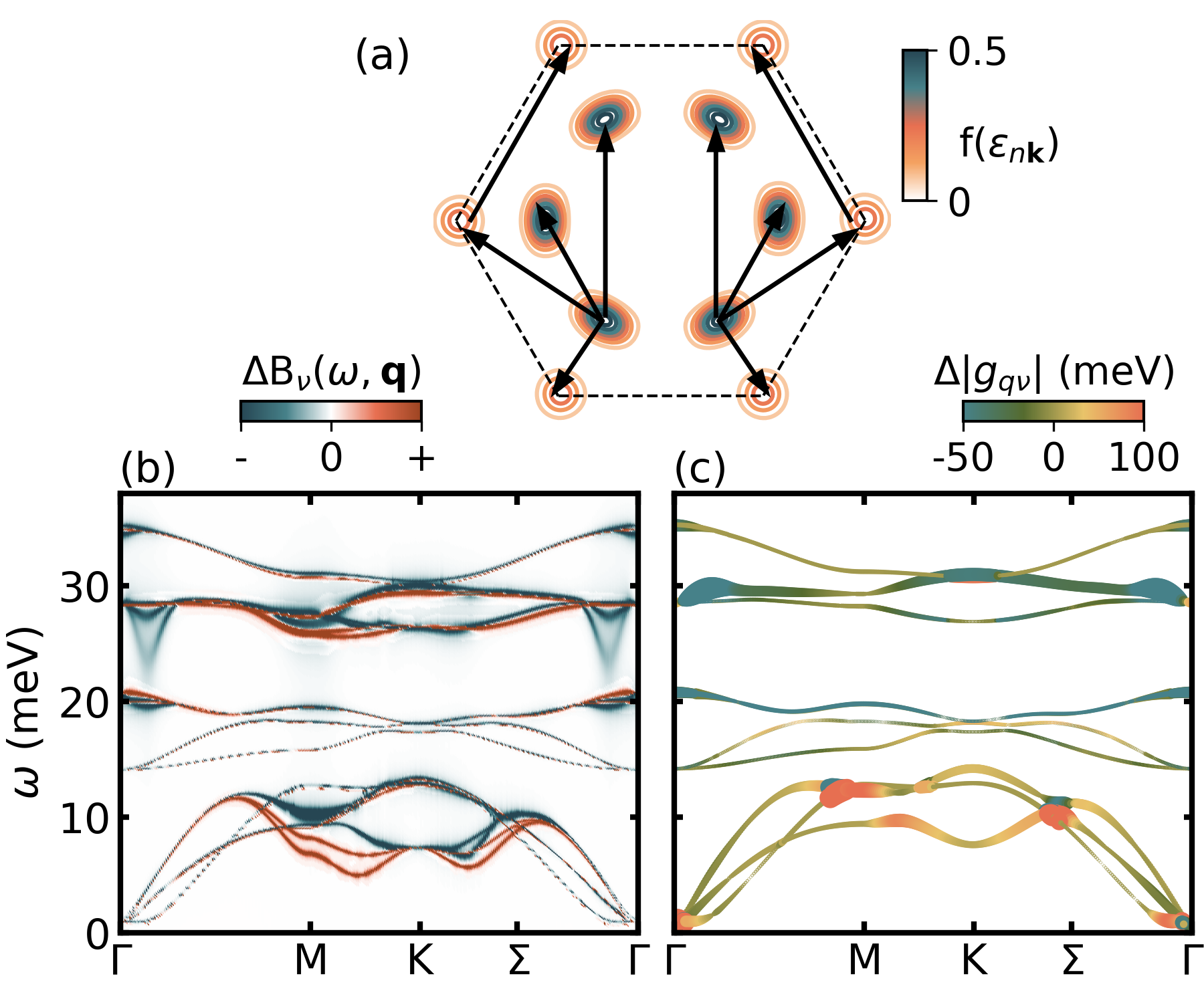}
\caption{\textbf{Nonequilibrium Phonon Softening and Renormalization of EPC Matrix Elements.} \textbf{(a)} Electron occupation factors projected onto constant energy contours near conduction band minima. Arrows represent intervalley scattering processes, which are responsible for phonon softening. \textbf{(b)} Energy- and momentum-resolved nonequilibrium renormalization of the phonon spectral function ($\Delta B_{\nu}(\omega,\textbf{q})$). $\Delta B_{\nu}(\omega,\textbf{q})$ is obtained by taking the difference between the phonon spectral function with equilibrium matrix elements and the phonon spectral function with nonequilibrium matrix elements. \textbf{(c)} Energy- and momentum-resolved nonequilibrium renormalization of EPC matrix elements ($\Delta \left | g_{q, \nu}\right |$).  $\Delta \left | g_{q, \nu}\right |$ is obtained by taking the difference between equilibrium and nonequilibrium EPC matrix elements and is projected on the equilibrium phonon bands.}
\label{fig:phonons}
\end{center}
\end{figure}

To further distinguish between the effects coming from nonequilibrium modifications of the available electronic scattering phase space and EPC matrix elements, we computed the phonon spectral function $B_{\nu}(\mathbf{q},\omega)$ (via the phonon self-energy $\pi_{\nu}\propto |g_{\rm q\nu}|^2\chi_{0}(\mathbf{q},\omega)$, where $\chi_{0}(\mathbf{q},\omega)$ is electronic susceptibility -- see Sec.\,S6 for more details\,\cite{SM}), using two different approaches: (i) the DFPT approach, where we use equilibrium unscreened $|g_{\rm q\nu}|^2$ and nonequilibrium screened $\chi_{0}(\mathbf{q},\omega)$ and (ii) the cDFPT approach, where both $|g_{\rm q\nu}|^2$ and $\chi_{0}(\mathbf{q},\omega)$ are screened by the photoexcited nonequilibrium electron distribution. Comparing DFPT and cDFPT results allows us to analyze the nonequilibrium renormalization of EPC matrix elements [$\Delta \left | g_{q, \nu}\right |$, Fig.~\ref{fig:phonons}(c)], as well as its impact on the renormalization of the spectral function [$\Delta B_{\nu}(\mathbf{q},\omega)$, Fig.~\ref{fig:phonons}(b)]. In the differential spectral function map [$\Delta B_{\nu}(\mathbf{q},\omega)$, Fig.~\ref{fig:phonons}(b)] the negative (blue) differential signal is thus associated with enhanced spectral function using a DFPT approach, thus capturing the role of photoexcited phase space for electronic transitions, while the positive (red) differential signal originates from enhanced spectral function calculated using a cDFPT approach, thus yielding information about the dynamical renormalization of the EPC matrix elements.

In Fig.~\ref{fig:phonons}(b), the primary effect of nonequilibrium phase-space modifications is the softening of strongly coupled long-wavelength optical phonon modes (blue negative signal below 30 meV near $\Gamma$). Conversely, the dominant spectral function renormalization driven by modifications in the EPC matrix elements appears in the acoustic phonons near the $\mathrm{M}$ and $\Sigma$ high-symmetry points (red positive signal below 10 meV). The nonequilibrium decrease in EPC strength around the Brillouin zone center (mainly optical phonons), along with its enhancement near the Brillouin zone edge (mainly acoustic phonons), is further confirmed by the explicit subtraction of the nonequilibrium-screened from the equilibrium-screened $|g_{\rm q\nu}|^2$ [see Fig.~\ref{fig:phonons}(c)]. Upon photoexcitation, the reduction in EPC strength for optical phonons is compensated by a significant increase in coupling for acoustic modes, leading to an overall nonequilibrium enhancement of the EPC strength. This mechanism underlies the observed band gap renormalization and shorter lifetimes with increasing photo-doping, as seen in experiments.

In equilibrium, strong coupling to optical phonon branches near $\Gamma$ is a known property of undoped TMDCs~\cite{sohier2019}. 
Moreover, in some cases, static carrier doping can promote coupling to the acoustic phonons~\cite{cui2023,Zhao2018}, their subsequent softening, and the occurrence of dynamical instabilities~\cite{BinSubhan2021,Girotto2023b,marini2021b}. Here, we observe a similar behavior for the case of nonequilibrium photodoping. Indeed, the increase of EPC matrix elements ($\Delta |g_{\rm q\nu}|$) for acoustic phonon modes (near 10~meV) around M and $\Sigma$ points (Fig.~\ref{fig:phonons} (c) and Sec.\,S4~\cite{SM}) and their subsequent softening and heating highlight their important role in the nonequilibrium electron-phonon physics of photoexcited 2H-MoTe$_2$. Similar observations were reported in time-resolved electron diffraction studies of 2H-WSe$_2$~\cite{waldecker17} and 2H-MoTe$_2$~\cite{Krishnamoorthy2019}, where a significant build-up of phonon population at the M and $\Sigma$ points was observed.

The effects of screening were also investigated from first-principles in semiconducting 2H-MoS$_2$ and were found to be crucial for describing the increase of ultrafast electron diffuse scattering signals at K, M, and $\Sigma$ points~\cite{Britt2022} and the reduction of the main Bragg peaks' intensity~\cite{pan2025}. Not only do we confirm screening-induced reduction of EPC strengths for long-wavelength optical phonons\,\cite{pan2025}, but we also find significant photo-induced renormalization of the acoustic and optical modes coupling at finite $\mathbf{q}$ high-symmetry points (see also Fig.\,S3~\cite{SM}), leading to the increase of the total EPC strength (see Fig.\,S4~\cite{SM}) exemplified in the measured transient band gap and carrier lifetime decrease. We claim that photodoping, similarly to conventional doping in TMDCs\,\cite{sohier2019}, is responsible for the redistribution of EPC matrix elements and is crucial to take into account in an \emph{ab initio} self-consistent and fully momentum-resolved manner when describing carrier relaxation and phonon dynamics.

In conclusion, we have demonstrated that excitation density plays a critical role in enhancing EPC in photoexcited 2H-MoTe$_2$. By combining time- and angle-resolved XUV photoemission spectroscopy with cDFPT, we provided direct evidence of band gap renormalizations and modified carrier lifetimes as a function of photoexcited carrier density. Our results reveal that increasing carrier densities enhances electron–phonon scattering rates, leading to a pronounced renormalization of phonon energies and linewidths. This effect is driven by modifications of EPC matrix elements, as confirmed by our theoretical calculations. Our results thus resolve the debate over the dominant microscopic mechanism governing ultrafast electron-phonon dynamics (modified scattering phase space versus photoinduced alteration of EPC matrix elements) in photoexcited semiconducting TMDCs.

\hspace{1cm}

\begin{acknowledgments}
\textbf{Acknowledgments} We thank Nikita Fedorov, Romain Delos, Pierre Hericourt, Rodrigue Bouillaud, Laurent Merzeau, Frank Blais, and Yiming Pan for technical assistance. S.B. gratefully acknowledges Yann Mairesse and Baptiste Fabre for insightful and stimulating discussions. M.S. acknowledges support from SNSF Ambizione Grant No. PZ00P2-193527. We acknowledge the financial support of the IdEx University of Bordeaux/Grand Research Program "GPR LIGHT". We acknowledge support from ERC Starting Grant ERC-2022-STG No.101076639, Quantum Matter Bordeaux, AAP CNRS Tremplin and AAP SMR from Université de Bordeaux. S.\,F. acknowledges funding from the European Union’s Horizon Europe research and innovation programme under the Marie Skłodowska-Curie 2024 Postdoctoral Fellowship No 101198277 (TopQMat). Funded by the European Union. Views and opinions expressed are however those of the author(s) only and do not necessarily reflect those of the European Union. Neither the European Union nor the granting authority can be held responsible for them. N.\,G.\,E. and D.\,N. acknowledge financial support from the project "Podizanje znanstvene izvrsnosti Centra za napredne laserske tehnike (CALTboost)" financed by the European Union through the National Recovery and Resilience Plan 2021-2026 (NRPP). This work was supported by the Fonds de la Recherche Scientifique - FNRS under Grants number T.0183.23 (PDR) and T.W011.23 (PDR-WEAVE). Computational resources have been provided by the Consortium des Équipements de Calcul Intensif (CÉCI), funded by the FRS-FNRS under Grant No. 2.5020.11.

\end{acknowledgments}

\textbf{Data Availability Statement} The data that support the findings of this article are openly available on Zenodo 
 \url{https://doi.org/10.5281/zenodo.15791674}. 

\providecommand{\noopsort}[1]{}\providecommand{\singleletter}[1]{#1}%
\renewcommand{\thesection}{S\arabic{section}}  
\renewcommand{\thefigure}{S\arabic{figure}}
\renewcommand{\theequation}{S\arabic{equation}}
\renewcommand{\bibnumfmt}[1]{[S#1]}
\renewcommand{\citenumfont}[1]{S#1}

\setcounter{figure}{0}

\newpage

\begin{center}
\Large \textbf{\textsc{Supplemental Material}}\\    
\end{center}

\section{Experimental setup}\label{sec:exp}
The experimental setup revolves around a polarization-tunable ultrafast extreme ultraviolet (XUV) beamline~\cite{Comby22SM}. In essence, we employ a commercial high-repetition-rate Yb fiber laser (166~kHz, 1030~nm, 135~fs, 50~W, Amplitude Laser Group). For the (XUV) probe arm, a portion of the laser beam is frequency-doubled in a BBO crystal, generating 5~W of 515~nm pulses, which are then focused into a high-density argon gas jet in an annular beam geometry to drive high-order harmonic generation. To spatially separate the driver from the resulting XUV beamlet, we use spatial filtering using pinholes. The 9th harmonic (21.6eV) is spectrally selected through a combination of reflections from Sc/SiC multilayer XUV mirrors (NTTAT) and transmission through a 200~nm thick Sn metallic filter (Luxel). In the pump arm, a minor fraction of the fundamental laser pulse (1030~nm, 135~fs) is utilized. The IR pump and XUV probe pulses are collinearly recombined using a drilled mirror and then focused onto the sample, achieving typical spot sizes of 69 $\mu$m $\times$ 138 $\mu$m for the IR pump and 45 $\mu$m $\times$ 33 $\mu$m for the XUV probe, respectively. Bulk $\mathrm{2H}$-$\mathrm{MoTe_2}$ sample (HQ Graphene) is cleaved in an ultrahigh vacuum environment at a base pressure of 1$\times$10$^{-10}$ mbar before being introduced into a motorized 6-axis hexapod for precise sample alignment within the main chamber (base pressure: 2$\times$10$^{-10}$ mbar). Photoemission measurements are performed using a custom time-of-flight momentum microscope equipped with an advanced front lens offering multiple operational modes (GST mbH)~\cite{tkach24SM, tkach24-2SM}. This detector facilitates the simultaneous acquisition of the entire surface Brillouin zone across an extended binding energy range. For post-processing, we employ an open-source data workflow~\cite{Xian20SM, Xian19_2SM} to efficiently transform raw single-event datasets into calibrated, binned hypervolumes of the desired dimensions. More details about the experimental setup can be found in Ref.~\cite{Fragkos2025SM}. 

The photoexcited carrier densities (carriers/cm$^{-2}$ or simply cm$^{-2}$) are calculated using the in-situ pump laser fluence and the Beer-Lambert law using reported optical constants for 2H-MoTe$_2$~\cite{Munkhbat2022SM}.

\section{Experimental determination of direct and indirect band gaps}\label{sec:fit}

In Fig. 1(c) of the main text, at K point, the valence band exhibits a spin-orbit (SO) splitting of $\Delta_{SO}$ = 218 ± 18~meV. Upon photoexcitation, valence bands broaden and shift toward higher energies. Simultaneously, one can see the emergence of photoemission intensity from conduction bands at K and $\Sigma$ valleys, due to direct interband optical transition and subsequent intervalley scattering, respectively. Using nonequilibrium electronic band structure mapping at the pump-probe temporal overlap, we can fit the position of the valence band maximum (at K) and conduction bands at K and $\Sigma$ and extract the associated direct ($\Delta_g^d$) and indirect ($\Delta_g^i$) band gap values, as a function photoexcited carrier density (n). 

Indeed, the momentum-resolved valence band position and the associated spin-orbit (SO) splitting were determined by employing a multi-Gaussian fitting of energy distribution curves (EDCs) near K points. The momentum-resolved energy position of the two spin-orbit-splitted valence bands are superimposed to the associated photoemission intensity on the left subpanel of Fig.~1(b) of the main text. At K point, the valence band exhibits a spin-orbit splitting of $\Delta_{SO}$ = 218$\pm$18~meV. 

\begin{figure}[t]
\begin{center}
\includegraphics[width=8.6cm]{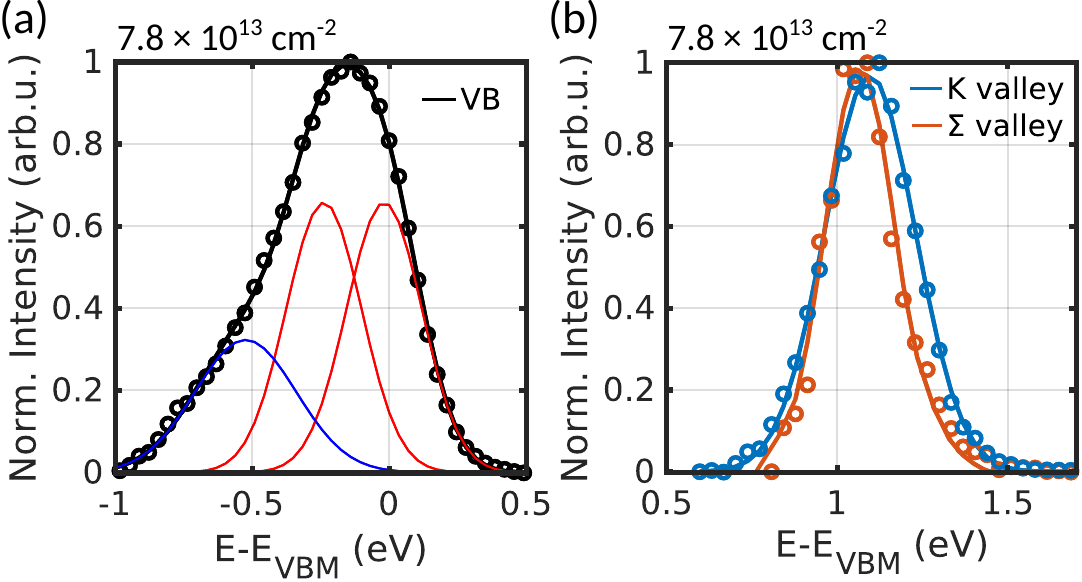}
\caption{\textbf{Direct and Indirect Band Gaps} \textbf{(a)} Background-free EDC of the valence band at K and associated fitting, using the procedure described in Section~\ref{sec:fit}. \textbf{(b)} EDCs of the conduction band pockets at K (blue) and $\Sigma$ (orange), and associated fitting, using the procedure described in Section~\ref{sec:fit}.}
\label{figS1}
\end{center}
\end{figure}

For the fluence-dependent measurements, the direct and indirect band gap values were determined by fitting the energy position of the valence band maximum at the K point and of the conduction band pockets at the K and $\Sigma$ points. Figure~\ref{figS1} presents the energy distribution curves (EDCs) and associated fits for the valence band at K (Fig.~\ref{figS1}(a)) and the conduction band pockets at K and $\Sigma$ (Fig.~\ref{figS1}(b)), for the highest photoexcitation carrier density (7.8$\times$10$^{13}$~cm$^{-2}$). To extract the energy position of the VB maximum, we first remove a Shirley background from the EDC and subsequently fit the background-free EDC by a function defined as a sum of three Gaussian, with the two topmost Gaussian constrained to have equal amplitude and width and separated in energy by the value of the spin-orbit splitting ($\Delta_{SO}$ = 218~meV, as determined from static scan measurements). The equation used to fit the background-free EDC of the valence band around the K point is thus given by:

\begin{equation}
\begin{split}
I_{EDC}(E) &= A \exp\left(-\frac{(E -E_{VBM})^2}{2\sigma^2}  \right) \\
&\quad + A \exp\left(-\frac{(E - (E_{VBM} - \Delta_{SO}))^2}{2\sigma^2}  \right) \\
&\quad + B \exp\left(-\frac{(E - \mu)^2}{2\sigma'^2} \right)
\end{split}
\end{equation}

For the conduction band pockets at K and $\Sigma$, following a Shirley background subtraction, a single Gaussian function was sufficient to determine their energy positions (Fig.~\ref{figS1}(b), here for the highest photoexcitation carrier density 7.8$\times$10$^{13}$~cm$^{-2}$). Performing the fitting for the minimum photoexcitation carrier density (2.7$\times$10$^{13}$~cm$^{-2}$) reveals a direct band gap of $\Delta_{g}^{d}$ = 1.19±0.01~eV at K and an indirect band gap of $\Delta_{g}^{i}$ = 1.14±0.01~eV at $\Sigma$ valley. This fitting procedure was used for all investigated pump fluences. 

\section{Valley-resolved electron population lifetimes}
\begin{figure}[t] \begin{center} \includegraphics[width=8.6cm]{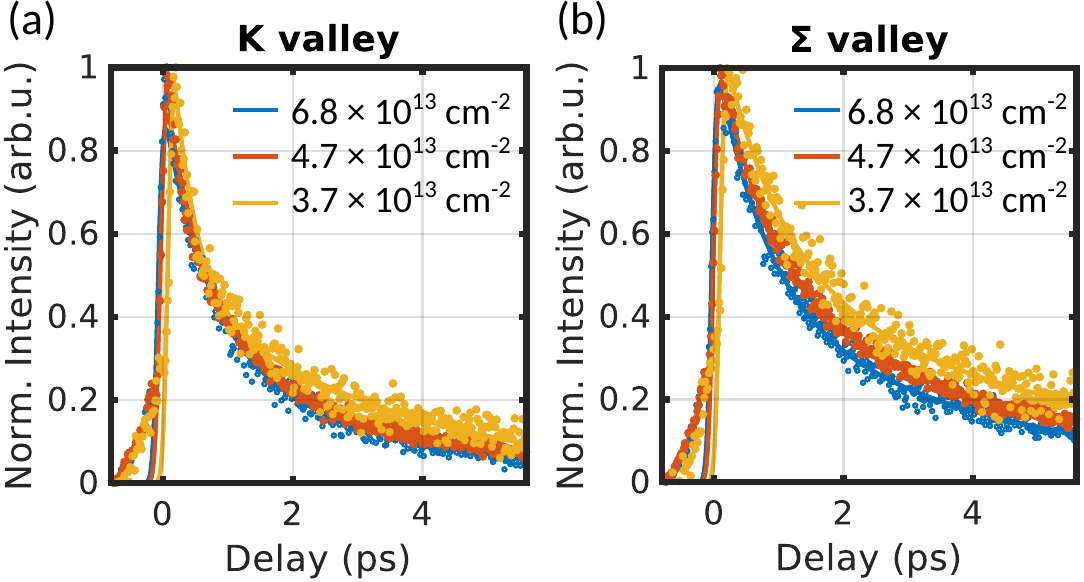} \caption{\textbf{Valley-Resolved Photoexcited Carrier Density-Dependent Electron Lifetimes.} Time-resolved population dynamics in \textbf{(a)} K and \textbf{(b)} $\Sigma$ conduction band valleys, for three intermediate photoexcited carrier densities (3.7$\times$10$^{13}$~cm$^{-2}$, 4.7$\times$10$^{13}$~cm$^{-2}$, and 6.8$\times$10$^{13}$~cm$^{-2}$) and associated fit using the procedure described in Section~\ref{sec:fit}.} 
\label{figS2} \end{center} \end{figure}

In Figure~\ref{figS2}, we show time-resolved population dynamics of photoexcited conduction band electrons in the K and $\Sigma$ valleys (Fig.~\ref{figS2}(a) and (b), respectively) for the the three intermediate photoexcited carrier densities (3.7$\times$10$^{13}$~cm$^{-2}$, 4.7$\times$10$^{13}$~cm$^{-2}$, and 6.8$\times$10$^{13}$~cm$^{-2}$) which are not presented in Fig.~3(b) of the main text. The valley-resolved electron populations, measured as a function of pump-probe delay, exhibit a rapid rise followed by a slower decay. To extract the valley-resolved electron lifetimes, we fitted the data using a double-exponential decay function multiplied by a step function that accounts for initial photoexcitation using the IR pump with a finite pulse duration. The fitting model is given by:

\begin{equation} 
\begin{split}
I_{exc}(\Delta t) = \frac{1}{2} \left(1 + \operatorname{erf} \left( \frac{\Delta t - t_0}{\sqrt{2} \sigma} \right) \right) \\ \times \left( A_1 e^{-\Delta t/\tau_1} + A_2 e^{-\Delta t/\tau_2} \right) 
\end{split}
\end{equation}

where $\tau_1$ and $\tau_2$ correspond to the fast and slow relaxation components, respectively. From the fits (solid lines), we observe that electron relaxation dynamics strongly depend on both valley pseudospin and photoexcited carrier density. The valley-resolved $\tau_1$ and $\tau_2$ for the five different investigated photoexcited carrier densities are shown in Fig.~3(c) and (d) of the manuscript. 

Note that the ultrafast population dynamics and their carrier density-dependence measured in our 2H-MoTe$_2$ single crystals differ significantly from those reported in previous time-resolved optical spectroscopic studies on 2H-MoTe$_2$ grown by chemical vapor deposition (CVD)~\cite{Chi2019SM}. In CVD-grown 2H-MoTe$_2$ films, the high concentration of defect states has been identified as the primary factor governing ultrafast dynamics, which is not prevalent in our single crystals.

\section{Details on EPC matrix elements}
\begin{figure}[t]
    \centering
    \includegraphics[width=\linewidth]{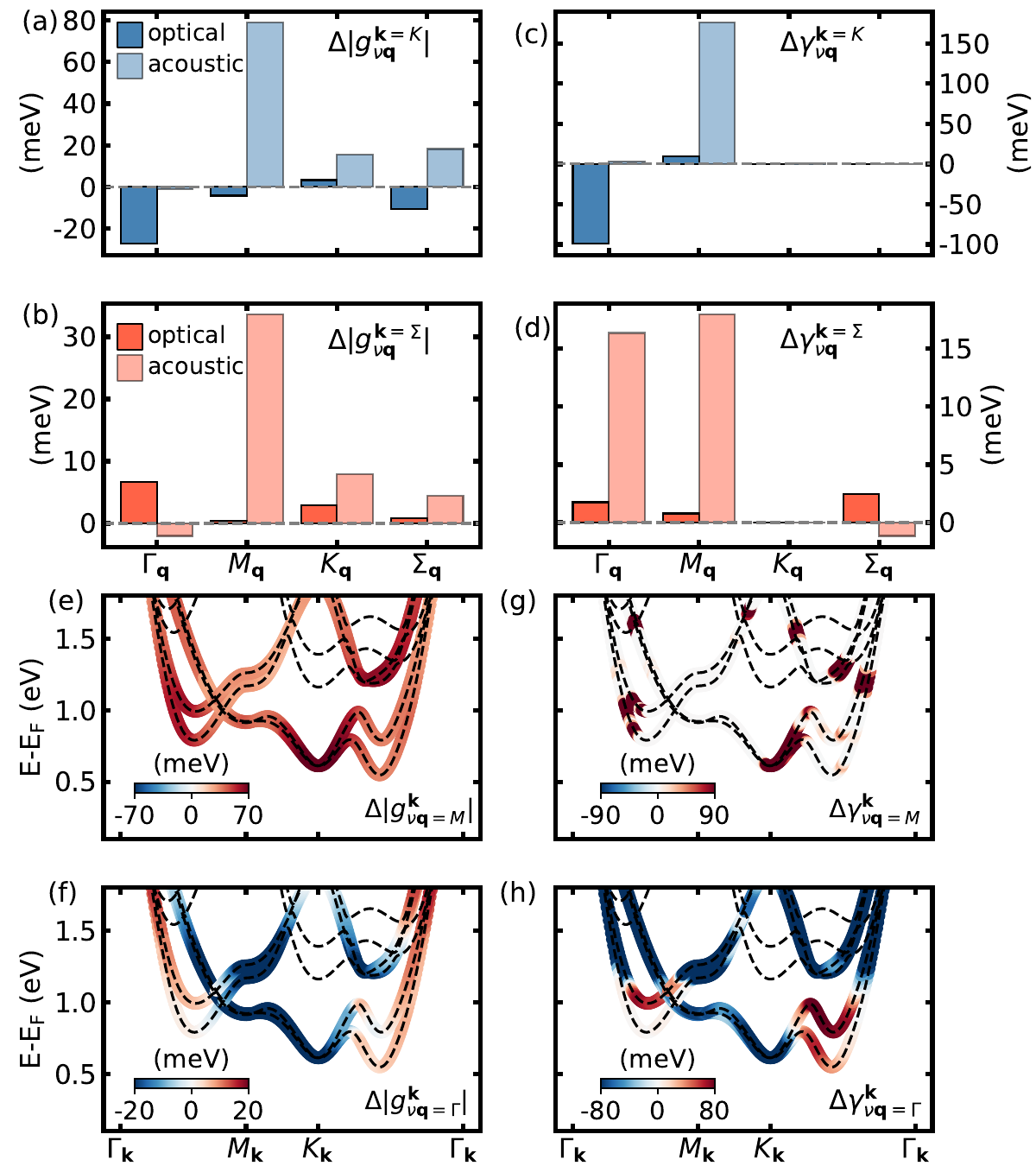}
    \caption{\textbf{Details on the Changes in the EPC Matrix Elements and Their Electron Linewidth Contributions.} The difference between photoexcited (cDFPT) and equilibrium (DFPT) EPC matrix elements at the high-symmetry $\mathbf{q}$ points at the bottom of the \textbf{(a)} K and \textbf{(b)} $\Sigma$ conduction band valleys summed for optical (bright color) and acoustic (faded color) modes separately. With the same resolution, we show the linewidth contribution at the bottom of \textbf{(c)} K and \textbf{(d)} $\Sigma$ valleys. In the lower left panels are the EPC matrix elements (electron linewidths) for acoustic and optical modes projected on the electron band structure evaluated for \textbf{(e)} [\textbf{(g)}]  $\mathbf{q}$ = M and \textbf{(f)} [\textbf{(h)}]  $\mathbf{q}$ = $\Gamma$.}
    \label{figS3}
\end{figure}

Figure~\ref{figS3} presents the full-resolution EPC matrix elements, highlighting the phonon modes and electronic states with the most significant photoinduced changes and quantifying their contributions to electron linewidths. In the first two panels [(a)-(b)], we subtract the DFPT matrix elements from their cDFPT counterparts at high-symmetry $\mathbf{q}$ points, revealing distinct photoinduced effects. We employ cDFPT with a carrier concentration of $7.85 \times 10^{13}$ cm$^{-2}$ as the starting point for these calculations.  

For $\mathbf{q}=\Gamma$, the most pronounced effect is a reduction in the optical phonon coupling to electronic states at the K valley bottom. While the suppression of long-wavelength optical phonon coupling has already been discussed in the main text, here we additionally provide details on the involved electronic states. A striking feature in panels (a) and (b) is the notable increase in matrix elements connecting $\mathbf{q}=\mathrm{M}$ point acoustic phonons with electronic states at the bottom of both conduction band valleys. This enhancement is particularly strong for the K valley, reaching values of 80\,meV.  

At $\mathbf{q}=\mathrm{K}$, we observe a slight increase in the coupling of both valleys to acoustic and optical phonons, in contrast to the reduction shown in Fig.~5(c) of the main text. This discrepancy arises from other electronic states, located away from the valley bottoms analyzed here.  

In the additional panels [(e) and (f)], we present the EPC matrix elements for $\mathbf{q}=\mathrm{M}$ and $\mathbf{q}=\Gamma$ projected onto the electronic band structure. For $\mathbf{q}=\mathrm{M}$, the photoinduced increase in matrix elements affects a broad range of conduction band states. In the case of $\mathbf{q}=\Gamma$, besides a strong reduction primarily affecting the Brillouin zone edge, we also observe a slightly weaker increase throughout the entire $\Sigma$ valley.

\begin{figure}[t]
    \centering
    \includegraphics[width=8.6cm]{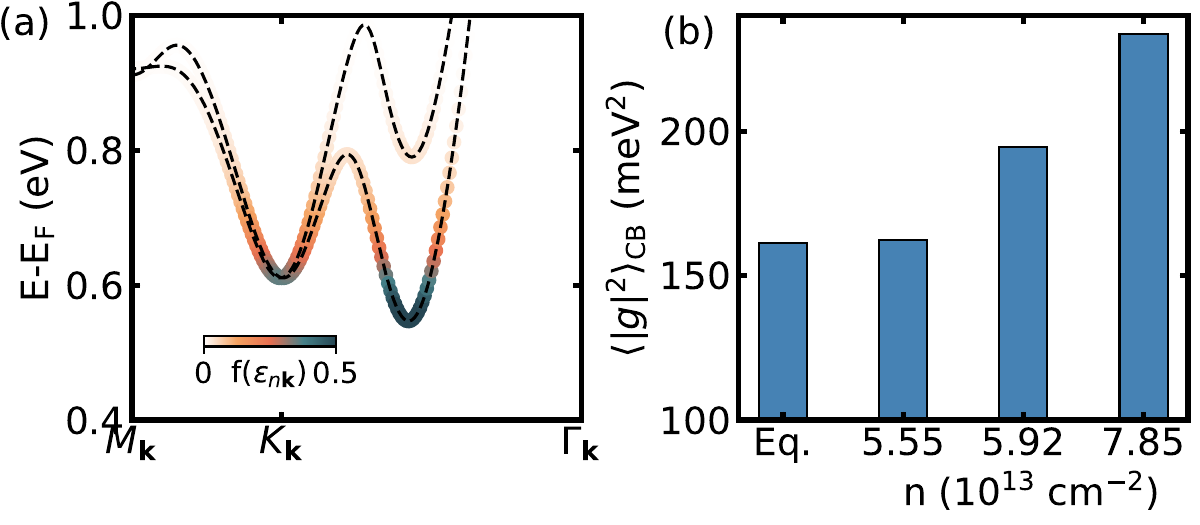}
    \caption{\textbf{Conduction Band Averaged Electron-Phonon Scattering Strength.} \textbf{(a)} The Fermi-Dirac electron distribution in the conduction band for the chemical potential of 0.57 eV and 800 K. \textbf{(b)} Conduction band averaged EPC strength calculated by restricting the available conduction band scattering phase space as in \textbf{(a)} and by using the matrix elements from a DFPT and various cDFPT calculations. }
    \label{figS4}
\end{figure}

\begin{figure}[t]
    \centering
    \includegraphics[width=8.6cm]{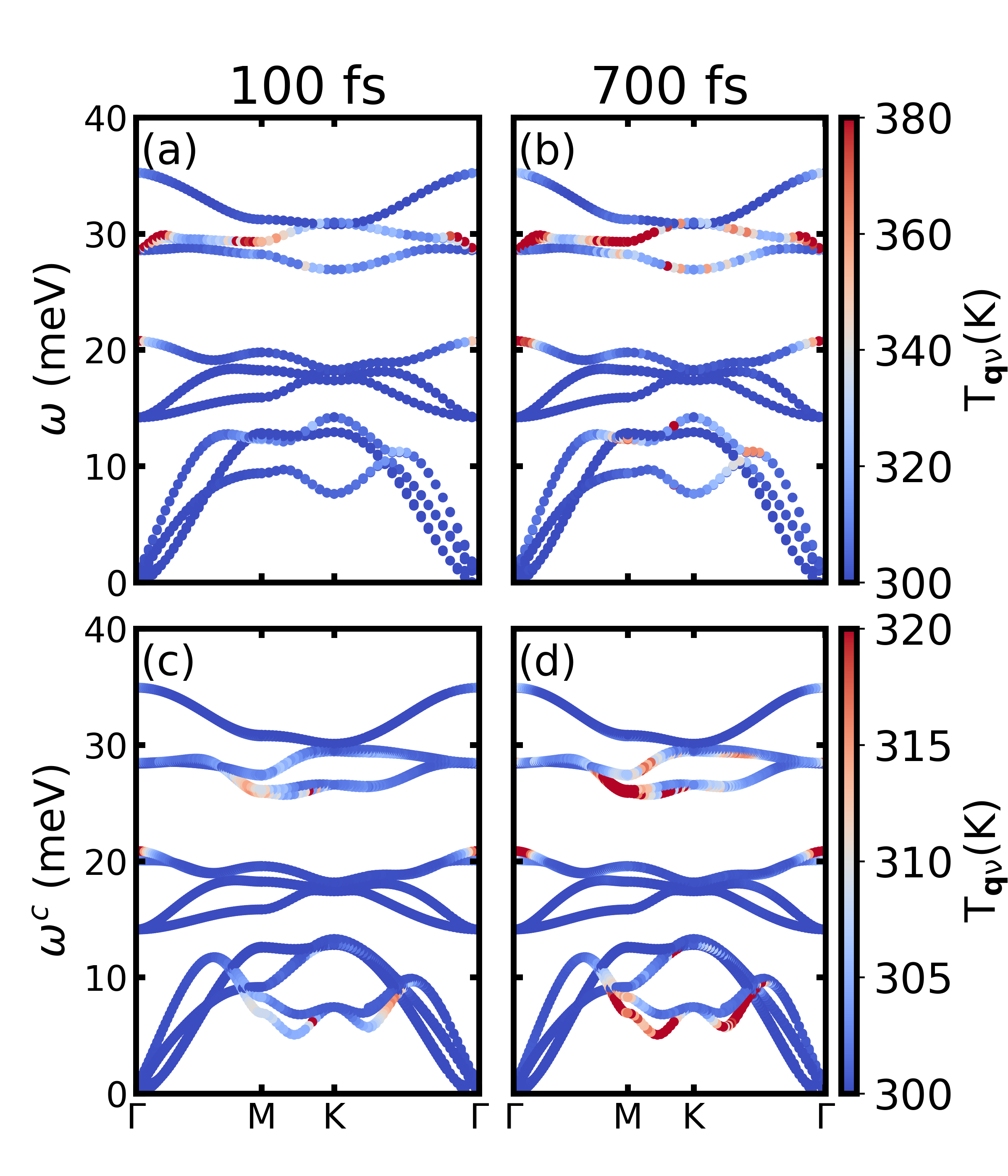}
    \caption{\textbf{Hot-Phonon Mode Redistribution Due to Changes in EPC.} Time-dependent Boltzmann equations (TDBE) results illustrating the time evolution of hot-phonon modes. \textbf{(a)} corresponds to the calculation using photoexcited matrix elements, which leads to the heating of optical phonon modes at $\mathbf{q}=$M and the soft acoustic phonon branch. In \textbf{(b)}, the equilibrium matrix elements are used, leading to excessive heating of the optical phonon branch predominantly for $\mathbf{q}=0$ and $\mathbf{q}=$M. }
    \label{figS5}
\end{figure}

In a similar manner, we analyze the electronic linewidth contributions from the high-symmetry $\mathbf{q}$ points for optical and acoustic modes separately in panels (c) and (d), and project them onto the electronic band structure in panels (g) and (h). Following the trends similar to those observed in the matrix elements, we find that the largest contribution to the electron linewidth at the K valley originates from $\mathbf{q}=\mathrm{M}$ point acoustic phonons, while the contribution from long-wavelength optical phonons significantly decreases. In the $\Sigma$ valley, the linewidth contributions are smaller and less affected by photoexcitation. Alongside the expected large contribution from $\mathbf{q}=\mathrm{M}$ point acoustic phonons, the long-wavelength optical modes appear to contribute equally. The linewidth increase arising from these long-wavelength modes is confined to the $\Sigma$ valley states [see panel (h)], whereas the photoinduced enhancement of contributions from $\mathbf{q}=\mathrm{M}$ point modes affects all electronic states connected by the $\mathbf{q}=\mathrm{M}$ vector.

In order to provide a straightforward measure of the photoinduced increase of the total EPC strength, we calculate the EPC matrix elements' conduction band average as 
\begin{equation}
\begin{aligned}
        &\langle |g^{(c)}|^2 \rangle_{\mathrm{CB}} =\frac{1}{\mathcal{N}}\\ &\sum_{\mathbf{q}\nu\mathbf{k} n_c m_c} |g^{(c)}_{\nu n_c m_c}(\mathbf{k},\mathbf{q})|^2 \delta(\varepsilon_{\mathbf{k}n_c}-\mu_c) \delta(\varepsilon_{\mathbf{k}+\mathbf{q} m_c}-\mu_c)
\end{aligned}
\end{equation}
where $n_c$ and $m_c$ are the conduction bands' indices and $\mu_c$ is the chemical potential set in the conduction band, in this case chosen to be 0.57 eV. This value of $\mu_c$ creates an electron distribution shown in Fig.~\ref{figS4} (a). The norm, $\mathcal{N}$ is calculated as $\sum_{\mathbf{q}\nu\mathbf{k} n_c m_c}\delta(\varepsilon_{\mathbf{k}n_c}-\mu_c) \delta(\varepsilon_{\mathbf{k}+\mathbf{q} m_c}-\mu_c)$. The average value of EPC matrix elements is shown in panel (b), showing a gradual increase with the density of photoexcited electrons.

\section{Hot Phonon Bottleneck Effect}

The discussed changes in the EPC matrix elements influence the distribution of the effective phonon temperature. To investigate this effect, we employ time-dependent Boltzmann equations (TDBE) to identify the modes exhibiting the most significant temperature changes, starting from an equilibrium phonon subsystem at room temperature.  In Fig.~\ref{figS5} we show the TDBE results obtained using two approaches; (i) starting from a DFPT phonon dispersion and (ii) starting from a cDFPT phonon dispersion containing the effects of phonon softening due to nonequilibrium screening of the matrix elements and electronic susceptibilty. In Fig.~\ref{figS5} we observe that solving the TDBE equations for a carrier concentration of $7.85 \times 10^{13}$ cm$^{-2}$ with equilibrium screened EPC matrix elements [panels (a) and (b)] leads to a rapid heating of long-wavelength optical modes immediately after photoexcitation. These modes eventually reach temperatures of about 380 K, while other modes remain largely unaffected. Alternatively, using cDFPT with a carrier concentration of $7.85 \times 10^{13}$ cm$^{-2}$ as a starting point results in a gradual and weak increase in the temperature of acoustic modes at the Brillouin zone edge [panels (c) and (d)], along with certain optical modes. Since the result in panels (c) and (d) corresponds to a high photoexcited carrier concentration, an experiment with gradually increasing fluence would transition the system from the DFPT to the cDFPT regime. At lower fluences, energy absorption primarily occurs in optical phonons, leading to their heating. At higher fluences, energy transfer to lower-frequency acoustic modes becomes more significant, potentially contributing to the observed second timescale in carrier relaxation.  This supports the hypothesis of a hot-phonon bottleneck effect, where the slowing of the electron-phonon cooling rate is associated with the accumulation of nonequilibrium phonons—particularly in optical modes at lower fluences and in acoustic modes at higher fluences.

\section{Computational details}
The first-order on-the-mass-shell electron self-energy can be computed using the expression\,\cite{Giustino2017SM}
\begin{equation}
\begin{aligned}
  &\Sigma_{n\mathbf{k}} = \sum_{m\mathbf{q}\nu} |g_{\nu nm}(\mathbf{k},\mathbf{q})|^2 \times \\
  & \Bigg(\frac{n_{\mathbf{q}\nu}+f_{m\mathbf{k}+\mathbf{q}}}
    {\varepsilon_{n\mathbf{k}} + i\eta+\omega_{\mathbf{q}\nu}-\varepsilon_{\mathbf{k}+\mathbf{q}}} 
    + \frac{n_{\mathbf{q}\nu}+1-f_{m\mathbf{k}+\mathbf{q}}}
    {\varepsilon_{n\mathbf{k}} + i\eta-\omega_{\mathbf{q}\nu}-\varepsilon_{m\mathbf{k}+\mathbf{q}}}\Bigg),
\end{aligned}
\end{equation}
where $\mathbf{k}$ and $n,m$ are electronic momentum and band indices and $\mathbf{q}$ and $\nu$ are phonon wavevector and band indices. $g_{\nu mn}(\mathbf{k},\mathbf{q})$ are equilibrium EPC matrix elements, $\varepsilon_{n\mathbf{k}}$ are Kohn-Sham energies, $f_{n\mathbf{k}}$ are Fermi-Dirac distribution functions, $n_{\mathbf{q}\nu}$ are Bose-Einstein functions, and $\omega_{\mathbf{q}\nu}$ are phonon frequencies from density functional perturbation theory (DFPT)~\cite{Baroni2001SM, Giannozzi2017SM}. 
Its imaginary part corresponds to the electron linewidth 
\begin{equation}
    \gamma_{n\mathbf{k}} = -2 \text{Im}\Sigma_{n\mathbf{k}}\ ,
\end{equation}
which in turn equals the inverse of single-electron lifetime. 
Together with the real part of the electron self-energy the electron spectral function can be constructed
\begin{equation}
    A_{n\mathbf{k}}(\varepsilon) =\frac{\mathrm{Im}\Sigma_{n\mathbf{k}}(\varepsilon)}{[\varepsilon-\varepsilon_{n\mathbf{k}}-\mathrm{Re}\Sigma_{n\mathbf{k}}(\varepsilon)+\mu]^2 + [\mathrm{Im}\Sigma_{n\mathbf{k}}(\varepsilon)]^2}.
\end{equation}
We use the electron spectral function calculations in order to simulate the band gap renormalizations in Fig.\,4(d) of the main text.
Similarly, phonon spectral function can be written as
\begin{equation}
B_{\nu}(\mathbf{q},\omega)=-\frac{1}{\pi}\mathrm{Im}\left[\frac{2\omega_{\mathbf{q}\nu}}{\omega^2-(\omega_{\mathbf{q}\nu})^2-2\omega_{\mathbf{q}\nu}\pi_{\nu}(\mathbf{q},\omega)} \right],
\label{eq:ph_spec}
\end{equation}
where $\pi_{\nu}(\mathbf{q},\omega)$ is the phonon self-energy due to electron-phonon coupling
\begin{equation}
    \pi_{\nu}(\mathbf{q},\omega) = \sum_{nm\mathbf{k}}|g_{\nu nm}(\mathbf{k},\mathbf{q})|^2\frac{f_{n\mathbf{k}} - f_{m\mathbf{k}+\mathbf{q}}}{\varepsilon_{n\mathbf{k}} -\varepsilon_{m\mathbf{k}+\mathbf{q}}+\omega + i\eta}.
    \label{eq:pi_dyn}
\end{equation}
All the above-written equations, including the electron and phonon eigenstates, are calculated from first-principles.  DFT and DFPT~\cite{Baroni2001SM} calculations were done using \textsc{Quantum ESPRESSO}~\cite{Giannozzi2017SM}. EPC properties were calculated by Wannier interpolation~\cite{wan90SM} of EPC matrix elements as implemented in the EPW code~\cite{giustino07SM, Noffsinger2010SM, Ponce2016SM}. 

All these quantities, such as EPC matrix elements, phonon frequencies, and relaxation rates, represent the standard equilibrium values, for the system where the electronic subsystem is described with the Fermi-Dirac distribution function (i.e., all the valence bands are occupied, while conduction bands are empty).

In order to simulate the nonequilibrium distribution of electrons and study the impact of this distribution on electron and phonon properties, we perform constrained density functional perturbation theory (cDFPT)\,\cite{Murray2007SM,liu2022SM,Girotto2023SM} calculations, where we constrain the electronic transitions contributing to the screening of the EPC matrix elements and DFPT frequencies. Specifically, when performing the self-consistent DFPT calculations instead of using a single Fermi-Dirac distribution, we impose two of them, with separate chemical potentials. One distribution describes the empty hole states in the valence band, while the other occupies states at the bottom of the conduction band. By performing cDFPT with nonequilibrium distribution we obtain new, additionally screened EPC matrix elements $g^{(c)}_{\nu mn}(\mathbf{k},\mathbf{q})$ and phonon frequencies $\omega_{\mathbf{q}\nu}^{(c)}$. We have implemented cDFPT in the \textsc{Quantum ESPRESSO} software in the routine which solves the Sternheimer equation\,\cite{Girotto2023SM}. With cDFPT as a starting point, we again perform calculations of EPC properties (within EPW), which then include the changes in the screening due to nonequilibrium distribution. With this we obtain the constrained electron and phonon self-energies $\Sigma_{n\mathbf{k}}^{(c)}$ and $\pi_{\nu}^{(c)}(\mathbf{q,\omega})$, and the corresponding spectral functions, where we use modified nonequilibrium occupation functions $f^{(c)}_{n\mathbf{k}}$ as described above. Namely, 
\begin{equation}
\begin{aligned}
  &\Sigma^{(c)}_{n\mathbf{k}} = \sum_{m\mathbf{q}\nu} |g^{(c)}_{\nu nm}(\mathbf{k},\mathbf{q})|^2 \times \\
  & \Bigg(\frac{n_{\mathbf{q}\nu}+f^{(c)}_{m\mathbf{k}+\mathbf{q}}}
    {\varepsilon_{n\mathbf{k}} + i\eta+\omega^{(c)}_{\mathbf{q}\nu}-\varepsilon_{\mathbf{k}+\mathbf{q}}} 
    + \frac{n_{\mathbf{q}\nu}+1-f^{(c)}_{m\mathbf{k}+\mathbf{q}}}
    {\varepsilon_{n\mathbf{k}} + i\eta-\omega^{(c)}_{\mathbf{q}\nu}-\varepsilon_{m\mathbf{k}+\mathbf{q}}}\Bigg),
\end{aligned}
\end{equation}
and
\begin{equation}
    \pi^{(c)}_{\nu}(\mathbf{q},\omega) = \sum_{nm\mathbf{k}}|g^{(c)}_{\nu nm}(\mathbf{k},\mathbf{q})|^2\frac{f^{(c)}_{n\mathbf{k}} - f^{(c)}_{m\mathbf{k}+\mathbf{q}}}{\varepsilon_{n\mathbf{k}} -\varepsilon_{m\mathbf{k}+\mathbf{q}}+\omega + i\eta}.
    \label{eq:pi_dync}
\end{equation}
With these equations one can analyze whether the nonequilibrium-induced modifications in the relaxation rates, electron energies (e.g., band gap), and phonon frequencies come dominantly from the changes in the scattering phase space (i.e., $f^{(c)}_{n\mathbf{k}}$), or from the modifications in the EPC matrix element strength $|g^{(c)}_{\nu nm}(\mathbf{k},\mathbf{q})|^2$\,\cite{Girotto2023SM}. For instance, the band gap reduction due to nonequilibrium distribution via $\mathrm{Re}\,\Sigma^{(c)}_{n\mathbf{k}}$ is only due to $|g^{(c)}_{\nu nm}(\mathbf{k},\mathbf{q})|^2$, while no band gap variations with the increase of photo-doped carriers is obtained when we use equilibrium $|g_{\nu nm}(\mathbf{k},\mathbf{q})|^2$ and $f^{(c)}_{n\mathbf{k}}$. For the phonons, using equilibrium $|g_{\nu nm}(\mathbf{k},\mathbf{q})|^2$ and $f^{(c)}_{n\mathbf{k}}$ induces optical phonon softening near $\Gamma$ and softening of the acoustic branches near $\mathrm{M}$ and between $\mathrm{K}$ and $\mathrm{\Sigma}$. The former effect vanishes once $|g_{\nu nm}(\mathbf{k},\mathbf{q})|^2$ are exchanged with $|g^{(c)}_{\nu nm}(\mathbf{k},\mathbf{q})|^2$, while the latter effect is largely enhanced and the softening region is broadened in the $\mathbf{q}$ space.

We use the fully-relativistic norm-conserving Perdew-Burke-Ernzerhof pseudopotentials from \textsc{Pseudo Dojo}~\cite{vanSetten2018SM} with a kinetic energy cutoff of 100 Ry. The relaxed lattice constant for 2H-MoTe$_2$ bilayer is found to be  3.530\,\AA, while the Mo-Mo inter-layer distance is 7.693\,\AA~and the neighboring sheets are separated by 15\,\AA~of vacuum. The self-consistent electron density calculation is done on a $24 \times 24\times 1$ k-point grid with Fermi-Dirac smearing of 0.005 Ry, corresponding to 800 K, and the phonon calculation on a $6 \times 6 \times 1$ q-point grid. In EPW, we use 22 maximally localized Wannier functions~\cite{Marzari2012SM} with the initial projections of d-orbitals on the Mo sites and p-orbitals on the Te atom sites. The fine sampling of the Brillouin zone for the electron-phonon interpolation depends on the calculation. To evaluate the expressions which require a q-grid summation, the q-grid is set to $60 \times 60\times 1$, while for the phonon spectral function the k-grid summation is done on a $120 \times 120\times 1$ grid. Smearing in the EPW calculation is set to 40\,meV. The electron linewidths are obtained at 300 K, while the phonon spectral functions are calculated at 800 K.
For TDBE simulations presented in Fig.~\ref{figS3} we used the EPW implementation obtained on request from Yiming Pan. The TDBE implementation uses the fine-grid interpolated matrix elements and then solves the TDBE by including electron-phonon scattering as a relaxation mechanism. The input parameters determine the carrier concentration through the elevated electronic temperature (T$_{el}$) and by specifying two chemical potentials. We use T$_{el}$=1500 K, while the phonon subsystem is at 300 K. We calculate the time-propagation only in the conduction band and set the conduction band chemical potential to 20 meV below the conduction band bottom leading to the net concentration of carriers used in our cDFPT calculation. We run the simulation for 1 ps and perform a fine grid summation on $60 \times 60\times 1$ k- and q-grids. 

\providecommand{\noopsort}[1]{}\providecommand{\singleletter}[1]{#1}%


\begin{thebibliography}{78}%
\makeatletter
\providecommand \@ifxundefined [1]{%
 \@ifx{#1\undefined}
}%
\providecommand \@ifnum [1]{%
 \ifnum #1\expandafter \@firstoftwo
 \else \expandafter \@secondoftwo
 \fi
}%
\providecommand \@ifx [1]{%
 \ifx #1\expandafter \@firstoftwo
 \else \expandafter \@secondoftwo
 \fi
}%
\providecommand \natexlab [1]{#1}%
\providecommand \enquote  [1]{``#1''}%
\providecommand \bibnamefont  [1]{#1}%
\providecommand \bibfnamefont [1]{#1}%
\providecommand \citenamefont [1]{#1}%
\providecommand \href@noop [0]{\@secondoftwo}%
\providecommand \href [0]{\begingroup \@sanitize@url \@href}%
\providecommand \@href[1]{\@@startlink{#1}\@@href}%
\providecommand \@@href[1]{\endgroup#1\@@endlink}%
\providecommand \@sanitize@url [0]{\catcode `\\12\catcode `\$12\catcode `\&12\catcode `\#12\catcode `\^12\catcode `\_12\catcode `\%12\relax}%
\providecommand \@@startlink[1]{}%
\providecommand \@@endlink[0]{}%
\providecommand \url  [0]{\begingroup\@sanitize@url \@url }%
\providecommand \@url [1]{\endgroup\@href {#1}{\urlprefix }}%
\providecommand \urlprefix  [0]{URL }%
\providecommand \Eprint [0]{\href }%
\providecommand \doibase [0]{http://dx.doi.org/}%
\providecommand \selectlanguage [0]{\@gobble}%
\providecommand \bibinfo  [0]{\@secondoftwo}%
\providecommand \bibfield  [0]{\@secondoftwo}%
\providecommand \translation [1]{[#1]}%
\providecommand \BibitemOpen [0]{}%
\providecommand \bibitemStop [0]{}%
\providecommand \bibitemNoStop [0]{.\EOS\space}%
\providecommand \EOS [0]{\spacefactor3000\relax}%
\providecommand \BibitemShut  [1]{\csname bibitem#1\endcsname}%
\let\auto@bib@innerbib\@empty
\bibitem [{\citenamefont {Giustino}(2017)}]{Giustino2017}%
  \BibitemOpen
  \bibfield  {author} {\bibinfo {author} {\bibfnamefont {F.}~\bibnamefont {Giustino}},\ }\href {\doibase 10.1103/RevModPhys.89.015003} {\bibfield  {journal} {\bibinfo  {journal} {Rev. Mod. Phys.}\ }\textbf {\bibinfo {volume} {89}},\ \bibinfo {pages} {015003} (\bibinfo {year} {2017})}\BibitemShut {NoStop}%
\bibitem [{\citenamefont {Zhu}\ \emph {et~al.}(2015)\citenamefont {Zhu}, \citenamefont {Cao}, \citenamefont {Zhang}, \citenamefont {Plummer},\ and\ \citenamefont {Guo}}]{zhu2015}%
  \BibitemOpen
  \bibfield  {author} {\bibinfo {author} {\bibfnamefont {X.}~\bibnamefont {Zhu}}, \bibinfo {author} {\bibfnamefont {Y.}~\bibnamefont {Cao}}, \bibinfo {author} {\bibfnamefont {J.}~\bibnamefont {Zhang}}, \bibinfo {author} {\bibfnamefont {E.~W.}\ \bibnamefont {Plummer}}, \ and\ \bibinfo {author} {\bibfnamefont {J.}~\bibnamefont {Guo}},\ }\href {\doibase 10.1073/pnas.1424791112} {\bibfield  {journal} {\bibinfo  {journal} {Proceedings of the National Academy of Sciences}\ }\textbf {\bibinfo {volume} {112}},\ \bibinfo {pages} {2367} (\bibinfo {year} {2015})}\BibitemShut {NoStop}%
\bibitem [{\citenamefont {Otto}\ \emph {et~al.}(2021)\citenamefont {Otto}, \citenamefont {Pöhls}, \citenamefont {de~Cotret}, \citenamefont {Stern}, \citenamefont {Sutton},\ and\ \citenamefont {Siwick}}]{otto21}%
  \BibitemOpen
  \bibfield  {author} {\bibinfo {author} {\bibfnamefont {M.~R.}\ \bibnamefont {Otto}}, \bibinfo {author} {\bibfnamefont {J.-H.}\ \bibnamefont {Pöhls}}, \bibinfo {author} {\bibfnamefont {L.~P.~R.}\ \bibnamefont {de~Cotret}}, \bibinfo {author} {\bibfnamefont {M.~J.}\ \bibnamefont {Stern}}, \bibinfo {author} {\bibfnamefont {M.}~\bibnamefont {Sutton}}, \ and\ \bibinfo {author} {\bibfnamefont {B.~J.}\ \bibnamefont {Siwick}},\ }\href {\doibase 10.1126/sciadv.abf2810} {\bibfield  {journal} {\bibinfo  {journal} {Science Advances}\ }\textbf {\bibinfo {volume} {7}},\ \bibinfo {pages} {eabf2810} (\bibinfo {year} {2021})}\BibitemShut {NoStop}%
\bibitem [{\citenamefont {Bin~Subhan}\ \emph {et~al.}(2021)\citenamefont {Bin~Subhan}, \citenamefont {Suleman}, \citenamefont {Moore}, \citenamefont {Phu}, \citenamefont {Hoesch}, \citenamefont {Kurebayashi}, \citenamefont {Howard},\ and\ \citenamefont {Schofield}}]{BinSubhan2021}%
  \BibitemOpen
  \bibfield  {author} {\bibinfo {author} {\bibfnamefont {M.~K.}\ \bibnamefont {Bin~Subhan}}, \bibinfo {author} {\bibfnamefont {A.}~\bibnamefont {Suleman}}, \bibinfo {author} {\bibfnamefont {G.}~\bibnamefont {Moore}}, \bibinfo {author} {\bibfnamefont {P.}~\bibnamefont {Phu}}, \bibinfo {author} {\bibfnamefont {M.}~\bibnamefont {Hoesch}}, \bibinfo {author} {\bibfnamefont {H.}~\bibnamefont {Kurebayashi}}, \bibinfo {author} {\bibfnamefont {C.~A.}\ \bibnamefont {Howard}}, \ and\ \bibinfo {author} {\bibfnamefont {S.~R.}\ \bibnamefont {Schofield}},\ }\href {\doibase 10.1021/acs.nanolett.1c00677} {\bibfield  {journal} {\bibinfo  {journal} {Nano Letters}\ }\textbf {\bibinfo {volume} {21}},\ \bibinfo {pages} {5516–5521} (\bibinfo {year} {2021})}\BibitemShut {NoStop}%
\bibitem [{\citenamefont {Kang}\ \emph {et~al.}(2018)\citenamefont {Kang}, \citenamefont {Jung}, \citenamefont {Shin}, \citenamefont {Sohn}, \citenamefont {Ryu}, \citenamefont {Kim}, \citenamefont {Hoesch},\ and\ \citenamefont {Kim}}]{kang2018}%
  \BibitemOpen
  \bibfield  {author} {\bibinfo {author} {\bibfnamefont {M.}~\bibnamefont {Kang}}, \bibinfo {author} {\bibfnamefont {S.~W.}\ \bibnamefont {Jung}}, \bibinfo {author} {\bibfnamefont {W.}~\bibnamefont {Shin}}, \bibinfo {author} {\bibfnamefont {Y.}~\bibnamefont {Sohn}}, \bibinfo {author} {\bibfnamefont {S.~H.}\ \bibnamefont {Ryu}}, \bibinfo {author} {\bibfnamefont {T.}~\bibnamefont {Kim}}, \bibinfo {author} {\bibfnamefont {M.}~\bibnamefont {Hoesch}}, \ and\ \bibinfo {author} {\bibfnamefont {K.}~\bibnamefont {Kim}},\ }\href {\doibase 10.1038/s41563-018-0092-7} {\bibfield  {journal} {\bibinfo  {journal} {Nature Materials}\ }\textbf {\bibinfo {volume} {17}},\ \bibinfo {pages} {676} (\bibinfo {year} {2018})}\BibitemShut {NoStop}%
\bibitem [{\citenamefont {Jung}\ \emph {et~al.}(2024)\citenamefont {Jung}, \citenamefont {Watson}, \citenamefont {Mukherjee}, \citenamefont {Evtushinsky}, \citenamefont {Cacho}, \citenamefont {Martino}, \citenamefont {Berger},\ and\ \citenamefont {Kim}}]{Jung2024}%
  \BibitemOpen
  \bibfield  {author} {\bibinfo {author} {\bibfnamefont {S.~W.}\ \bibnamefont {Jung}}, \bibinfo {author} {\bibfnamefont {M.~D.}\ \bibnamefont {Watson}}, \bibinfo {author} {\bibfnamefont {S.}~\bibnamefont {Mukherjee}}, \bibinfo {author} {\bibfnamefont {D.~V.}\ \bibnamefont {Evtushinsky}}, \bibinfo {author} {\bibfnamefont {C.}~\bibnamefont {Cacho}}, \bibinfo {author} {\bibfnamefont {E.}~\bibnamefont {Martino}}, \bibinfo {author} {\bibfnamefont {H.}~\bibnamefont {Berger}}, \ and\ \bibinfo {author} {\bibfnamefont {T.~K.}\ \bibnamefont {Kim}},\ }\href {\doibase 10.1021/acsnano.4c07805} {\bibfield  {journal} {\bibinfo  {journal} {ACS Nano}\ }\textbf {\bibinfo {volume} {18}},\ \bibinfo {pages} {33359} (\bibinfo {year} {2024})}\BibitemShut {NoStop}%
\bibitem [{\citenamefont {Costanzo}\ \emph {et~al.}(2015)\citenamefont {Costanzo}, \citenamefont {Jo}, \citenamefont {Berger},\ and\ \citenamefont {Morpurgo}}]{Costanzo2015}%
  \BibitemOpen
  \bibfield  {author} {\bibinfo {author} {\bibfnamefont {D.}~\bibnamefont {Costanzo}}, \bibinfo {author} {\bibfnamefont {S.}~\bibnamefont {Jo}}, \bibinfo {author} {\bibfnamefont {H.}~\bibnamefont {Berger}}, \ and\ \bibinfo {author} {\bibfnamefont {A.~F.}\ \bibnamefont {Morpurgo}},\ }\href {https://api.semanticscholar.org/CorpusID:205453752} {\bibfield  {journal} {\bibinfo  {journal} {Nature nanotechnology}\ }\textbf {\bibinfo {volume} {11 4}},\ \bibinfo {pages} {339} (\bibinfo {year} {2015})}\BibitemShut {NoStop}%
\bibitem [{\citenamefont {Weber}\ \emph {et~al.}(2011)\citenamefont {Weber}, \citenamefont {Rosenkranz}, \citenamefont {Castellan}, \citenamefont {Osborn}, \citenamefont {Hott}, \citenamefont {Heid}, \citenamefont {Bohnen}, \citenamefont {Egami}, \citenamefont {Said},\ and\ \citenamefont {Reznik}}]{weber2011}%
  \BibitemOpen
  \bibfield  {author} {\bibinfo {author} {\bibfnamefont {F.}~\bibnamefont {Weber}}, \bibinfo {author} {\bibfnamefont {S.}~\bibnamefont {Rosenkranz}}, \bibinfo {author} {\bibfnamefont {J.-P.}\ \bibnamefont {Castellan}}, \bibinfo {author} {\bibfnamefont {R.}~\bibnamefont {Osborn}}, \bibinfo {author} {\bibfnamefont {R.}~\bibnamefont {Hott}}, \bibinfo {author} {\bibfnamefont {R.}~\bibnamefont {Heid}}, \bibinfo {author} {\bibfnamefont {K.-P.}\ \bibnamefont {Bohnen}}, \bibinfo {author} {\bibfnamefont {T.}~\bibnamefont {Egami}}, \bibinfo {author} {\bibfnamefont {A.~H.}\ \bibnamefont {Said}}, \ and\ \bibinfo {author} {\bibfnamefont {D.}~\bibnamefont {Reznik}},\ }\href {\doibase 10.1103/PhysRevLett.107.107403} {\bibfield  {journal} {\bibinfo  {journal} {Phys. Rev. Lett.}\ }\textbf {\bibinfo {volume} {107}},\ \bibinfo {pages} {107403} (\bibinfo {year} {2011})}\BibitemShut {NoStop}%
\bibitem [{\citenamefont {Varma}\ and\ \citenamefont {Weber}(1977)}]{varma1977}%
  \BibitemOpen
  \bibfield  {author} {\bibinfo {author} {\bibfnamefont {C.~M.}\ \bibnamefont {Varma}}\ and\ \bibinfo {author} {\bibfnamefont {W.}~\bibnamefont {Weber}},\ }\href {\doibase 10.1103/PhysRevLett.39.1094} {\bibfield  {journal} {\bibinfo  {journal} {Phys. Rev. Lett.}\ }\textbf {\bibinfo {volume} {39}},\ \bibinfo {pages} {1094} (\bibinfo {year} {1977})}\BibitemShut {NoStop}%
\bibitem [{\citenamefont {Weber}\ \emph {et~al.}(2018)\citenamefont {Weber}, \citenamefont {Hott}, \citenamefont {Heid}, \citenamefont {Lev}, \citenamefont {Caputo}, \citenamefont {Schmitt},\ and\ \citenamefont {Strocov}}]{weber2018}%
  \BibitemOpen
  \bibfield  {author} {\bibinfo {author} {\bibfnamefont {F.}~\bibnamefont {Weber}}, \bibinfo {author} {\bibfnamefont {R.}~\bibnamefont {Hott}}, \bibinfo {author} {\bibfnamefont {R.}~\bibnamefont {Heid}}, \bibinfo {author} {\bibfnamefont {L.~L.}\ \bibnamefont {Lev}}, \bibinfo {author} {\bibfnamefont {M.}~\bibnamefont {Caputo}}, \bibinfo {author} {\bibfnamefont {T.}~\bibnamefont {Schmitt}}, \ and\ \bibinfo {author} {\bibfnamefont {V.~N.}\ \bibnamefont {Strocov}},\ }\href {\doibase 10.1103/PhysRevB.97.235122} {\bibfield  {journal} {\bibinfo  {journal} {Phys. Rev. B}\ }\textbf {\bibinfo {volume} {97}},\ \bibinfo {pages} {235122} (\bibinfo {year} {2018})}\BibitemShut {NoStop}%
\bibitem [{\citenamefont {Sohier}\ \emph {et~al.}(2019)\citenamefont {Sohier}, \citenamefont {Ponomarev}, \citenamefont {Gibertini}, \citenamefont {Berger}, \citenamefont {Marzari}, \citenamefont {Ubrig},\ and\ \citenamefont {Morpurgo}}]{sohier2019}%
  \BibitemOpen
  \bibfield  {author} {\bibinfo {author} {\bibfnamefont {T.}~\bibnamefont {Sohier}}, \bibinfo {author} {\bibfnamefont {E.}~\bibnamefont {Ponomarev}}, \bibinfo {author} {\bibfnamefont {M.}~\bibnamefont {Gibertini}}, \bibinfo {author} {\bibfnamefont {H.}~\bibnamefont {Berger}}, \bibinfo {author} {\bibfnamefont {N.}~\bibnamefont {Marzari}}, \bibinfo {author} {\bibfnamefont {N.}~\bibnamefont {Ubrig}}, \ and\ \bibinfo {author} {\bibfnamefont {A.~F.}\ \bibnamefont {Morpurgo}},\ }\href {\doibase 10.1103/PhysRevX.9.031019} {\bibfield  {journal} {\bibinfo  {journal} {Phys. Rev. X}\ }\textbf {\bibinfo {volume} {9}},\ \bibinfo {pages} {031019} (\bibinfo {year} {2019})}\BibitemShut {NoStop}%
\bibitem [{\citenamefont {Chi}\ \emph {et~al.}(2019)\citenamefont {Chi}, \citenamefont {Chen}, \citenamefont {Zhao},\ and\ \citenamefont {Weng}}]{Chi2019}%
  \BibitemOpen
  \bibfield  {author} {\bibinfo {author} {\bibfnamefont {Z.}~\bibnamefont {Chi}}, \bibinfo {author} {\bibfnamefont {H.}~\bibnamefont {Chen}}, \bibinfo {author} {\bibfnamefont {Q.}~\bibnamefont {Zhao}}, \ and\ \bibinfo {author} {\bibfnamefont {Y.-X.}\ \bibnamefont {Weng}},\ }\href {\doibase 10.1063/1.5115467} {\bibfield  {journal} {\bibinfo  {journal} {The Journal of Chemical Physics}\ }\textbf {\bibinfo {volume} {151}},\ \bibinfo {pages} {114704} (\bibinfo {year} {2019})}\BibitemShut {NoStop}%
\bibitem [{\citenamefont {Sayers}\ \emph {et~al.}(2023)\citenamefont {Sayers}, \citenamefont {Genco}, \citenamefont {Trovatello}, \citenamefont {Conte}, \citenamefont {Khaustov}, \citenamefont {Cervantes-Villanueva}, \citenamefont {Sangalli}, \citenamefont {Molina-Sanchez}, \citenamefont {Coletti}, \citenamefont {Gadermaier},\ and\ \citenamefont {Cerullo}}]{Sayers2023}%
  \BibitemOpen
  \bibfield  {author} {\bibinfo {author} {\bibfnamefont {C.~J.}\ \bibnamefont {Sayers}}, \bibinfo {author} {\bibfnamefont {A.}~\bibnamefont {Genco}}, \bibinfo {author} {\bibfnamefont {C.}~\bibnamefont {Trovatello}}, \bibinfo {author} {\bibfnamefont {S.~D.}\ \bibnamefont {Conte}}, \bibinfo {author} {\bibfnamefont {V.~O.}\ \bibnamefont {Khaustov}}, \bibinfo {author} {\bibfnamefont {J.}~\bibnamefont {Cervantes-Villanueva}}, \bibinfo {author} {\bibfnamefont {D.}~\bibnamefont {Sangalli}}, \bibinfo {author} {\bibfnamefont {A.}~\bibnamefont {Molina-Sanchez}}, \bibinfo {author} {\bibfnamefont {C.}~\bibnamefont {Coletti}}, \bibinfo {author} {\bibfnamefont {C.}~\bibnamefont {Gadermaier}}, \ and\ \bibinfo {author} {\bibfnamefont {G.}~\bibnamefont {Cerullo}},\ }\href {\doibase 10.1021/acs.nanolett.3c01936} {\bibfield  {journal} {\bibinfo  {journal} {Nano Letters}\ }\textbf {\bibinfo {volume} {23}},\ \bibinfo {pages} {9235} (\bibinfo {year} {2023})}\BibitemShut {NoStop}%
\bibitem [{\citenamefont {Chatelain}\ \emph {et~al.}(2014)\citenamefont {Chatelain}, \citenamefont {Morrison}, \citenamefont {Klarenaar},\ and\ \citenamefont {Siwick}}]{Chatelin14}%
  \BibitemOpen
  \bibfield  {author} {\bibinfo {author} {\bibfnamefont {R.~P.}\ \bibnamefont {Chatelain}}, \bibinfo {author} {\bibfnamefont {V.~R.}\ \bibnamefont {Morrison}}, \bibinfo {author} {\bibfnamefont {B.~L.~M.}\ \bibnamefont {Klarenaar}}, \ and\ \bibinfo {author} {\bibfnamefont {B.~J.}\ \bibnamefont {Siwick}},\ }\href {\doibase 10.1103/PhysRevLett.113.235502} {\bibfield  {journal} {\bibinfo  {journal} {Phys. Rev. Lett.}\ }\textbf {\bibinfo {volume} {113}},\ \bibinfo {pages} {235502} (\bibinfo {year} {2014})}\BibitemShut {NoStop}%
\bibitem [{\citenamefont {Seiler}\ \emph {et~al.}(2021{\natexlab{a}})\citenamefont {Seiler}, \citenamefont {Zahn}, \citenamefont {Zacharias}, \citenamefont {Hildebrandt}, \citenamefont {Vasileiadis}, \citenamefont {Windsor}, \citenamefont {Qi}, \citenamefont {Carbogno}, \citenamefont {Draxl}, \citenamefont {Ernstorfer},\ and\ \citenamefont {Caruso}}]{Seiler2021}%
  \BibitemOpen
  \bibfield  {author} {\bibinfo {author} {\bibfnamefont {H.}~\bibnamefont {Seiler}}, \bibinfo {author} {\bibfnamefont {D.}~\bibnamefont {Zahn}}, \bibinfo {author} {\bibfnamefont {M.}~\bibnamefont {Zacharias}}, \bibinfo {author} {\bibfnamefont {P.-N.}\ \bibnamefont {Hildebrandt}}, \bibinfo {author} {\bibfnamefont {T.}~\bibnamefont {Vasileiadis}}, \bibinfo {author} {\bibfnamefont {Y.~W.}\ \bibnamefont {Windsor}}, \bibinfo {author} {\bibfnamefont {Y.}~\bibnamefont {Qi}}, \bibinfo {author} {\bibfnamefont {C.}~\bibnamefont {Carbogno}}, \bibinfo {author} {\bibfnamefont {C.}~\bibnamefont {Draxl}}, \bibinfo {author} {\bibfnamefont {R.}~\bibnamefont {Ernstorfer}}, \ and\ \bibinfo {author} {\bibfnamefont {F.}~\bibnamefont {Caruso}},\ }\href {\doibase 10.1021/acs.nanolett.1c01786} {\bibfield  {journal} {\bibinfo  {journal} {Nano Letters}\ }\textbf {\bibinfo {volume} {21}},\ \bibinfo {pages} {6171} (\bibinfo {year} {2021}{\natexlab{a}})}\BibitemShut {NoStop}%
\bibitem [{\citenamefont {Fukuda}\ \emph {et~al.}(2023)\citenamefont {Fukuda}, \citenamefont {Ozaki}, \citenamefont {Jeong}, \citenamefont {Arashida}, \citenamefont {En-ya}, \citenamefont {Yoshida}, \citenamefont {Fons}, \citenamefont {Fujita}, \citenamefont {Ueno}, \citenamefont {Hase},\ and\ \citenamefont {Hada}}]{Fukuda2023}%
  \BibitemOpen
  \bibfield  {author} {\bibinfo {author} {\bibfnamefont {T.}~\bibnamefont {Fukuda}}, \bibinfo {author} {\bibfnamefont {U.}~\bibnamefont {Ozaki}}, \bibinfo {author} {\bibfnamefont {S.}~\bibnamefont {Jeong}}, \bibinfo {author} {\bibfnamefont {Y.}~\bibnamefont {Arashida}}, \bibinfo {author} {\bibfnamefont {K.}~\bibnamefont {En-ya}}, \bibinfo {author} {\bibfnamefont {S.}~\bibnamefont {Yoshida}}, \bibinfo {author} {\bibfnamefont {P.~J.}\ \bibnamefont {Fons}}, \bibinfo {author} {\bibfnamefont {J.-i.}\ \bibnamefont {Fujita}}, \bibinfo {author} {\bibfnamefont {K.}~\bibnamefont {Ueno}}, \bibinfo {author} {\bibfnamefont {M.}~\bibnamefont {Hase}}, \ and\ \bibinfo {author} {\bibfnamefont {M.}~\bibnamefont {Hada}},\ }\href {\doibase 10.1021/acs.jpcc.3c02838} {\bibfield  {journal} {\bibinfo  {journal} {The Journal of Physical Chemistry C}\ }\textbf {\bibinfo {volume} {127}},\ \bibinfo {pages} {13149} (\bibinfo {year} {2023})}\BibitemShut {NoStop}%
\bibitem [{\citenamefont {Boschini}\ \emph {et~al.}(2024)\citenamefont {Boschini}, \citenamefont {Zonno},\ and\ \citenamefont {Damascelli}}]{Boschini2024}%
  \BibitemOpen
  \bibfield  {author} {\bibinfo {author} {\bibfnamefont {F.}~\bibnamefont {Boschini}}, \bibinfo {author} {\bibfnamefont {M.}~\bibnamefont {Zonno}}, \ and\ \bibinfo {author} {\bibfnamefont {A.}~\bibnamefont {Damascelli}},\ }\href {\doibase 10.1103/RevModPhys.96.015003} {\bibfield  {journal} {\bibinfo  {journal} {Rev. Mod. Phys.}\ }\textbf {\bibinfo {volume} {96}},\ \bibinfo {pages} {015003} (\bibinfo {year} {2024})}\BibitemShut {NoStop}%
\bibitem [{\citenamefont {Sidiropoulos}\ \emph {et~al.}(2021)\citenamefont {Sidiropoulos}, \citenamefont {Di~Palo}, \citenamefont {Rivas}, \citenamefont {Severino}, \citenamefont {Reduzzi}, \citenamefont {Nandy}, \citenamefont {Bauerhenne}, \citenamefont {Krylow}, \citenamefont {Vasileiadis}, \citenamefont {Danz}, \citenamefont {Elliott}, \citenamefont {Sharma}, \citenamefont {Dewhurst}, \citenamefont {Ropers}, \citenamefont {Joly}, \citenamefont {Garcia}, \citenamefont {Wolf}, \citenamefont {Ernstorfer},\ and\ \citenamefont {Biegert}}]{sidiropoulos21}%
  \BibitemOpen
  \bibfield  {author} {\bibinfo {author} {\bibfnamefont {T.~P.~H.}\ \bibnamefont {Sidiropoulos}}, \bibinfo {author} {\bibfnamefont {N.}~\bibnamefont {Di~Palo}}, \bibinfo {author} {\bibfnamefont {D.~E.}\ \bibnamefont {Rivas}}, \bibinfo {author} {\bibfnamefont {S.}~\bibnamefont {Severino}}, \bibinfo {author} {\bibfnamefont {M.}~\bibnamefont {Reduzzi}}, \bibinfo {author} {\bibfnamefont {B.}~\bibnamefont {Nandy}}, \bibinfo {author} {\bibfnamefont {B.}~\bibnamefont {Bauerhenne}}, \bibinfo {author} {\bibfnamefont {S.}~\bibnamefont {Krylow}}, \bibinfo {author} {\bibfnamefont {T.}~\bibnamefont {Vasileiadis}}, \bibinfo {author} {\bibfnamefont {T.}~\bibnamefont {Danz}}, \bibinfo {author} {\bibfnamefont {P.}~\bibnamefont {Elliott}}, \bibinfo {author} {\bibfnamefont {S.}~\bibnamefont {Sharma}}, \bibinfo {author} {\bibfnamefont {K.}~\bibnamefont {Dewhurst}}, \bibinfo {author} {\bibfnamefont {C.}~\bibnamefont {Ropers}}, \bibinfo {author} {\bibfnamefont {Y.}~\bibnamefont {Joly}}, \bibinfo {author} {\bibfnamefont {M.~E.}\
  \bibnamefont {Garcia}}, \bibinfo {author} {\bibfnamefont {M.}~\bibnamefont {Wolf}}, \bibinfo {author} {\bibfnamefont {R.}~\bibnamefont {Ernstorfer}}, \ and\ \bibinfo {author} {\bibfnamefont {J.}~\bibnamefont {Biegert}},\ }\href {\doibase 10.1103/PhysRevX.11.041060} {\bibfield  {journal} {\bibinfo  {journal} {Phys. Rev. X}\ }\textbf {\bibinfo {volume} {11}},\ \bibinfo {pages} {041060} (\bibinfo {year} {2021})}\BibitemShut {NoStop}%
\bibitem [{\citenamefont {Kampfrath}\ \emph {et~al.}(2005)\citenamefont {Kampfrath}, \citenamefont {Perfetti}, \citenamefont {Schapper}, \citenamefont {Frischkorn},\ and\ \citenamefont {Wolf}}]{kampfrath05}%
  \BibitemOpen
  \bibfield  {author} {\bibinfo {author} {\bibfnamefont {T.}~\bibnamefont {Kampfrath}}, \bibinfo {author} {\bibfnamefont {L.}~\bibnamefont {Perfetti}}, \bibinfo {author} {\bibfnamefont {F.}~\bibnamefont {Schapper}}, \bibinfo {author} {\bibfnamefont {C.}~\bibnamefont {Frischkorn}}, \ and\ \bibinfo {author} {\bibfnamefont {M.}~\bibnamefont {Wolf}},\ }\href {\doibase 10.1103/PhysRevLett.95.187403} {\bibfield  {journal} {\bibinfo  {journal} {Phys. Rev. Lett.}\ }\textbf {\bibinfo {volume} {95}},\ \bibinfo {pages} {187403} (\bibinfo {year} {2005})}\BibitemShut {NoStop}%
\bibitem [{\citenamefont {Cappelluti}\ \emph {et~al.}(2022)\citenamefont {Cappelluti}, \citenamefont {Caruso},\ and\ \citenamefont {Novko}}]{cappelluti2022}%
  \BibitemOpen
  \bibfield  {author} {\bibinfo {author} {\bibfnamefont {E.}~\bibnamefont {Cappelluti}}, \bibinfo {author} {\bibfnamefont {F.}~\bibnamefont {Caruso}}, \ and\ \bibinfo {author} {\bibfnamefont {D.}~\bibnamefont {Novko}},\ }\href {\doibase https://doi.org/10.1016/j.progsurf.2022.100664} {\bibfield  {journal} {\bibinfo  {journal} {Progress in Surface Science}\ }\textbf {\bibinfo {volume} {97}},\ \bibinfo {pages} {100664} (\bibinfo {year} {2022})}\BibitemShut {NoStop}%
\bibitem [{\citenamefont {Seiler}\ \emph {et~al.}(2021{\natexlab{b}})\citenamefont {Seiler}, \citenamefont {Zahn}, \citenamefont {Zacharias}, \citenamefont {Hildebrandt}, \citenamefont {Vasileiadis}, \citenamefont {Windsor}, \citenamefont {Qi}, \citenamefont {Carbogno}, \citenamefont {Draxl}, \citenamefont {Ernstorfer},\ and\ \citenamefont {et~al.}}]{seiler21}%
  \BibitemOpen
  \bibfield  {author} {\bibinfo {author} {\bibfnamefont {H.}~\bibnamefont {Seiler}}, \bibinfo {author} {\bibfnamefont {D.}~\bibnamefont {Zahn}}, \bibinfo {author} {\bibfnamefont {M.}~\bibnamefont {Zacharias}}, \bibinfo {author} {\bibfnamefont {P.-N.}\ \bibnamefont {Hildebrandt}}, \bibinfo {author} {\bibfnamefont {T.}~\bibnamefont {Vasileiadis}}, \bibinfo {author} {\bibfnamefont {Y.~W.}\ \bibnamefont {Windsor}}, \bibinfo {author} {\bibfnamefont {Y.}~\bibnamefont {Qi}}, \bibinfo {author} {\bibfnamefont {C.}~\bibnamefont {Carbogno}}, \bibinfo {author} {\bibfnamefont {C.}~\bibnamefont {Draxl}}, \bibinfo {author} {\bibfnamefont {R.}~\bibnamefont {Ernstorfer}}, \ and\ \bibinfo {author} {\bibnamefont {et~al.}},\ }\href {\doibase 10.1021/acs.nanolett.1c01786} {\bibfield  {journal} {\bibinfo  {journal} {Nano Letters}\ }\textbf {\bibinfo {volume} {21}},\ \bibinfo {pages} {6171} (\bibinfo {year} {2021}{\natexlab{b}})}\BibitemShut {NoStop}%
\bibitem [{\citenamefont {Perfetti}\ \emph {et~al.}(2007)\citenamefont {Perfetti}, \citenamefont {Loukakos}, \citenamefont {Lisowski}, \citenamefont {Bovensiepen}, \citenamefont {Eisaki},\ and\ \citenamefont {Wolf}}]{Perfetti2007}%
  \BibitemOpen
  \bibfield  {author} {\bibinfo {author} {\bibfnamefont {L.}~\bibnamefont {Perfetti}}, \bibinfo {author} {\bibfnamefont {P.~A.}\ \bibnamefont {Loukakos}}, \bibinfo {author} {\bibfnamefont {M.}~\bibnamefont {Lisowski}}, \bibinfo {author} {\bibfnamefont {U.}~\bibnamefont {Bovensiepen}}, \bibinfo {author} {\bibfnamefont {H.}~\bibnamefont {Eisaki}}, \ and\ \bibinfo {author} {\bibfnamefont {M.}~\bibnamefont {Wolf}},\ }\href {\doibase 10.1103/PhysRevLett.99.197001} {\bibfield  {journal} {\bibinfo  {journal} {Phys. Rev. Lett.}\ }\textbf {\bibinfo {volume} {99}},\ \bibinfo {pages} {197001} (\bibinfo {year} {2007})}\BibitemShut {NoStop}%
\bibitem [{\citenamefont {Johannsen}\ \emph {et~al.}(2013)\citenamefont {Johannsen}, \citenamefont {Ulstrup}, \citenamefont {Cilento}, \citenamefont {Crepaldi}, \citenamefont {Zacchigna}, \citenamefont {Cacho}, \citenamefont {Turcu}, \citenamefont {Springate}, \citenamefont {Fromm}, \citenamefont {Raidel}, \citenamefont {Seyller}, \citenamefont {Parmigiani}, \citenamefont {Grioni},\ and\ \citenamefont {Hofmann}}]{Johannsen2013}%
  \BibitemOpen
  \bibfield  {author} {\bibinfo {author} {\bibfnamefont {J.~C.}\ \bibnamefont {Johannsen}}, \bibinfo {author} {\bibfnamefont {S.}~\bibnamefont {Ulstrup}}, \bibinfo {author} {\bibfnamefont {F.}~\bibnamefont {Cilento}}, \bibinfo {author} {\bibfnamefont {A.}~\bibnamefont {Crepaldi}}, \bibinfo {author} {\bibfnamefont {M.}~\bibnamefont {Zacchigna}}, \bibinfo {author} {\bibfnamefont {C.}~\bibnamefont {Cacho}}, \bibinfo {author} {\bibfnamefont {I.~C.~E.}\ \bibnamefont {Turcu}}, \bibinfo {author} {\bibfnamefont {E.}~\bibnamefont {Springate}}, \bibinfo {author} {\bibfnamefont {F.}~\bibnamefont {Fromm}}, \bibinfo {author} {\bibfnamefont {C.}~\bibnamefont {Raidel}}, \bibinfo {author} {\bibfnamefont {T.}~\bibnamefont {Seyller}}, \bibinfo {author} {\bibfnamefont {F.}~\bibnamefont {Parmigiani}}, \bibinfo {author} {\bibfnamefont {M.}~\bibnamefont {Grioni}}, \ and\ \bibinfo {author} {\bibfnamefont {P.}~\bibnamefont {Hofmann}},\ }\href {\doibase 10.1103/PhysRevLett.111.027403} {\bibfield  {journal} {\bibinfo  {journal} {Phys.
  Rev. Lett.}\ }\textbf {\bibinfo {volume} {111}},\ \bibinfo {pages} {027403} (\bibinfo {year} {2013})}\BibitemShut {NoStop}%
\bibitem [{\citenamefont {Novko}\ \emph {et~al.}(2020)\citenamefont {Novko}, \citenamefont {Caruso}, \citenamefont {Draxl},\ and\ \citenamefont {Cappelluti}}]{novko2020}%
  \BibitemOpen
  \bibfield  {author} {\bibinfo {author} {\bibfnamefont {D.}~\bibnamefont {Novko}}, \bibinfo {author} {\bibfnamefont {F.}~\bibnamefont {Caruso}}, \bibinfo {author} {\bibfnamefont {C.}~\bibnamefont {Draxl}}, \ and\ \bibinfo {author} {\bibfnamefont {E.}~\bibnamefont {Cappelluti}},\ }\href {\doibase 10.1103/PhysRevLett.124.077001} {\bibfield  {journal} {\bibinfo  {journal} {Phys. Rev. Lett.}\ }\textbf {\bibinfo {volume} {124}},\ \bibinfo {pages} {077001} (\bibinfo {year} {2020})}\BibitemShut {NoStop}%
\bibitem [{\citenamefont {Pomarico}\ \emph {et~al.}(2017)\citenamefont {Pomarico}, \citenamefont {Mitrano}, \citenamefont {Bromberger}, \citenamefont {Sentef}, \citenamefont {Al-Temimy}, \citenamefont {Coletti}, \citenamefont {St\"ohr}, \citenamefont {Link}, \citenamefont {Starke}, \citenamefont {Cacho},\ and\ \citenamefont {et~al.}}]{pomarico17}%
  \BibitemOpen
  \bibfield  {author} {\bibinfo {author} {\bibfnamefont {E.}~\bibnamefont {Pomarico}}, \bibinfo {author} {\bibfnamefont {M.}~\bibnamefont {Mitrano}}, \bibinfo {author} {\bibfnamefont {H.}~\bibnamefont {Bromberger}}, \bibinfo {author} {\bibfnamefont {M.~A.}\ \bibnamefont {Sentef}}, \bibinfo {author} {\bibfnamefont {A.}~\bibnamefont {Al-Temimy}}, \bibinfo {author} {\bibfnamefont {C.}~\bibnamefont {Coletti}}, \bibinfo {author} {\bibfnamefont {A.}~\bibnamefont {St\"ohr}}, \bibinfo {author} {\bibfnamefont {S.}~\bibnamefont {Link}}, \bibinfo {author} {\bibfnamefont {U.}~\bibnamefont {Starke}}, \bibinfo {author} {\bibfnamefont {C.}~\bibnamefont {Cacho}}, \ and\ \bibinfo {author} {\bibnamefont {et~al.}},\ }\href {\doibase 10.1103/PhysRevB.95.024304} {\bibfield  {journal} {\bibinfo  {journal} {Phys. Rev. B}\ }\textbf {\bibinfo {volume} {95}},\ \bibinfo {pages} {024304} (\bibinfo {year} {2017})}\BibitemShut {NoStop}%
\bibitem [{\citenamefont {Ishioka}\ \emph {et~al.}(2008)\citenamefont {Ishioka}, \citenamefont {Hase}, \citenamefont {Kitajima}, \citenamefont {Wirtz}, \citenamefont {Rubio},\ and\ \citenamefont {Petek}}]{ishioka08}%
  \BibitemOpen
  \bibfield  {author} {\bibinfo {author} {\bibfnamefont {K.}~\bibnamefont {Ishioka}}, \bibinfo {author} {\bibfnamefont {M.}~\bibnamefont {Hase}}, \bibinfo {author} {\bibfnamefont {M.}~\bibnamefont {Kitajima}}, \bibinfo {author} {\bibfnamefont {L.}~\bibnamefont {Wirtz}}, \bibinfo {author} {\bibfnamefont {A.}~\bibnamefont {Rubio}}, \ and\ \bibinfo {author} {\bibfnamefont {H.}~\bibnamefont {Petek}},\ }\href {\doibase 10.1103/PhysRevB.77.121402} {\bibfield  {journal} {\bibinfo  {journal} {Phys. Rev. B}\ }\textbf {\bibinfo {volume} {77}},\ \bibinfo {pages} {121402} (\bibinfo {year} {2008})}\BibitemShut {NoStop}%
\bibitem [{\citenamefont {Hu}\ \emph {et~al.}(2022)\citenamefont {Hu}, \citenamefont {Zhao}, \citenamefont {Lian}, \citenamefont {Liu}, \citenamefont {Guan},\ and\ \citenamefont {Meng}}]{hu22}%
  \BibitemOpen
  \bibfield  {author} {\bibinfo {author} {\bibfnamefont {S.-Q.}\ \bibnamefont {Hu}}, \bibinfo {author} {\bibfnamefont {H.}~\bibnamefont {Zhao}}, \bibinfo {author} {\bibfnamefont {C.}~\bibnamefont {Lian}}, \bibinfo {author} {\bibfnamefont {X.-B.}\ \bibnamefont {Liu}}, \bibinfo {author} {\bibfnamefont {M.-X.}\ \bibnamefont {Guan}}, \ and\ \bibinfo {author} {\bibfnamefont {S.}~\bibnamefont {Meng}},\ }\href {\doibase 10.1038/s41535-021-00421-7} {\bibfield  {journal} {\bibinfo  {journal} {npj Quantum Materials}\ }\textbf {\bibinfo {volume} {7}},\ \bibinfo {pages} {14} (\bibinfo {year} {2022})}\BibitemShut {NoStop}%
\bibitem [{\citenamefont {Girotto}\ and\ \citenamefont {Novko}(2023)}]{Girotto2023}%
  \BibitemOpen
  \bibfield  {author} {\bibinfo {author} {\bibfnamefont {N.}~\bibnamefont {Girotto}}\ and\ \bibinfo {author} {\bibfnamefont {D.}~\bibnamefont {Novko}},\ }\href {\doibase 10.1021/acs.jpclett.3c01905} {\bibfield  {journal} {\bibinfo  {journal} {The Journal of Physical Chemistry Letters}\ }\textbf {\bibinfo {volume} {14}},\ \bibinfo {pages} {8709} (\bibinfo {year} {2023})}\BibitemShut {NoStop}%
\bibitem [{\citenamefont {Liu}\ \emph {et~al.}(2022{\natexlab{a}})\citenamefont {Liu}, \citenamefont {Hu}, \citenamefont {Chen}, \citenamefont {Guan}, \citenamefont {Chen},\ and\ \citenamefont {Meng}}]{liu22}%
  \BibitemOpen
  \bibfield  {author} {\bibinfo {author} {\bibfnamefont {X.-B.}\ \bibnamefont {Liu}}, \bibinfo {author} {\bibfnamefont {S.-Q.}\ \bibnamefont {Hu}}, \bibinfo {author} {\bibfnamefont {D.}~\bibnamefont {Chen}}, \bibinfo {author} {\bibfnamefont {M.}~\bibnamefont {Guan}}, \bibinfo {author} {\bibfnamefont {Q.}~\bibnamefont {Chen}}, \ and\ \bibinfo {author} {\bibfnamefont {S.}~\bibnamefont {Meng}},\ }\href {\doibase 10.1021/acs.nanolett.2c01105} {\bibfield  {journal} {\bibinfo  {journal} {Nano Letters}\ }\textbf {\bibinfo {volume} {22}},\ \bibinfo {pages} {4800} (\bibinfo {year} {2022}{\natexlab{a}})}\BibitemShut {NoStop}%
\bibitem [{\citenamefont {Pan}\ \emph {et~al.}(2025)\citenamefont {Pan}, \citenamefont {Hildebrandt}, \citenamefont {Zahn}, \citenamefont {Zacharias}, \citenamefont {Windsor}, \citenamefont {Ernstorfer}, \citenamefont {Caruso},\ and\ \citenamefont {Seiler}}]{pan2025}%
  \BibitemOpen
  \bibfield  {author} {\bibinfo {author} {\bibfnamefont {Y.}~\bibnamefont {Pan}}, \bibinfo {author} {\bibfnamefont {P.-N.}\ \bibnamefont {Hildebrandt}}, \bibinfo {author} {\bibfnamefont {D.}~\bibnamefont {Zahn}}, \bibinfo {author} {\bibfnamefont {M.}~\bibnamefont {Zacharias}}, \bibinfo {author} {\bibfnamefont {Y.~W.}\ \bibnamefont {Windsor}}, \bibinfo {author} {\bibfnamefont {R.}~\bibnamefont {Ernstorfer}}, \bibinfo {author} {\bibfnamefont {F.}~\bibnamefont {Caruso}}, \ and\ \bibinfo {author} {\bibfnamefont {H.}~\bibnamefont {Seiler}},\ }\href {\doibase 10.1021/acsnano.5c00744} {\bibfield  {journal} {\bibinfo  {journal} {ACS Nano}\ }\textbf {\bibinfo {volume} {19}},\ \bibinfo {pages} {11381} (\bibinfo {year} {2025})}\BibitemShut {NoStop}%
\bibitem [{\citenamefont {Fausti}\ \emph {et~al.}(2011)\citenamefont {Fausti}, \citenamefont {Tobey}, \citenamefont {Dean}, \citenamefont {Kaiser}, \citenamefont {Dienst}, \citenamefont {Hoffmann}, \citenamefont {Pyon}, \citenamefont {Takayama}, \citenamefont {Takagi},\ and\ \citenamefont {Cavalleri}}]{fausti11}%
  \BibitemOpen
  \bibfield  {author} {\bibinfo {author} {\bibfnamefont {D.}~\bibnamefont {Fausti}}, \bibinfo {author} {\bibfnamefont {R.~I.}\ \bibnamefont {Tobey}}, \bibinfo {author} {\bibfnamefont {N.}~\bibnamefont {Dean}}, \bibinfo {author} {\bibfnamefont {S.}~\bibnamefont {Kaiser}}, \bibinfo {author} {\bibfnamefont {A.}~\bibnamefont {Dienst}}, \bibinfo {author} {\bibfnamefont {M.~C.}\ \bibnamefont {Hoffmann}}, \bibinfo {author} {\bibfnamefont {S.}~\bibnamefont {Pyon}}, \bibinfo {author} {\bibfnamefont {T.}~\bibnamefont {Takayama}}, \bibinfo {author} {\bibfnamefont {H.}~\bibnamefont {Takagi}}, \ and\ \bibinfo {author} {\bibfnamefont {A.}~\bibnamefont {Cavalleri}},\ }\href {\doibase 10.1126/science.1197294} {\bibfield  {journal} {\bibinfo  {journal} {Science}\ }\textbf {\bibinfo {volume} {331}},\ \bibinfo {pages} {189} (\bibinfo {year} {2011})}\BibitemShut {NoStop}%
\bibitem [{\citenamefont {Zhang}\ \emph {et~al.}(2014)\citenamefont {Zhang}, \citenamefont {Liu}, \citenamefont {Luo}, \citenamefont {Freeman},\ and\ \citenamefont {Zunger}}]{zhang14}%
  \BibitemOpen
  \bibfield  {author} {\bibinfo {author} {\bibfnamefont {X.}~\bibnamefont {Zhang}}, \bibinfo {author} {\bibfnamefont {Q.}~\bibnamefont {Liu}}, \bibinfo {author} {\bibfnamefont {J.-W.}\ \bibnamefont {Luo}}, \bibinfo {author} {\bibfnamefont {A.~J.}\ \bibnamefont {Freeman}}, \ and\ \bibinfo {author} {\bibfnamefont {A.}~\bibnamefont {Zunger}},\ }\href {\doibase 10.1038/nphys2933} {\bibfield  {journal} {\bibinfo  {journal} {Nature Physics}\ }\textbf {\bibinfo {volume} {10}},\ \bibinfo {pages} {387} (\bibinfo {year} {2014})}\BibitemShut {NoStop}%
\bibitem [{\citenamefont {Rohwer}\ \emph {et~al.}(2011)\citenamefont {Rohwer}, \citenamefont {Hellmann}, \citenamefont {Wiesenmayer}, \citenamefont {Sohrt}, \citenamefont {Stange}, \citenamefont {Slomski}, \citenamefont {Carr}, \citenamefont {Liu}, \citenamefont {Avila}, \citenamefont {Kalläne},\ and\ \citenamefont {et~al.}}]{rohwer11}%
  \BibitemOpen
  \bibfield  {author} {\bibinfo {author} {\bibfnamefont {T.}~\bibnamefont {Rohwer}}, \bibinfo {author} {\bibfnamefont {S.}~\bibnamefont {Hellmann}}, \bibinfo {author} {\bibfnamefont {M.}~\bibnamefont {Wiesenmayer}}, \bibinfo {author} {\bibfnamefont {C.}~\bibnamefont {Sohrt}}, \bibinfo {author} {\bibfnamefont {A.}~\bibnamefont {Stange}}, \bibinfo {author} {\bibfnamefont {B.}~\bibnamefont {Slomski}}, \bibinfo {author} {\bibfnamefont {A.}~\bibnamefont {Carr}}, \bibinfo {author} {\bibfnamefont {Y.}~\bibnamefont {Liu}}, \bibinfo {author} {\bibfnamefont {L.~M.}\ \bibnamefont {Avila}}, \bibinfo {author} {\bibfnamefont {M.}~\bibnamefont {Kalläne}}, \ and\ \bibinfo {author} {\bibnamefont {et~al.}},\ }\href {\doibase 10.1038/nature09829} {\bibfield  {journal} {\bibinfo  {journal} {Nature}\ }\textbf {\bibinfo {volume} {471}},\ \bibinfo {pages} {490} (\bibinfo {year} {2011})}\BibitemShut {NoStop}%
\bibitem [{\citenamefont {Stojchevska}\ \emph {et~al.}(2014)\citenamefont {Stojchevska}, \citenamefont {Vaskivskyi}, \citenamefont {Mertelj}, \citenamefont {Kusar}, \citenamefont {Svetin}, \citenamefont {Brazovskii},\ and\ \citenamefont {Mihailovic}}]{stojchevska14}%
  \BibitemOpen
  \bibfield  {author} {\bibinfo {author} {\bibfnamefont {L.}~\bibnamefont {Stojchevska}}, \bibinfo {author} {\bibfnamefont {I.}~\bibnamefont {Vaskivskyi}}, \bibinfo {author} {\bibfnamefont {T.}~\bibnamefont {Mertelj}}, \bibinfo {author} {\bibfnamefont {P.}~\bibnamefont {Kusar}}, \bibinfo {author} {\bibfnamefont {D.}~\bibnamefont {Svetin}}, \bibinfo {author} {\bibfnamefont {S.}~\bibnamefont {Brazovskii}}, \ and\ \bibinfo {author} {\bibfnamefont {D.}~\bibnamefont {Mihailovic}},\ }\href {\doibase 10.1126/science.1241591} {\bibfield  {journal} {\bibinfo  {journal} {Science}\ }\textbf {\bibinfo {volume} {344}},\ \bibinfo {pages} {177} (\bibinfo {year} {2014})}\BibitemShut {NoStop}%
\bibitem [{\citenamefont {Murray}\ \emph {et~al.}(2007)\citenamefont {Murray}, \citenamefont {Fahy}, \citenamefont {Prendergast}, \citenamefont {Ogitsu}, \citenamefont {Fritz},\ and\ \citenamefont {Reis}}]{Murray2007}%
  \BibitemOpen
  \bibfield  {author} {\bibinfo {author} {\bibfnamefont {E.~D.}\ \bibnamefont {Murray}}, \bibinfo {author} {\bibfnamefont {S.}~\bibnamefont {Fahy}}, \bibinfo {author} {\bibfnamefont {D.}~\bibnamefont {Prendergast}}, \bibinfo {author} {\bibfnamefont {T.}~\bibnamefont {Ogitsu}}, \bibinfo {author} {\bibfnamefont {D.~M.}\ \bibnamefont {Fritz}}, \ and\ \bibinfo {author} {\bibfnamefont {D.~A.}\ \bibnamefont {Reis}},\ }\href {\doibase 10.1103/PhysRevB.75.184301} {\bibfield  {journal} {\bibinfo  {journal} {Phys. Rev. B}\ }\textbf {\bibinfo {volume} {75}},\ \bibinfo {pages} {184301} (\bibinfo {year} {2007})}\BibitemShut {NoStop}%
\bibitem [{\citenamefont {Liu}\ \emph {et~al.}(2022{\natexlab{b}})\citenamefont {Liu}, \citenamefont {Mao}, \citenamefont {Zhang},\ and\ \citenamefont {Zhou}}]{liu2022}%
  \BibitemOpen
  \bibfield  {author} {\bibinfo {author} {\bibfnamefont {K.}~\bibnamefont {Liu}}, \bibinfo {author} {\bibfnamefont {S.}~\bibnamefont {Mao}}, \bibinfo {author} {\bibfnamefont {S.}~\bibnamefont {Zhang}}, \ and\ \bibinfo {author} {\bibfnamefont {J.}~\bibnamefont {Zhou}},\ }\href {\doibase 10.1021/acs.nanolett.2c03238} {\bibfield  {journal} {\bibinfo  {journal} {Nano Letters}\ }\textbf {\bibinfo {volume} {22}},\ \bibinfo {pages} {9006} (\bibinfo {year} {2022}{\natexlab{b}})}\BibitemShut {NoStop}%
\bibitem [{\citenamefont {Waldecker}\ \emph {et~al.}(2016)\citenamefont {Waldecker}, \citenamefont {Bertoni}, \citenamefont {Ernstorfer},\ and\ \citenamefont {Vorberger}}]{Waldecker16}%
  \BibitemOpen
  \bibfield  {author} {\bibinfo {author} {\bibfnamefont {L.}~\bibnamefont {Waldecker}}, \bibinfo {author} {\bibfnamefont {R.}~\bibnamefont {Bertoni}}, \bibinfo {author} {\bibfnamefont {R.}~\bibnamefont {Ernstorfer}}, \ and\ \bibinfo {author} {\bibfnamefont {J.}~\bibnamefont {Vorberger}},\ }\href {\doibase 10.1103/PhysRevX.6.021003} {\bibfield  {journal} {\bibinfo  {journal} {Phys. Rev. X}\ }\textbf {\bibinfo {volume} {6}},\ \bibinfo {pages} {021003} (\bibinfo {year} {2016})}\BibitemShut {NoStop}%
\bibitem [{\citenamefont {Caruso}\ and\ \citenamefont {Novko}(2022)}]{caruso2022}%
  \BibitemOpen
  \bibfield  {author} {\bibinfo {author} {\bibfnamefont {F.}~\bibnamefont {Caruso}}\ and\ \bibinfo {author} {\bibfnamefont {D.}~\bibnamefont {Novko}},\ }\href {\doibase 10.1080/23746149.2022.2095925} {\bibfield  {journal} {\bibinfo  {journal} {Advances in Physics: X}\ }\textbf {\bibinfo {volume} {7}},\ \bibinfo {pages} {2095925} (\bibinfo {year} {2022})}\BibitemShut {NoStop}%
\bibitem [{\citenamefont {Comby}\ \emph {et~al.}(2022)\citenamefont {Comby}, \citenamefont {Rajak}, \citenamefont {Descamps}, \citenamefont {Petit}, \citenamefont {Blanchet}, \citenamefont {Mairesse}, \citenamefont {Gaudin},\ and\ \citenamefont {Beaulieu}}]{Comby22}%
  \BibitemOpen
  \bibfield  {author} {\bibinfo {author} {\bibfnamefont {A.}~\bibnamefont {Comby}}, \bibinfo {author} {\bibfnamefont {D.}~\bibnamefont {Rajak}}, \bibinfo {author} {\bibfnamefont {D.}~\bibnamefont {Descamps}}, \bibinfo {author} {\bibfnamefont {S.}~\bibnamefont {Petit}}, \bibinfo {author} {\bibfnamefont {V.}~\bibnamefont {Blanchet}}, \bibinfo {author} {\bibfnamefont {Y.}~\bibnamefont {Mairesse}}, \bibinfo {author} {\bibfnamefont {J.}~\bibnamefont {Gaudin}}, \ and\ \bibinfo {author} {\bibfnamefont {S.}~\bibnamefont {Beaulieu}},\ }\href {\doibase 10.1088/2040-8986/ac7a49} {\bibfield  {journal} {\bibinfo  {journal} {Journal of Optics}\ }\textbf {\bibinfo {volume} {24}},\ \bibinfo {pages} {084003} (\bibinfo {year} {2022})}\BibitemShut {NoStop}%
\bibitem [{\citenamefont {Tkach}\ and\ \citenamefont {Schönhense}(2025)}]{tkach24}%
  \BibitemOpen
  \bibfield  {author} {\bibinfo {author} {\bibfnamefont {O.}~\bibnamefont {Tkach}}\ and\ \bibinfo {author} {\bibfnamefont {G.}~\bibnamefont {Schönhense}},\ }\href {\doibase https://doi.org/10.1016/j.ultramic.2025.114167} {\bibfield  {journal} {\bibinfo  {journal} {Ultramicroscopy}\ }\textbf {\bibinfo {volume} {276}},\ \bibinfo {pages} {114167} (\bibinfo {year} {2025})}\BibitemShut {NoStop}%
\bibitem [{\citenamefont {Tkach}\ \emph {et~al.}(2024)\citenamefont {Tkach}, \citenamefont {Fragkos}, \citenamefont {Nguyen}, \citenamefont {Chernov}, \citenamefont {Scholz}, \citenamefont {Wind}, \citenamefont {Babenkov}, \citenamefont {Fedchenko}, \citenamefont {Lytvynenko}, \citenamefont {Zimmer}, \citenamefont {Hloskovskii}, \citenamefont {Kutnyakhov}, \citenamefont {Pressacco}, \citenamefont {Dilling}, \citenamefont {Bruckmeier}, \citenamefont {Heber}, \citenamefont {Scholz}, \citenamefont {Sobota}, \citenamefont {Koralek}, \citenamefont {Sirica}, \citenamefont {Kallmayer}, \citenamefont {Hoesch}, \citenamefont {Schlueter}, \citenamefont {Odnodvorets}, \citenamefont {Mairesse}, \citenamefont {Rossnagel}, \citenamefont {Elmers}, \citenamefont {Beaulieu},\ and\ \citenamefont {Schoenhense}}]{tkach24-2}%
  \BibitemOpen
  \bibfield  {author} {\bibinfo {author} {\bibfnamefont {O.}~\bibnamefont {Tkach}}, \bibinfo {author} {\bibfnamefont {S.}~\bibnamefont {Fragkos}}, \bibinfo {author} {\bibfnamefont {Q.}~\bibnamefont {Nguyen}}, \bibinfo {author} {\bibfnamefont {S.}~\bibnamefont {Chernov}}, \bibinfo {author} {\bibfnamefont {M.}~\bibnamefont {Scholz}}, \bibinfo {author} {\bibfnamefont {N.}~\bibnamefont {Wind}}, \bibinfo {author} {\bibfnamefont {S.}~\bibnamefont {Babenkov}}, \bibinfo {author} {\bibfnamefont {O.}~\bibnamefont {Fedchenko}}, \bibinfo {author} {\bibfnamefont {Y.}~\bibnamefont {Lytvynenko}}, \bibinfo {author} {\bibfnamefont {D.}~\bibnamefont {Zimmer}}, \bibinfo {author} {\bibfnamefont {A.}~\bibnamefont {Hloskovskii}}, \bibinfo {author} {\bibfnamefont {D.}~\bibnamefont {Kutnyakhov}}, \bibinfo {author} {\bibfnamefont {F.}~\bibnamefont {Pressacco}}, \bibinfo {author} {\bibfnamefont {J.}~\bibnamefont {Dilling}}, \bibinfo {author} {\bibfnamefont {L.}~\bibnamefont {Bruckmeier}}, \bibinfo {author} {\bibfnamefont
  {M.}~\bibnamefont {Heber}}, \bibinfo {author} {\bibfnamefont {F.}~\bibnamefont {Scholz}}, \bibinfo {author} {\bibfnamefont {J.}~\bibnamefont {Sobota}}, \bibinfo {author} {\bibfnamefont {J.}~\bibnamefont {Koralek}}, \bibinfo {author} {\bibfnamefont {N.}~\bibnamefont {Sirica}}, \bibinfo {author} {\bibfnamefont {M.}~\bibnamefont {Kallmayer}}, \bibinfo {author} {\bibfnamefont {M.}~\bibnamefont {Hoesch}}, \bibinfo {author} {\bibfnamefont {C.}~\bibnamefont {Schlueter}}, \bibinfo {author} {\bibfnamefont {L.~V.}\ \bibnamefont {Odnodvorets}}, \bibinfo {author} {\bibfnamefont {Y.}~\bibnamefont {Mairesse}}, \bibinfo {author} {\bibfnamefont {K.}~\bibnamefont {Rossnagel}}, \bibinfo {author} {\bibfnamefont {H.~J.}\ \bibnamefont {Elmers}}, \bibinfo {author} {\bibfnamefont {S.}~\bibnamefont {Beaulieu}}, \ and\ \bibinfo {author} {\bibfnamefont {G.}~\bibnamefont {Schoenhense}},\ }\href {https://arxiv.org/abs/2401.10084} {\enquote {\bibinfo {title} {Multi-mode front lens for momentum microscopy: Part {II} experiments},}\ }
  (\bibinfo {year} {2024}),\ \Eprint {http://arxiv.org/abs/2401.10084} {arXiv:2401.10084 [cond-mat.mtrl-sci]} \BibitemShut {NoStop}%
\bibitem [{SM()}]{SM}%
  \BibitemOpen
  \href@noop {} {}\bibinfo {note} {See Supplemental Material at [URL], which additional includes Refs.\,\cite{Xian20,Xian19_2,Munkhbat2022,Giannozzi2017,wan90,giustino07,Noffsinger2010,Ponce2016,vanSetten2018,Marzari2012}, for more details on experimental setup, determination of direct and indirect band gaps, valley-resolved fitting of electron population lifetimes, hot phonon bottleneck effect, more details on EPC matrix elements, and computational details.}\BibitemShut {Stop}%
\bibitem [{\citenamefont {Fragkos}\ \emph {et~al.}(2025)\citenamefont {Fragkos}, \citenamefont {Courtade}, \citenamefont {Tkach}, \citenamefont {Gaudin}, \citenamefont {Descamps}, \citenamefont {Barrette}, \citenamefont {Petit}, \citenamefont {Schönhense}, \citenamefont {Mairesse},\ and\ \citenamefont {Beaulieu}}]{Fragkos2025}%
  \BibitemOpen
  \bibfield  {author} {\bibinfo {author} {\bibfnamefont {S.}~\bibnamefont {Fragkos}}, \bibinfo {author} {\bibfnamefont {Q.}~\bibnamefont {Courtade}}, \bibinfo {author} {\bibfnamefont {O.}~\bibnamefont {Tkach}}, \bibinfo {author} {\bibfnamefont {J.}~\bibnamefont {Gaudin}}, \bibinfo {author} {\bibfnamefont {D.}~\bibnamefont {Descamps}}, \bibinfo {author} {\bibfnamefont {G.}~\bibnamefont {Barrette}}, \bibinfo {author} {\bibfnamefont {S.}~\bibnamefont {Petit}}, \bibinfo {author} {\bibfnamefont {G.}~\bibnamefont {Schönhense}}, \bibinfo {author} {\bibfnamefont {Y.}~\bibnamefont {Mairesse}}, \ and\ \bibinfo {author} {\bibfnamefont {S.}~\bibnamefont {Beaulieu}},\ }\href {https://arxiv.org/abs/2507.02371} {\enquote {\bibinfo {title} {Time- and polarization-resolved extreme ultraviolet momentum microscopy},}\ } (\bibinfo {year} {2025}),\ \Eprint {http://arxiv.org/abs/2507.02371} {arXiv:2507.02371 [cond-mat.mtrl-sci]} \BibitemShut {NoStop}%
\bibitem [{\citenamefont {Cunningham}\ \emph {et~al.}(2017)\citenamefont {Cunningham}, \citenamefont {Hanbicki}, \citenamefont {McCreary},\ and\ \citenamefont {Jonker}}]{Cunningham2017}%
  \BibitemOpen
  \bibfield  {author} {\bibinfo {author} {\bibfnamefont {P.~D.}\ \bibnamefont {Cunningham}}, \bibinfo {author} {\bibfnamefont {A.~T.}\ \bibnamefont {Hanbicki}}, \bibinfo {author} {\bibfnamefont {K.~M.}\ \bibnamefont {McCreary}}, \ and\ \bibinfo {author} {\bibfnamefont {B.~T.}\ \bibnamefont {Jonker}},\ }\href {\doibase 10.1021/acsnano.7b06885} {\bibfield  {journal} {\bibinfo  {journal} {ACS Nano}\ }\textbf {\bibinfo {volume} {11}},\ \bibinfo {pages} {12601} (\bibinfo {year} {2017})}\BibitemShut {NoStop}%
\bibitem [{\citenamefont {Liu}\ \emph {et~al.}(2019)\citenamefont {Liu}, \citenamefont {Ziffer}, \citenamefont {Hansen}, \citenamefont {Wang},\ and\ \citenamefont {Zhu}}]{Liu19}%
  \BibitemOpen
  \bibfield  {author} {\bibinfo {author} {\bibfnamefont {F.}~\bibnamefont {Liu}}, \bibinfo {author} {\bibfnamefont {M.~E.}\ \bibnamefont {Ziffer}}, \bibinfo {author} {\bibfnamefont {K.~R.}\ \bibnamefont {Hansen}}, \bibinfo {author} {\bibfnamefont {J.}~\bibnamefont {Wang}}, \ and\ \bibinfo {author} {\bibfnamefont {X.}~\bibnamefont {Zhu}},\ }\href {\doibase 10.1103/PhysRevLett.122.246803} {\bibfield  {journal} {\bibinfo  {journal} {Phys. Rev. Lett.}\ }\textbf {\bibinfo {volume} {122}},\ \bibinfo {pages} {246803} (\bibinfo {year} {2019})}\BibitemShut {NoStop}%
\bibitem [{\citenamefont {Meckbach}\ \emph {et~al.}(2020)\citenamefont {Meckbach}, \citenamefont {Hader}, \citenamefont {Huttner}, \citenamefont {Neuhaus}, \citenamefont {Steiner}, \citenamefont {Stroucken}, \citenamefont {Moloney},\ and\ \citenamefont {Koch}}]{Meckbach20}%
  \BibitemOpen
  \bibfield  {author} {\bibinfo {author} {\bibfnamefont {L.}~\bibnamefont {Meckbach}}, \bibinfo {author} {\bibfnamefont {J.}~\bibnamefont {Hader}}, \bibinfo {author} {\bibfnamefont {U.}~\bibnamefont {Huttner}}, \bibinfo {author} {\bibfnamefont {J.}~\bibnamefont {Neuhaus}}, \bibinfo {author} {\bibfnamefont {J.~T.}\ \bibnamefont {Steiner}}, \bibinfo {author} {\bibfnamefont {T.}~\bibnamefont {Stroucken}}, \bibinfo {author} {\bibfnamefont {J.~V.}\ \bibnamefont {Moloney}}, \ and\ \bibinfo {author} {\bibfnamefont {S.~W.}\ \bibnamefont {Koch}},\ }\href {\doibase 10.1103/PhysRevB.101.075401} {\bibfield  {journal} {\bibinfo  {journal} {Phys. Rev. B}\ }\textbf {\bibinfo {volume} {101}},\ \bibinfo {pages} {075401} (\bibinfo {year} {2020})}\BibitemShut {NoStop}%
\bibitem [{\citenamefont {Bertoni}\ \emph {et~al.}(2016)\citenamefont {Bertoni}, \citenamefont {Nicholson}, \citenamefont {Waldecker}, \citenamefont {H\"ubener}, \citenamefont {Monney}, \citenamefont {De~Giovannini}, \citenamefont {Puppin}, \citenamefont {Hoesch}, \citenamefont {Springate}, \citenamefont {Chapman}, \citenamefont {Cacho}, \citenamefont {Wolf}, \citenamefont {Rubio},\ and\ \citenamefont {Ernstorfer}}]{Bertoni16}%
  \BibitemOpen
  \bibfield  {author} {\bibinfo {author} {\bibfnamefont {R.}~\bibnamefont {Bertoni}}, \bibinfo {author} {\bibfnamefont {C.~W.}\ \bibnamefont {Nicholson}}, \bibinfo {author} {\bibfnamefont {L.}~\bibnamefont {Waldecker}}, \bibinfo {author} {\bibfnamefont {H.}~\bibnamefont {H\"ubener}}, \bibinfo {author} {\bibfnamefont {C.}~\bibnamefont {Monney}}, \bibinfo {author} {\bibfnamefont {U.}~\bibnamefont {De~Giovannini}}, \bibinfo {author} {\bibfnamefont {M.}~\bibnamefont {Puppin}}, \bibinfo {author} {\bibfnamefont {M.}~\bibnamefont {Hoesch}}, \bibinfo {author} {\bibfnamefont {E.}~\bibnamefont {Springate}}, \bibinfo {author} {\bibfnamefont {R.~T.}\ \bibnamefont {Chapman}}, \bibinfo {author} {\bibfnamefont {C.}~\bibnamefont {Cacho}}, \bibinfo {author} {\bibfnamefont {M.}~\bibnamefont {Wolf}}, \bibinfo {author} {\bibfnamefont {A.}~\bibnamefont {Rubio}}, \ and\ \bibinfo {author} {\bibfnamefont {R.}~\bibnamefont {Ernstorfer}},\ }\href {\doibase 10.1103/PhysRevLett.117.277201} {\bibfield  {journal} {\bibinfo  {journal}
  {Phys. Rev. Lett.}\ }\textbf {\bibinfo {volume} {117}},\ \bibinfo {pages} {277201} (\bibinfo {year} {2016})}\BibitemShut {NoStop}%
\bibitem [{\citenamefont {Hein}\ \emph {et~al.}(2016)\citenamefont {Hein}, \citenamefont {Stange}, \citenamefont {Hanff}, \citenamefont {Yang}, \citenamefont {Rohde}, \citenamefont {Rossnagel},\ and\ \citenamefont {Bauer}}]{Hein16}%
  \BibitemOpen
  \bibfield  {author} {\bibinfo {author} {\bibfnamefont {P.}~\bibnamefont {Hein}}, \bibinfo {author} {\bibfnamefont {A.}~\bibnamefont {Stange}}, \bibinfo {author} {\bibfnamefont {K.}~\bibnamefont {Hanff}}, \bibinfo {author} {\bibfnamefont {L.~X.}\ \bibnamefont {Yang}}, \bibinfo {author} {\bibfnamefont {G.}~\bibnamefont {Rohde}}, \bibinfo {author} {\bibfnamefont {K.}~\bibnamefont {Rossnagel}}, \ and\ \bibinfo {author} {\bibfnamefont {M.}~\bibnamefont {Bauer}},\ }\href {\doibase 10.1103/PhysRevB.94.205406} {\bibfield  {journal} {\bibinfo  {journal} {Phys. Rev. B}\ }\textbf {\bibinfo {volume} {94}},\ \bibinfo {pages} {205406} (\bibinfo {year} {2016})}\BibitemShut {NoStop}%
\bibitem [{\citenamefont {Wallauer}\ \emph {et~al.}(2016)\citenamefont {Wallauer}, \citenamefont {Reimann}, \citenamefont {Armbrust}, \citenamefont {Güdde},\ and\ \citenamefont {Höfer}}]{Wallauer16}%
  \BibitemOpen
  \bibfield  {author} {\bibinfo {author} {\bibfnamefont {R.}~\bibnamefont {Wallauer}}, \bibinfo {author} {\bibfnamefont {J.}~\bibnamefont {Reimann}}, \bibinfo {author} {\bibfnamefont {N.}~\bibnamefont {Armbrust}}, \bibinfo {author} {\bibfnamefont {J.}~\bibnamefont {Güdde}}, \ and\ \bibinfo {author} {\bibfnamefont {U.}~\bibnamefont {Höfer}},\ }\href {\doibase 10.1063/1.4965839} {\bibfield  {journal} {\bibinfo  {journal} {Applied Physics Letters}\ }\textbf {\bibinfo {volume} {109}},\ \bibinfo {pages} {162102} (\bibinfo {year} {2016})}\BibitemShut {NoStop}%
\bibitem [{\citenamefont {Dong}\ \emph {et~al.}(2021)\citenamefont {Dong}, \citenamefont {Puppin}, \citenamefont {Pincelli}, \citenamefont {Beaulieu}, \citenamefont {Christiansen}, \citenamefont {Hübener}, \citenamefont {Nicholson}, \citenamefont {Xian}, \citenamefont {Dendzik}, \citenamefont {Deng}, \citenamefont {Windsor}, \citenamefont {Selig}, \citenamefont {Malic}, \citenamefont {Rubio}, \citenamefont {Knorr}, \citenamefont {Wolf}, \citenamefont {Rettig},\ and\ \citenamefont {Ernstorfer}}]{Dong21}%
  \BibitemOpen
  \bibfield  {author} {\bibinfo {author} {\bibfnamefont {S.}~\bibnamefont {Dong}}, \bibinfo {author} {\bibfnamefont {M.}~\bibnamefont {Puppin}}, \bibinfo {author} {\bibfnamefont {T.}~\bibnamefont {Pincelli}}, \bibinfo {author} {\bibfnamefont {S.}~\bibnamefont {Beaulieu}}, \bibinfo {author} {\bibfnamefont {D.}~\bibnamefont {Christiansen}}, \bibinfo {author} {\bibfnamefont {H.}~\bibnamefont {Hübener}}, \bibinfo {author} {\bibfnamefont {C.~W.}\ \bibnamefont {Nicholson}}, \bibinfo {author} {\bibfnamefont {R.~P.}\ \bibnamefont {Xian}}, \bibinfo {author} {\bibfnamefont {M.}~\bibnamefont {Dendzik}}, \bibinfo {author} {\bibfnamefont {Y.}~\bibnamefont {Deng}}, \bibinfo {author} {\bibfnamefont {Y.~W.}\ \bibnamefont {Windsor}}, \bibinfo {author} {\bibfnamefont {M.}~\bibnamefont {Selig}}, \bibinfo {author} {\bibfnamefont {E.}~\bibnamefont {Malic}}, \bibinfo {author} {\bibfnamefont {A.}~\bibnamefont {Rubio}}, \bibinfo {author} {\bibfnamefont {A.}~\bibnamefont {Knorr}}, \bibinfo {author} {\bibfnamefont {M.}~\bibnamefont
  {Wolf}}, \bibinfo {author} {\bibfnamefont {L.}~\bibnamefont {Rettig}}, \ and\ \bibinfo {author} {\bibfnamefont {R.}~\bibnamefont {Ernstorfer}},\ }\href {\doibase https://doi.org/10.1002/ntls.10010} {\bibfield  {journal} {\bibinfo  {journal} {Natural Sciences}\ }\textbf {\bibinfo {volume} {1}},\ \bibinfo {pages} {e10010} (\bibinfo {year} {2021})}\BibitemShut {NoStop}%
\bibitem [{\citenamefont {Shi}\ \emph {et~al.}(2019)\citenamefont {Shi}, \citenamefont {You}, \citenamefont {Zhang}, \citenamefont {Tao}, \citenamefont {Oppeneer}, \citenamefont {Wu}, \citenamefont {Thomale}, \citenamefont {Rossnagel}, \citenamefont {Bauer}, \citenamefont {Kapteyn},\ and\ \citenamefont {Murnane}}]{shi2019}%
  \BibitemOpen
  \bibfield  {author} {\bibinfo {author} {\bibfnamefont {X.}~\bibnamefont {Shi}}, \bibinfo {author} {\bibfnamefont {W.}~\bibnamefont {You}}, \bibinfo {author} {\bibfnamefont {Y.}~\bibnamefont {Zhang}}, \bibinfo {author} {\bibfnamefont {Z.}~\bibnamefont {Tao}}, \bibinfo {author} {\bibfnamefont {P.~M.}\ \bibnamefont {Oppeneer}}, \bibinfo {author} {\bibfnamefont {X.}~\bibnamefont {Wu}}, \bibinfo {author} {\bibfnamefont {R.}~\bibnamefont {Thomale}}, \bibinfo {author} {\bibfnamefont {K.}~\bibnamefont {Rossnagel}}, \bibinfo {author} {\bibfnamefont {M.}~\bibnamefont {Bauer}}, \bibinfo {author} {\bibfnamefont {H.}~\bibnamefont {Kapteyn}}, \ and\ \bibinfo {author} {\bibfnamefont {M.}~\bibnamefont {Murnane}},\ }\href {\doibase 10.1126/sciadv.aav4449} {\bibfield  {journal} {\bibinfo  {journal} {Science Advances}\ }\textbf {\bibinfo {volume} {5}},\ \bibinfo {pages} {eaav4449} (\bibinfo {year} {2019})}\BibitemShut {NoStop}%
\bibitem [{\citenamefont {Chi}\ \emph {et~al.}(2020)\citenamefont {Chi}, \citenamefont {Chen}, \citenamefont {Zhao},\ and\ \citenamefont {Weng}}]{chi2020}%
  \BibitemOpen
  \bibfield  {author} {\bibinfo {author} {\bibfnamefont {Z.}~\bibnamefont {Chi}}, \bibinfo {author} {\bibfnamefont {H.}~\bibnamefont {Chen}}, \bibinfo {author} {\bibfnamefont {Q.}~\bibnamefont {Zhao}}, \ and\ \bibinfo {author} {\bibfnamefont {Y.-X.}\ \bibnamefont {Weng}},\ }\href {\doibase 10.1088/1361-6528/ab79ad} {\bibfield  {journal} {\bibinfo  {journal} {Nanotechnology}\ }\textbf {\bibinfo {volume} {31}},\ \bibinfo {pages} {235712} (\bibinfo {year} {2020})}\BibitemShut {NoStop}%
\bibitem [{\citenamefont {Baroni}\ \emph {et~al.}(2001)\citenamefont {Baroni}, \citenamefont {de~Gironcoli}, \citenamefont {Dal~Corso},\ and\ \citenamefont {Giannozzi}}]{Baroni2001}%
  \BibitemOpen
  \bibfield  {author} {\bibinfo {author} {\bibfnamefont {S.}~\bibnamefont {Baroni}}, \bibinfo {author} {\bibfnamefont {S.}~\bibnamefont {de~Gironcoli}}, \bibinfo {author} {\bibfnamefont {A.}~\bibnamefont {Dal~Corso}}, \ and\ \bibinfo {author} {\bibfnamefont {P.}~\bibnamefont {Giannozzi}},\ }\href {\doibase 10.1103/RevModPhys.73.515} {\bibfield  {journal} {\bibinfo  {journal} {Rev. Mod. Phys.}\ }\textbf {\bibinfo {volume} {73}},\ \bibinfo {pages} {515} (\bibinfo {year} {2001})}\BibitemShut {NoStop}%
\bibitem [{\citenamefont {Valla}\ \emph {et~al.}(1999)\citenamefont {Valla}, \citenamefont {Fedorov}, \citenamefont {Johnson},\ and\ \citenamefont {Hulbert}}]{valla1999}%
  \BibitemOpen
  \bibfield  {author} {\bibinfo {author} {\bibfnamefont {T.}~\bibnamefont {Valla}}, \bibinfo {author} {\bibfnamefont {A.~V.}\ \bibnamefont {Fedorov}}, \bibinfo {author} {\bibfnamefont {P.~D.}\ \bibnamefont {Johnson}}, \ and\ \bibinfo {author} {\bibfnamefont {S.~L.}\ \bibnamefont {Hulbert}},\ }\href {\doibase 10.1103/PhysRevLett.83.2085} {\bibfield  {journal} {\bibinfo  {journal} {Phys. Rev. Lett.}\ }\textbf {\bibinfo {volume} {83}},\ \bibinfo {pages} {2085} (\bibinfo {year} {1999})}\BibitemShut {NoStop}%
\bibitem [{\citenamefont {Valla}\ \emph {et~al.}(2000)\citenamefont {Valla}, \citenamefont {Fedorov}, \citenamefont {Johnson}, \citenamefont {Xue}, \citenamefont {Smith},\ and\ \citenamefont {DiSalvo}}]{valla2000}%
  \BibitemOpen
  \bibfield  {author} {\bibinfo {author} {\bibfnamefont {T.}~\bibnamefont {Valla}}, \bibinfo {author} {\bibfnamefont {A.~V.}\ \bibnamefont {Fedorov}}, \bibinfo {author} {\bibfnamefont {P.~D.}\ \bibnamefont {Johnson}}, \bibinfo {author} {\bibfnamefont {J.}~\bibnamefont {Xue}}, \bibinfo {author} {\bibfnamefont {K.~E.}\ \bibnamefont {Smith}}, \ and\ \bibinfo {author} {\bibfnamefont {F.~J.}\ \bibnamefont {DiSalvo}},\ }\href {\doibase 10.1103/PhysRevLett.85.4759} {\bibfield  {journal} {\bibinfo  {journal} {Phys. Rev. Lett.}\ }\textbf {\bibinfo {volume} {85}},\ \bibinfo {pages} {4759} (\bibinfo {year} {2000})}\BibitemShut {NoStop}%
\bibitem [{\citenamefont {Bernardi}\ \emph {et~al.}(2014)\citenamefont {Bernardi}, \citenamefont {Vigil-Fowler}, \citenamefont {Lischner}, \citenamefont {Neaton},\ and\ \citenamefont {Louie}}]{bernardi2014}%
  \BibitemOpen
  \bibfield  {author} {\bibinfo {author} {\bibfnamefont {M.}~\bibnamefont {Bernardi}}, \bibinfo {author} {\bibfnamefont {D.}~\bibnamefont {Vigil-Fowler}}, \bibinfo {author} {\bibfnamefont {J.}~\bibnamefont {Lischner}}, \bibinfo {author} {\bibfnamefont {J.~B.}\ \bibnamefont {Neaton}}, \ and\ \bibinfo {author} {\bibfnamefont {S.~G.}\ \bibnamefont {Louie}},\ }\href {\doibase 10.1103/PhysRevLett.112.257402} {\bibfield  {journal} {\bibinfo  {journal} {Phys. Rev. Lett.}\ }\textbf {\bibinfo {volume} {112}},\ \bibinfo {pages} {257402} (\bibinfo {year} {2014})}\BibitemShut {NoStop}%
\bibitem [{\citenamefont {D{\"u}vel}\ \emph {et~al.}(2022)\citenamefont {D{\"u}vel}, \citenamefont {Merboldt}, \citenamefont {Bange}, \citenamefont {Strauch}, \citenamefont {Stellbrink}, \citenamefont {Pierz}, \citenamefont {Schumacher}, \citenamefont {Momeni}, \citenamefont {Steil}, \citenamefont {Jansen}, \citenamefont {Steil}, \citenamefont {Novko}, \citenamefont {Mathias},\ and\ \citenamefont {Reutzel}}]{duvel2022}%
  \BibitemOpen
  \bibfield  {author} {\bibinfo {author} {\bibfnamefont {M.}~\bibnamefont {D{\"u}vel}}, \bibinfo {author} {\bibfnamefont {M.}~\bibnamefont {Merboldt}}, \bibinfo {author} {\bibfnamefont {J.~P.}\ \bibnamefont {Bange}}, \bibinfo {author} {\bibfnamefont {H.}~\bibnamefont {Strauch}}, \bibinfo {author} {\bibfnamefont {M.}~\bibnamefont {Stellbrink}}, \bibinfo {author} {\bibfnamefont {K.}~\bibnamefont {Pierz}}, \bibinfo {author} {\bibfnamefont {H.~W.}\ \bibnamefont {Schumacher}}, \bibinfo {author} {\bibfnamefont {D.}~\bibnamefont {Momeni}}, \bibinfo {author} {\bibfnamefont {D.}~\bibnamefont {Steil}}, \bibinfo {author} {\bibfnamefont {G.~S.~M.}\ \bibnamefont {Jansen}}, \bibinfo {author} {\bibfnamefont {S.}~\bibnamefont {Steil}}, \bibinfo {author} {\bibfnamefont {D.}~\bibnamefont {Novko}}, \bibinfo {author} {\bibfnamefont {S.}~\bibnamefont {Mathias}}, \ and\ \bibinfo {author} {\bibfnamefont {M.}~\bibnamefont {Reutzel}},\ }\href {\doibase 10.1021/acs.nanolett.2c01325} {\bibfield  {journal} {\bibinfo  {journal} {Nano
  Letters}\ }\textbf {\bibinfo {volume} {22}},\ \bibinfo {pages} {4897} (\bibinfo {year} {2022})}\BibitemShut {NoStop}%
\bibitem [{\citenamefont {Bauer}\ \emph {et~al.}(2015)\citenamefont {Bauer}, \citenamefont {Marienfeld},\ and\ \citenamefont {Aeschlimann}}]{bauer2015}%
  \BibitemOpen
  \bibfield  {author} {\bibinfo {author} {\bibfnamefont {M.}~\bibnamefont {Bauer}}, \bibinfo {author} {\bibfnamefont {A.}~\bibnamefont {Marienfeld}}, \ and\ \bibinfo {author} {\bibfnamefont {M.}~\bibnamefont {Aeschlimann}},\ }\href {\doibase https://doi.org/10.1016/j.progsurf.2015.05.001} {\bibfield  {journal} {\bibinfo  {journal} {Progress in Surface Science}\ }\textbf {\bibinfo {volume} {90}},\ \bibinfo {pages} {319} (\bibinfo {year} {2015})}\BibitemShut {NoStop}%
\bibitem [{\citenamefont {Yang}\ \emph {et~al.}(2015)\citenamefont {Yang}, \citenamefont {Sobota}, \citenamefont {Leuenberger}, \citenamefont {He}, \citenamefont {Hashimoto}, \citenamefont {Lu}, \citenamefont {Eisaki}, \citenamefont {Kirchmann},\ and\ \citenamefont {Shen}}]{yang2015}%
  \BibitemOpen
  \bibfield  {author} {\bibinfo {author} {\bibfnamefont {S.-L.}\ \bibnamefont {Yang}}, \bibinfo {author} {\bibfnamefont {J.~A.}\ \bibnamefont {Sobota}}, \bibinfo {author} {\bibfnamefont {D.}~\bibnamefont {Leuenberger}}, \bibinfo {author} {\bibfnamefont {Y.}~\bibnamefont {He}}, \bibinfo {author} {\bibfnamefont {M.}~\bibnamefont {Hashimoto}}, \bibinfo {author} {\bibfnamefont {D.~H.}\ \bibnamefont {Lu}}, \bibinfo {author} {\bibfnamefont {H.}~\bibnamefont {Eisaki}}, \bibinfo {author} {\bibfnamefont {P.~S.}\ \bibnamefont {Kirchmann}}, \ and\ \bibinfo {author} {\bibfnamefont {Z.-X.}\ \bibnamefont {Shen}},\ }\href {\doibase 10.1103/PhysRevLett.114.247001} {\bibfield  {journal} {\bibinfo  {journal} {Phys. Rev. Lett.}\ }\textbf {\bibinfo {volume} {114}},\ \bibinfo {pages} {247001} (\bibinfo {year} {2015})}\BibitemShut {NoStop}%
\bibitem [{\citenamefont {Picano}\ \emph {et~al.}(2021)\citenamefont {Picano}, \citenamefont {Li},\ and\ \citenamefont {Eckstein}}]{picano21}%
  \BibitemOpen
  \bibfield  {author} {\bibinfo {author} {\bibfnamefont {A.}~\bibnamefont {Picano}}, \bibinfo {author} {\bibfnamefont {J.}~\bibnamefont {Li}}, \ and\ \bibinfo {author} {\bibfnamefont {M.}~\bibnamefont {Eckstein}},\ }\href {\doibase 10.1103/PhysRevB.104.085108} {\bibfield  {journal} {\bibinfo  {journal} {Phys. Rev. B}\ }\textbf {\bibinfo {volume} {104}},\ \bibinfo {pages} {085108} (\bibinfo {year} {2021})}\BibitemShut {NoStop}%
\bibitem [{\citenamefont {Champagne}\ \emph {et~al.}(2023)\citenamefont {Champagne}, \citenamefont {Haber}, \citenamefont {Pokawanvit}, \citenamefont {Qiu}, \citenamefont {Biswas}, \citenamefont {Atwater}, \citenamefont {da~Jornada},\ and\ \citenamefont {Neaton}}]{Champagne2023}%
  \BibitemOpen
  \bibfield  {author} {\bibinfo {author} {\bibfnamefont {A.}~\bibnamefont {Champagne}}, \bibinfo {author} {\bibfnamefont {J.~B.}\ \bibnamefont {Haber}}, \bibinfo {author} {\bibfnamefont {S.}~\bibnamefont {Pokawanvit}}, \bibinfo {author} {\bibfnamefont {D.~Y.}\ \bibnamefont {Qiu}}, \bibinfo {author} {\bibfnamefont {S.}~\bibnamefont {Biswas}}, \bibinfo {author} {\bibfnamefont {H.~A.}\ \bibnamefont {Atwater}}, \bibinfo {author} {\bibfnamefont {F.~H.}\ \bibnamefont {da~Jornada}}, \ and\ \bibinfo {author} {\bibfnamefont {J.~B.}\ \bibnamefont {Neaton}},\ }\href {\doibase 10.1021/acs.nanolett.3c00386} {\bibfield  {journal} {\bibinfo  {journal} {Nano Letters}\ }\textbf {\bibinfo {volume} {23}},\ \bibinfo {pages} {4274} (\bibinfo {year} {2023})}\BibitemShut {NoStop}%
\bibitem [{\citenamefont {Cui}\ \emph {et~al.}(2023)\citenamefont {Cui}, \citenamefont {Yan}, \citenamefont {Yan}, \citenamefont {Zhou},\ and\ \citenamefont {Cai}}]{cui2023}%
  \BibitemOpen
  \bibfield  {author} {\bibinfo {author} {\bibfnamefont {X.}~\bibnamefont {Cui}}, \bibinfo {author} {\bibfnamefont {H.}~\bibnamefont {Yan}}, \bibinfo {author} {\bibfnamefont {X.}~\bibnamefont {Yan}}, \bibinfo {author} {\bibfnamefont {K.}~\bibnamefont {Zhou}}, \ and\ \bibinfo {author} {\bibfnamefont {Y.}~\bibnamefont {Cai}},\ }\href {\doibase 10.1021/acsnano.3c01229} {\bibfield  {journal} {\bibinfo  {journal} {ACS Nano}\ }\textbf {\bibinfo {volume} {17}},\ \bibinfo {pages} {16530} (\bibinfo {year} {2023})}\BibitemShut {NoStop}%
\bibitem [{\citenamefont {Zhao}\ \emph {et~al.}(2018)\citenamefont {Zhao}, \citenamefont {Dai}, \citenamefont {Zhang}, \citenamefont {Lian}, \citenamefont {Zeng}, \citenamefont {Li}, \citenamefont {Meng},\ and\ \citenamefont {Ni}}]{Zhao2018}%
  \BibitemOpen
  \bibfield  {author} {\bibinfo {author} {\bibfnamefont {Y.}~\bibnamefont {Zhao}}, \bibinfo {author} {\bibfnamefont {Z.}~\bibnamefont {Dai}}, \bibinfo {author} {\bibfnamefont {C.}~\bibnamefont {Zhang}}, \bibinfo {author} {\bibfnamefont {C.}~\bibnamefont {Lian}}, \bibinfo {author} {\bibfnamefont {S.}~\bibnamefont {Zeng}}, \bibinfo {author} {\bibfnamefont {G.}~\bibnamefont {Li}}, \bibinfo {author} {\bibfnamefont {S.}~\bibnamefont {Meng}}, \ and\ \bibinfo {author} {\bibfnamefont {J.}~\bibnamefont {Ni}},\ }\href {\doibase 10.1088/1367-2630/aab338} {\bibfield  {journal} {\bibinfo  {journal} {New Journal of Physics}\ }\textbf {\bibinfo {volume} {20}},\ \bibinfo {pages} {043009} (\bibinfo {year} {2018})}\BibitemShut {NoStop}%
\bibitem [{\citenamefont {Girotto}\ \emph {et~al.}(2023)\citenamefont {Girotto}, \citenamefont {Caruso},\ and\ \citenamefont {Novko}}]{Girotto2023b}%
  \BibitemOpen
  \bibfield  {author} {\bibinfo {author} {\bibfnamefont {N.}~\bibnamefont {Girotto}}, \bibinfo {author} {\bibfnamefont {F.}~\bibnamefont {Caruso}}, \ and\ \bibinfo {author} {\bibfnamefont {D.}~\bibnamefont {Novko}},\ }\href {\doibase 10.1021/acs.jpcc.3c03664} {\bibfield  {journal} {\bibinfo  {journal} {The Journal of Physical Chemistry C}\ }\textbf {\bibinfo {volume} {127}},\ \bibinfo {pages} {16515} (\bibinfo {year} {2023})}\BibitemShut {NoStop}%
\bibitem [{\citenamefont {Marini}\ and\ \citenamefont {Calandra}(2021)}]{marini2021b}%
  \BibitemOpen
  \bibfield  {author} {\bibinfo {author} {\bibfnamefont {G.}~\bibnamefont {Marini}}\ and\ \bibinfo {author} {\bibfnamefont {M.}~\bibnamefont {Calandra}},\ }\href {\doibase 10.1103/PhysRevLett.127.257401} {\bibfield  {journal} {\bibinfo  {journal} {Phys. Rev. Lett.}\ }\textbf {\bibinfo {volume} {127}},\ \bibinfo {pages} {257401} (\bibinfo {year} {2021})}\BibitemShut {NoStop}%
\bibitem [{\citenamefont {Waldecker}\ \emph {et~al.}(2017)\citenamefont {Waldecker}, \citenamefont {Bertoni}, \citenamefont {H\"ubener}, \citenamefont {Brumme}, \citenamefont {Vasileiadis}, \citenamefont {Zahn}, \citenamefont {Rubio},\ and\ \citenamefont {Ernstorfer}}]{waldecker17}%
  \BibitemOpen
  \bibfield  {author} {\bibinfo {author} {\bibfnamefont {L.}~\bibnamefont {Waldecker}}, \bibinfo {author} {\bibfnamefont {R.}~\bibnamefont {Bertoni}}, \bibinfo {author} {\bibfnamefont {H.}~\bibnamefont {H\"ubener}}, \bibinfo {author} {\bibfnamefont {T.}~\bibnamefont {Brumme}}, \bibinfo {author} {\bibfnamefont {T.}~\bibnamefont {Vasileiadis}}, \bibinfo {author} {\bibfnamefont {D.}~\bibnamefont {Zahn}}, \bibinfo {author} {\bibfnamefont {A.}~\bibnamefont {Rubio}}, \ and\ \bibinfo {author} {\bibfnamefont {R.}~\bibnamefont {Ernstorfer}},\ }\href {\doibase 10.1103/PhysRevLett.119.036803} {\bibfield  {journal} {\bibinfo  {journal} {Phys. Rev. Lett.}\ }\textbf {\bibinfo {volume} {119}},\ \bibinfo {pages} {036803} (\bibinfo {year} {2017})}\BibitemShut {NoStop}%
\bibitem [{\citenamefont {Krishnamoorthy}\ \emph {et~al.}(2019)\citenamefont {Krishnamoorthy}, \citenamefont {Lin}, \citenamefont {Zhang}, \citenamefont {Weninger}, \citenamefont {Ma}, \citenamefont {Britz}, \citenamefont {Tiwary}, \citenamefont {Kochat}, \citenamefont {Apte}, \citenamefont {Yang}, \citenamefont {Park}, \citenamefont {Li}, \citenamefont {Shen}, \citenamefont {Wang}, \citenamefont {Kalia}, \citenamefont {Nakano}, \citenamefont {Shimojo}, \citenamefont {Fritz}, \citenamefont {Bergmann}, \citenamefont {Ajayan},\ and\ \citenamefont {Vashishta}}]{Krishnamoorthy2019}%
  \BibitemOpen
  \bibfield  {author} {\bibinfo {author} {\bibfnamefont {A.}~\bibnamefont {Krishnamoorthy}}, \bibinfo {author} {\bibfnamefont {M.-F.}\ \bibnamefont {Lin}}, \bibinfo {author} {\bibfnamefont {X.}~\bibnamefont {Zhang}}, \bibinfo {author} {\bibfnamefont {C.}~\bibnamefont {Weninger}}, \bibinfo {author} {\bibfnamefont {R.}~\bibnamefont {Ma}}, \bibinfo {author} {\bibfnamefont {A.}~\bibnamefont {Britz}}, \bibinfo {author} {\bibfnamefont {C.~S.}\ \bibnamefont {Tiwary}}, \bibinfo {author} {\bibfnamefont {V.}~\bibnamefont {Kochat}}, \bibinfo {author} {\bibfnamefont {A.}~\bibnamefont {Apte}}, \bibinfo {author} {\bibfnamefont {J.}~\bibnamefont {Yang}}, \bibinfo {author} {\bibfnamefont {S.}~\bibnamefont {Park}}, \bibinfo {author} {\bibfnamefont {R.}~\bibnamefont {Li}}, \bibinfo {author} {\bibfnamefont {X.}~\bibnamefont {Shen}}, \bibinfo {author} {\bibfnamefont {X.}~\bibnamefont {Wang}}, \bibinfo {author} {\bibfnamefont {R.}~\bibnamefont {Kalia}}, \bibinfo {author} {\bibfnamefont {A.}~\bibnamefont {Nakano}}, \bibinfo
  {author} {\bibfnamefont {F.}~\bibnamefont {Shimojo}}, \bibinfo {author} {\bibfnamefont {D.}~\bibnamefont {Fritz}}, \bibinfo {author} {\bibfnamefont {U.}~\bibnamefont {Bergmann}}, \bibinfo {author} {\bibfnamefont {P.}~\bibnamefont {Ajayan}}, \ and\ \bibinfo {author} {\bibfnamefont {P.}~\bibnamefont {Vashishta}},\ }\href {\doibase 10.1021/acs.nanolett.9b01179} {\bibfield  {journal} {\bibinfo  {journal} {Nano Letters}\ }\textbf {\bibinfo {volume} {19}},\ \bibinfo {pages} {4981} (\bibinfo {year} {2019})}\BibitemShut {NoStop}%
\bibitem [{\citenamefont {Britt}\ \emph {et~al.}(2022)\citenamefont {Britt}, \citenamefont {Li}, \citenamefont {Ren{\'e}~de Cotret}, \citenamefont {Olsen}, \citenamefont {Otto}, \citenamefont {Hassan}, \citenamefont {Zacharias}, \citenamefont {Caruso}, \citenamefont {Zhu},\ and\ \citenamefont {Siwick}}]{Britt2022}%
  \BibitemOpen
  \bibfield  {author} {\bibinfo {author} {\bibfnamefont {T.~L.}\ \bibnamefont {Britt}}, \bibinfo {author} {\bibfnamefont {Q.}~\bibnamefont {Li}}, \bibinfo {author} {\bibfnamefont {L.~P.}\ \bibnamefont {Ren{\'e}~de Cotret}}, \bibinfo {author} {\bibfnamefont {N.}~\bibnamefont {Olsen}}, \bibinfo {author} {\bibfnamefont {M.}~\bibnamefont {Otto}}, \bibinfo {author} {\bibfnamefont {S.~A.}\ \bibnamefont {Hassan}}, \bibinfo {author} {\bibfnamefont {M.}~\bibnamefont {Zacharias}}, \bibinfo {author} {\bibfnamefont {F.}~\bibnamefont {Caruso}}, \bibinfo {author} {\bibfnamefont {X.}~\bibnamefont {Zhu}}, \ and\ \bibinfo {author} {\bibfnamefont {B.~J.}\ \bibnamefont {Siwick}},\ }\href {\doibase 10.1021/acs.nanolett.2c00850} {\bibfield  {journal} {\bibinfo  {journal} {Nano Letters}\ }\textbf {\bibinfo {volume} {22}},\ \bibinfo {pages} {4718} (\bibinfo {year} {2022})}\BibitemShut {NoStop}%
\bibitem [{\citenamefont {Xian}\ \emph {et~al.}(2020)\citenamefont {Xian}, \citenamefont {Acremann}, \citenamefont {Agustsson}, \citenamefont {Dendzik}, \citenamefont {B{\"u}hlmann}, \citenamefont {Curcio}, \citenamefont {Kutnyakhov}, \citenamefont {Pressacco}, \citenamefont {Heber}, \citenamefont {Dong}, \citenamefont {Pincelli}, \citenamefont {Demsar}, \citenamefont {Wurth}, \citenamefont {Hofmann}, \citenamefont {Wolf}, \citenamefont {Scheidgen}, \citenamefont {Rettig},\ and\ \citenamefont {Ernstorfer}}]{Xian20}%
  \BibitemOpen
  \bibfield  {author} {\bibinfo {author} {\bibfnamefont {R.~P.}\ \bibnamefont {Xian}}, \bibinfo {author} {\bibfnamefont {Y.}~\bibnamefont {Acremann}}, \bibinfo {author} {\bibfnamefont {S.~Y.}\ \bibnamefont {Agustsson}}, \bibinfo {author} {\bibfnamefont {M.}~\bibnamefont {Dendzik}}, \bibinfo {author} {\bibfnamefont {K.}~\bibnamefont {B{\"u}hlmann}}, \bibinfo {author} {\bibfnamefont {D.}~\bibnamefont {Curcio}}, \bibinfo {author} {\bibfnamefont {D.}~\bibnamefont {Kutnyakhov}}, \bibinfo {author} {\bibfnamefont {F.}~\bibnamefont {Pressacco}}, \bibinfo {author} {\bibfnamefont {M.}~\bibnamefont {Heber}}, \bibinfo {author} {\bibfnamefont {S.}~\bibnamefont {Dong}}, \bibinfo {author} {\bibfnamefont {T.}~\bibnamefont {Pincelli}}, \bibinfo {author} {\bibfnamefont {J.}~\bibnamefont {Demsar}}, \bibinfo {author} {\bibfnamefont {W.}~\bibnamefont {Wurth}}, \bibinfo {author} {\bibfnamefont {P.}~\bibnamefont {Hofmann}}, \bibinfo {author} {\bibfnamefont {M.}~\bibnamefont {Wolf}}, \bibinfo {author} {\bibfnamefont
  {M.}~\bibnamefont {Scheidgen}}, \bibinfo {author} {\bibfnamefont {L.}~\bibnamefont {Rettig}}, \ and\ \bibinfo {author} {\bibfnamefont {R.}~\bibnamefont {Ernstorfer}},\ }\href {\doibase 10.1038/s41597-020-00769-8} {\bibfield  {journal} {\bibinfo  {journal} {Scientific Data}\ }\textbf {\bibinfo {volume} {7}},\ \bibinfo {pages} {442} (\bibinfo {year} {2020})}\BibitemShut {NoStop}%
\bibitem [{\citenamefont {Xian}\ \emph {et~al.}(2019)\citenamefont {Xian}, \citenamefont {Rettig},\ and\ \citenamefont {Ernstorfer}}]{Xian19_2}%
  \BibitemOpen
  \bibfield  {author} {\bibinfo {author} {\bibfnamefont {R.~P.}\ \bibnamefont {Xian}}, \bibinfo {author} {\bibfnamefont {L.}~\bibnamefont {Rettig}}, \ and\ \bibinfo {author} {\bibfnamefont {R.}~\bibnamefont {Ernstorfer}},\ }\href {\doibase https://doi.org/10.1016/j.ultramic.2019.04.004} {\bibfield  {journal} {\bibinfo  {journal} {Ultramicroscopy}\ }\textbf {\bibinfo {volume} {202}},\ \bibinfo {pages} {133 } (\bibinfo {year} {2019})}\BibitemShut {NoStop}%
\bibitem [{\citenamefont {Munkhbat}\ \emph {et~al.}(2022)\citenamefont {Munkhbat}, \citenamefont {Wr{\'o}bel}, \citenamefont {Antosiewicz},\ and\ \citenamefont {Shegai}}]{Munkhbat2022}%
  \BibitemOpen
  \bibfield  {author} {\bibinfo {author} {\bibfnamefont {B.}~\bibnamefont {Munkhbat}}, \bibinfo {author} {\bibfnamefont {P.}~\bibnamefont {Wr{\'o}bel}}, \bibinfo {author} {\bibfnamefont {T.~J.}\ \bibnamefont {Antosiewicz}}, \ and\ \bibinfo {author} {\bibfnamefont {T.~O.}\ \bibnamefont {Shegai}},\ }\href {\doibase 10.1021/acsphotonics.2c00433} {\bibfield  {journal} {\bibinfo  {journal} {ACS Photonics}\ }\textbf {\bibinfo {volume} {9}},\ \bibinfo {pages} {2398} (\bibinfo {year} {2022})}\BibitemShut {NoStop}%
\bibitem [{\citenamefont {Giannozzi}\ \emph {et~al.}(2017)\citenamefont {Giannozzi}, \citenamefont {Andreussi}, \citenamefont {Brumme}, \citenamefont {Bunau}, \citenamefont {Nardelli}, \citenamefont {Calandra}, \citenamefont {Car}, \citenamefont {Cavazzoni}, \citenamefont {Ceresoli}, \citenamefont {Cococcioni},\ and\ \citenamefont {et~al.}}]{Giannozzi2017}%
  \BibitemOpen
  \bibfield  {author} {\bibinfo {author} {\bibfnamefont {P.}~\bibnamefont {Giannozzi}}, \bibinfo {author} {\bibfnamefont {O.}~\bibnamefont {Andreussi}}, \bibinfo {author} {\bibfnamefont {T.}~\bibnamefont {Brumme}}, \bibinfo {author} {\bibfnamefont {O.}~\bibnamefont {Bunau}}, \bibinfo {author} {\bibfnamefont {M.~B.}\ \bibnamefont {Nardelli}}, \bibinfo {author} {\bibfnamefont {M.}~\bibnamefont {Calandra}}, \bibinfo {author} {\bibfnamefont {R.}~\bibnamefont {Car}}, \bibinfo {author} {\bibfnamefont {C.}~\bibnamefont {Cavazzoni}}, \bibinfo {author} {\bibfnamefont {D.}~\bibnamefont {Ceresoli}}, \bibinfo {author} {\bibfnamefont {M.}~\bibnamefont {Cococcioni}}, \ and\ \bibinfo {author} {\bibnamefont {et~al.}},\ }\href {\doibase 10.1088/1361-648X/aa8f79} {\bibfield  {journal} {\bibinfo  {journal} {J. Phys. Condens. Matter}\ }\textbf {\bibinfo {volume} {29}},\ \bibinfo {pages} {465901} (\bibinfo {year} {2017})}\BibitemShut {NoStop}%
\bibitem [{\citenamefont {Mostofi}\ \emph {et~al.}(2008)\citenamefont {Mostofi}, \citenamefont {Yates}, \citenamefont {Lee}, \citenamefont {Souza}, \citenamefont {Vanderbilt},\ and\ \citenamefont {Marzari}}]{wan90}%
  \BibitemOpen
  \bibfield  {author} {\bibinfo {author} {\bibfnamefont {A.~A.}\ \bibnamefont {Mostofi}}, \bibinfo {author} {\bibfnamefont {J.~R.}\ \bibnamefont {Yates}}, \bibinfo {author} {\bibfnamefont {Y.-S.}\ \bibnamefont {Lee}}, \bibinfo {author} {\bibfnamefont {I.}~\bibnamefont {Souza}}, \bibinfo {author} {\bibfnamefont {D.}~\bibnamefont {Vanderbilt}}, \ and\ \bibinfo {author} {\bibfnamefont {N.}~\bibnamefont {Marzari}},\ }\href {\doibase https://doi.org/10.1016/j.cpc.2007.11.016} {\bibfield  {journal} {\bibinfo  {journal} {Computer Physics Communications}\ }\textbf {\bibinfo {volume} {178}},\ \bibinfo {pages} {685} (\bibinfo {year} {2008})}\BibitemShut {NoStop}%
\bibitem [{\citenamefont {Giustino}\ \emph {et~al.}(2007)\citenamefont {Giustino}, \citenamefont {Cohen},\ and\ \citenamefont {Louie}}]{giustino07}%
  \BibitemOpen
  \bibfield  {author} {\bibinfo {author} {\bibfnamefont {F.}~\bibnamefont {Giustino}}, \bibinfo {author} {\bibfnamefont {M.~L.}\ \bibnamefont {Cohen}}, \ and\ \bibinfo {author} {\bibfnamefont {S.~G.}\ \bibnamefont {Louie}},\ }\href {\doibase 10.1103/PhysRevB.76.165108} {\bibfield  {journal} {\bibinfo  {journal} {Phys. Rev. B}\ }\textbf {\bibinfo {volume} {76}},\ \bibinfo {pages} {165108} (\bibinfo {year} {2007})}\BibitemShut {NoStop}%
\bibitem [{\citenamefont {Noffsinger}\ \emph {et~al.}(2010)\citenamefont {Noffsinger}, \citenamefont {Giustino}, \citenamefont {Malone}, \citenamefont {Park}, \citenamefont {Louie},\ and\ \citenamefont {Cohen}}]{Noffsinger2010}%
  \BibitemOpen
  \bibfield  {author} {\bibinfo {author} {\bibfnamefont {J.}~\bibnamefont {Noffsinger}}, \bibinfo {author} {\bibfnamefont {F.}~\bibnamefont {Giustino}}, \bibinfo {author} {\bibfnamefont {B.~D.}\ \bibnamefont {Malone}}, \bibinfo {author} {\bibfnamefont {C.-H.}\ \bibnamefont {Park}}, \bibinfo {author} {\bibfnamefont {S.~G.}\ \bibnamefont {Louie}}, \ and\ \bibinfo {author} {\bibfnamefont {M.~L.}\ \bibnamefont {Cohen}},\ }\href {\doibase 10.1016/j.cpc.2010.08.027} {\bibfield  {journal} {\bibinfo  {journal} {Comput. Phys. Commun.}\ }\textbf {\bibinfo {volume} {181}},\ \bibinfo {pages} {2140} (\bibinfo {year} {2010})}\BibitemShut {NoStop}%
\bibitem [{\citenamefont {Ponc\'e}\ \emph {et~al.}(2016)\citenamefont {Ponc\'e}, \citenamefont {Margine}, \citenamefont {Verdi},\ and\ \citenamefont {Giustino}}]{Ponce2016}%
  \BibitemOpen
  \bibfield  {author} {\bibinfo {author} {\bibfnamefont {S.}~\bibnamefont {Ponc\'e}}, \bibinfo {author} {\bibfnamefont {E.}~\bibnamefont {Margine}}, \bibinfo {author} {\bibfnamefont {C.}~\bibnamefont {Verdi}}, \ and\ \bibinfo {author} {\bibfnamefont {F.}~\bibnamefont {Giustino}},\ }\href {\doibase 10.1016/j.cpc.2016.07.028} {\bibfield  {journal} {\bibinfo  {journal} {Comput. Phys. Commun.}\ }\textbf {\bibinfo {volume} {209}},\ \bibinfo {pages} {116} (\bibinfo {year} {2016})}\BibitemShut {NoStop}%
\bibitem [{\citenamefont {van Setten}\ \emph {et~al.}(2018)\citenamefont {van Setten}, \citenamefont {Giantomassi}, \citenamefont {Bousquet}, \citenamefont {Verstraete}, \citenamefont {Hamann}, \citenamefont {Gonze},\ and\ \citenamefont {Rignanese}}]{vanSetten2018}%
  \BibitemOpen
  \bibfield  {author} {\bibinfo {author} {\bibfnamefont {M.~J.}\ \bibnamefont {van Setten}}, \bibinfo {author} {\bibfnamefont {M.}~\bibnamefont {Giantomassi}}, \bibinfo {author} {\bibfnamefont {E.}~\bibnamefont {Bousquet}}, \bibinfo {author} {\bibfnamefont {M.~J.}\ \bibnamefont {Verstraete}}, \bibinfo {author} {\bibfnamefont {D.~R.}\ \bibnamefont {Hamann}}, \bibinfo {author} {\bibfnamefont {X.}~\bibnamefont {Gonze}}, \ and\ \bibinfo {author} {\bibfnamefont {G.~M.}\ \bibnamefont {Rignanese}},\ }\href {\doibase 10.1016/j.cpc.2018.01.012} {\bibfield  {journal} {\bibinfo  {journal} {Comput. Phys. Commun.}\ }\textbf {\bibinfo {volume} {226}},\ \bibinfo {pages} {39} (\bibinfo {year} {2018})}\BibitemShut {NoStop}%
\bibitem [{\citenamefont {Marzari}\ \emph {et~al.}(2012)\citenamefont {Marzari}, \citenamefont {Mostofi}, \citenamefont {Yates}, \citenamefont {Souza},\ and\ \citenamefont {Vanderbilt}}]{Marzari2012}%
  \BibitemOpen
  \bibfield  {author} {\bibinfo {author} {\bibfnamefont {N.}~\bibnamefont {Marzari}}, \bibinfo {author} {\bibfnamefont {A.~A.}\ \bibnamefont {Mostofi}}, \bibinfo {author} {\bibfnamefont {J.~R.}\ \bibnamefont {Yates}}, \bibinfo {author} {\bibfnamefont {I.}~\bibnamefont {Souza}}, \ and\ \bibinfo {author} {\bibfnamefont {D.}~\bibnamefont {Vanderbilt}},\ }\href {\doibase 10.1103/RevModPhys.84.1419} {\bibfield  {journal} {\bibinfo  {journal} {Rev. Mod. Phys.}\ }\textbf {\bibinfo {volume} {84}},\ \bibinfo {pages} {1419} (\bibinfo {year} {2012})}\BibitemShut {NoStop}%
\end{thebibliography}

\begin{thebibliography}{20}%
\makeatletter
\providecommand \@ifxundefined [1]{%
 \@ifx{#1\undefined}
}%
\providecommand \@ifnum [1]{%
 \ifnum #1\expandafter \@firstoftwo
 \else \expandafter \@secondoftwo
 \fi
}%
\providecommand \@ifx [1]{%
 \ifx #1\expandafter \@firstoftwo
 \else \expandafter \@secondoftwo
 \fi
}%
\providecommand \natexlab [1]{#1}%
\providecommand \enquote  [1]{``#1''}%
\providecommand \bibnamefont  [1]{#1}%
\providecommand \bibfnamefont [1]{#1}%
\providecommand \citenamefont [1]{#1}%
\providecommand \href@noop [0]{\@secondoftwo}%
\providecommand \href [0]{\begingroup \@sanitize@url \@href}%
\providecommand \@href[1]{\@@startlink{#1}\@@href}%
\providecommand \@@href[1]{\endgroup#1\@@endlink}%
\providecommand \@sanitize@url [0]{\catcode `\\12\catcode `\$12\catcode `\&12\catcode `\#12\catcode `\^12\catcode `\_12\catcode `\%12\relax}%
\providecommand \@@startlink[1]{}%
\providecommand \@@endlink[0]{}%
\providecommand \url  [0]{\begingroup\@sanitize@url \@url }%
\providecommand \@url [1]{\endgroup\@href {#1}{\urlprefix }}%
\providecommand \urlprefix  [0]{URL }%
\providecommand \Eprint [0]{\href }%
\providecommand \doibase [0]{http://dx.doi.org/}%
\providecommand \selectlanguage [0]{\@gobble}%
\providecommand \bibinfo  [0]{\@secondoftwo}%
\providecommand \bibfield  [0]{\@secondoftwo}%
\providecommand \translation [1]{[#1]}%
\providecommand \BibitemOpen [0]{}%
\providecommand \bibitemStop [0]{}%
\providecommand \bibitemNoStop [0]{.\EOS\space}%
\providecommand \EOS [0]{\spacefactor3000\relax}%
\providecommand \BibitemShut  [1]{\csname bibitem#1\endcsname}%
\let\auto@bib@innerbib\@empty
\bibitem [{\citenamefont {Comby}\ \emph {et~al.}(2022)\citenamefont {Comby}, \citenamefont {Rajak}, \citenamefont {Descamps}, \citenamefont {Petit}, \citenamefont {Blanchet}, \citenamefont {Mairesse}, \citenamefont {Gaudin},\ and\ \citenamefont {Beaulieu}}]{Comby22SM}%
  \BibitemOpen
  \bibfield  {author} {\bibinfo {author} {\bibfnamefont {A.}~\bibnamefont {Comby}}, \bibinfo {author} {\bibfnamefont {D.}~\bibnamefont {Rajak}}, \bibinfo {author} {\bibfnamefont {D.}~\bibnamefont {Descamps}}, \bibinfo {author} {\bibfnamefont {S.}~\bibnamefont {Petit}}, \bibinfo {author} {\bibfnamefont {V.}~\bibnamefont {Blanchet}}, \bibinfo {author} {\bibfnamefont {Y.}~\bibnamefont {Mairesse}}, \bibinfo {author} {\bibfnamefont {J.}~\bibnamefont {Gaudin}}, \ and\ \bibinfo {author} {\bibfnamefont {S.}~\bibnamefont {Beaulieu}},\ }\href {\doibase 10.1088/2040-8986/ac7a49} {\bibfield  {journal} {\bibinfo  {journal} {Journal of Optics}\ }\textbf {\bibinfo {volume} {24}},\ \bibinfo {pages} {084003} (\bibinfo {year} {2022})}\BibitemShut {NoStop}%
\bibitem [{\citenamefont {Tkach}\ and\ \citenamefont {Schönhense}(2025)}]{tkach24SM}%
  \BibitemOpen
  \bibfield  {author} {\bibinfo {author} {\bibfnamefont {O.}~\bibnamefont {Tkach}}\ and\ \bibinfo {author} {\bibfnamefont {G.}~\bibnamefont {Schönhense}},\ }\href {\doibase https://doi.org/10.1016/j.ultramic.2025.114167} {\bibfield  {journal} {\bibinfo  {journal} {Ultramicroscopy}\ }\textbf {\bibinfo {volume} {276}},\ \bibinfo {pages} {114167} (\bibinfo {year} {2025})}\BibitemShut {NoStop}%
\bibitem [{\citenamefont {Tkach}\ \emph {et~al.}(2024)\citenamefont {Tkach}, \citenamefont {Fragkos}, \citenamefont {Nguyen}, \citenamefont {Chernov}, \citenamefont {Scholz}, \citenamefont {Wind}, \citenamefont {Babenkov}, \citenamefont {Fedchenko}, \citenamefont {Lytvynenko}, \citenamefont {Zimmer}, \citenamefont {Hloskovskii}, \citenamefont {Kutnyakhov}, \citenamefont {Pressacco}, \citenamefont {Dilling}, \citenamefont {Bruckmeier}, \citenamefont {Heber}, \citenamefont {Scholz}, \citenamefont {Sobota}, \citenamefont {Koralek}, \citenamefont {Sirica}, \citenamefont {Kallmayer}, \citenamefont {Hoesch}, \citenamefont {Schlueter}, \citenamefont {Odnodvorets}, \citenamefont {Mairesse}, \citenamefont {Rossnagel}, \citenamefont {Elmers}, \citenamefont {Beaulieu},\ and\ \citenamefont {Schoenhense}}]{tkach24-2SM}%
  \BibitemOpen
  \bibfield  {author} {\bibinfo {author} {\bibfnamefont {O.}~\bibnamefont {Tkach}}, \bibinfo {author} {\bibfnamefont {S.}~\bibnamefont {Fragkos}}, \bibinfo {author} {\bibfnamefont {Q.}~\bibnamefont {Nguyen}}, \bibinfo {author} {\bibfnamefont {S.}~\bibnamefont {Chernov}}, \bibinfo {author} {\bibfnamefont {M.}~\bibnamefont {Scholz}}, \bibinfo {author} {\bibfnamefont {N.}~\bibnamefont {Wind}}, \bibinfo {author} {\bibfnamefont {S.}~\bibnamefont {Babenkov}}, \bibinfo {author} {\bibfnamefont {O.}~\bibnamefont {Fedchenko}}, \bibinfo {author} {\bibfnamefont {Y.}~\bibnamefont {Lytvynenko}}, \bibinfo {author} {\bibfnamefont {D.}~\bibnamefont {Zimmer}}, \bibinfo {author} {\bibfnamefont {A.}~\bibnamefont {Hloskovskii}}, \bibinfo {author} {\bibfnamefont {D.}~\bibnamefont {Kutnyakhov}}, \bibinfo {author} {\bibfnamefont {F.}~\bibnamefont {Pressacco}}, \bibinfo {author} {\bibfnamefont {J.}~\bibnamefont {Dilling}}, \bibinfo {author} {\bibfnamefont {L.}~\bibnamefont {Bruckmeier}}, \bibinfo {author} {\bibfnamefont
  {M.}~\bibnamefont {Heber}}, \bibinfo {author} {\bibfnamefont {F.}~\bibnamefont {Scholz}}, \bibinfo {author} {\bibfnamefont {J.}~\bibnamefont {Sobota}}, \bibinfo {author} {\bibfnamefont {J.}~\bibnamefont {Koralek}}, \bibinfo {author} {\bibfnamefont {N.}~\bibnamefont {Sirica}}, \bibinfo {author} {\bibfnamefont {M.}~\bibnamefont {Kallmayer}}, \bibinfo {author} {\bibfnamefont {M.}~\bibnamefont {Hoesch}}, \bibinfo {author} {\bibfnamefont {C.}~\bibnamefont {Schlueter}}, \bibinfo {author} {\bibfnamefont {L.~V.}\ \bibnamefont {Odnodvorets}}, \bibinfo {author} {\bibfnamefont {Y.}~\bibnamefont {Mairesse}}, \bibinfo {author} {\bibfnamefont {K.}~\bibnamefont {Rossnagel}}, \bibinfo {author} {\bibfnamefont {H.~J.}\ \bibnamefont {Elmers}}, \bibinfo {author} {\bibfnamefont {S.}~\bibnamefont {Beaulieu}}, \ and\ \bibinfo {author} {\bibfnamefont {G.}~\bibnamefont {Schoenhense}},\ }\href {https://arxiv.org/abs/2401.10084} {\enquote {\bibinfo {title} {Multi-mode front lens for momentum microscopy: Part {II} experiments},}\ }
  (\bibinfo {year} {2024}),\ \Eprint {http://arxiv.org/abs/2401.10084} {arXiv:2401.10084 [cond-mat.mtrl-sci]} \BibitemShut {NoStop}%
\bibitem [{\citenamefont {Xian}\ \emph {et~al.}(2020)\citenamefont {Xian}, \citenamefont {Acremann}, \citenamefont {Agustsson}, \citenamefont {Dendzik}, \citenamefont {B{\"u}hlmann}, \citenamefont {Curcio}, \citenamefont {Kutnyakhov}, \citenamefont {Pressacco}, \citenamefont {Heber}, \citenamefont {Dong}, \citenamefont {Pincelli}, \citenamefont {Demsar}, \citenamefont {Wurth}, \citenamefont {Hofmann}, \citenamefont {Wolf}, \citenamefont {Scheidgen}, \citenamefont {Rettig},\ and\ \citenamefont {Ernstorfer}}]{Xian20SM}%
  \BibitemOpen
  \bibfield  {author} {\bibinfo {author} {\bibfnamefont {R.~P.}\ \bibnamefont {Xian}}, \bibinfo {author} {\bibfnamefont {Y.}~\bibnamefont {Acremann}}, \bibinfo {author} {\bibfnamefont {S.~Y.}\ \bibnamefont {Agustsson}}, \bibinfo {author} {\bibfnamefont {M.}~\bibnamefont {Dendzik}}, \bibinfo {author} {\bibfnamefont {K.}~\bibnamefont {B{\"u}hlmann}}, \bibinfo {author} {\bibfnamefont {D.}~\bibnamefont {Curcio}}, \bibinfo {author} {\bibfnamefont {D.}~\bibnamefont {Kutnyakhov}}, \bibinfo {author} {\bibfnamefont {F.}~\bibnamefont {Pressacco}}, \bibinfo {author} {\bibfnamefont {M.}~\bibnamefont {Heber}}, \bibinfo {author} {\bibfnamefont {S.}~\bibnamefont {Dong}}, \bibinfo {author} {\bibfnamefont {T.}~\bibnamefont {Pincelli}}, \bibinfo {author} {\bibfnamefont {J.}~\bibnamefont {Demsar}}, \bibinfo {author} {\bibfnamefont {W.}~\bibnamefont {Wurth}}, \bibinfo {author} {\bibfnamefont {P.}~\bibnamefont {Hofmann}}, \bibinfo {author} {\bibfnamefont {M.}~\bibnamefont {Wolf}}, \bibinfo {author} {\bibfnamefont
  {M.}~\bibnamefont {Scheidgen}}, \bibinfo {author} {\bibfnamefont {L.}~\bibnamefont {Rettig}}, \ and\ \bibinfo {author} {\bibfnamefont {R.}~\bibnamefont {Ernstorfer}},\ }\href {\doibase 10.1038/s41597-020-00769-8} {\bibfield  {journal} {\bibinfo  {journal} {Scientific Data}\ }\textbf {\bibinfo {volume} {7}},\ \bibinfo {pages} {442} (\bibinfo {year} {2020})}\BibitemShut {NoStop}%
\bibitem [{\citenamefont {Xian}\ \emph {et~al.}(2019)\citenamefont {Xian}, \citenamefont {Rettig},\ and\ \citenamefont {Ernstorfer}}]{Xian19_2SM}%
  \BibitemOpen
  \bibfield  {author} {\bibinfo {author} {\bibfnamefont {R.~P.}\ \bibnamefont {Xian}}, \bibinfo {author} {\bibfnamefont {L.}~\bibnamefont {Rettig}}, \ and\ \bibinfo {author} {\bibfnamefont {R.}~\bibnamefont {Ernstorfer}},\ }\href {\doibase https://doi.org/10.1016/j.ultramic.2019.04.004} {\bibfield  {journal} {\bibinfo  {journal} {Ultramicroscopy}\ }\textbf {\bibinfo {volume} {202}},\ \bibinfo {pages} {133 } (\bibinfo {year} {2019})}\BibitemShut {NoStop}%
\bibitem [{\citenamefont {Fragkos}\ \emph {et~al.}(2025)\citenamefont {Fragkos}, \citenamefont {Courtade}, \citenamefont {Tkach}, \citenamefont {Gaudin}, \citenamefont {Descamps}, \citenamefont {Barrette}, \citenamefont {Petit}, \citenamefont {Schönhense}, \citenamefont {Mairesse},\ and\ \citenamefont {Beaulieu}}]{Fragkos2025SM}%
  \BibitemOpen
  \bibfield  {author} {\bibinfo {author} {\bibfnamefont {S.}~\bibnamefont {Fragkos}}, \bibinfo {author} {\bibfnamefont {Q.}~\bibnamefont {Courtade}}, \bibinfo {author} {\bibfnamefont {O.}~\bibnamefont {Tkach}}, \bibinfo {author} {\bibfnamefont {J.}~\bibnamefont {Gaudin}}, \bibinfo {author} {\bibfnamefont {D.}~\bibnamefont {Descamps}}, \bibinfo {author} {\bibfnamefont {G.}~\bibnamefont {Barrette}}, \bibinfo {author} {\bibfnamefont {S.}~\bibnamefont {Petit}}, \bibinfo {author} {\bibfnamefont {G.}~\bibnamefont {Schönhense}}, \bibinfo {author} {\bibfnamefont {Y.}~\bibnamefont {Mairesse}}, \ and\ \bibinfo {author} {\bibfnamefont {S.}~\bibnamefont {Beaulieu}},\ }\href {https://arxiv.org/abs/2507.02371} {\enquote {\bibinfo {title} {Time- and polarization-resolved extreme ultraviolet momentum microscopy},}\ } (\bibinfo {year} {2025}),\ \Eprint {http://arxiv.org/abs/2507.02371} {arXiv:2507.02371 [cond-mat.mtrl-sci]} \BibitemShut {NoStop}%
\bibitem [{\citenamefont {Munkhbat}\ \emph {et~al.}(2022)\citenamefont {Munkhbat}, \citenamefont {Wr{\'o}bel}, \citenamefont {Antosiewicz},\ and\ \citenamefont {Shegai}}]{Munkhbat2022SM}%
  \BibitemOpen
  \bibfield  {author} {\bibinfo {author} {\bibfnamefont {B.}~\bibnamefont {Munkhbat}}, \bibinfo {author} {\bibfnamefont {P.}~\bibnamefont {Wr{\'o}bel}}, \bibinfo {author} {\bibfnamefont {T.~J.}\ \bibnamefont {Antosiewicz}}, \ and\ \bibinfo {author} {\bibfnamefont {T.~O.}\ \bibnamefont {Shegai}},\ }\href {\doibase 10.1021/acsphotonics.2c00433} {\bibfield  {journal} {\bibinfo  {journal} {ACS Photonics}\ }\textbf {\bibinfo {volume} {9}},\ \bibinfo {pages} {2398} (\bibinfo {year} {2022})}\BibitemShut {NoStop}%
\bibitem [{\citenamefont {Chi}\ \emph {et~al.}(2019)\citenamefont {Chi}, \citenamefont {Chen}, \citenamefont {Zhao},\ and\ \citenamefont {Weng}}]{Chi2019SM}%
  \BibitemOpen
  \bibfield  {author} {\bibinfo {author} {\bibfnamefont {Z.}~\bibnamefont {Chi}}, \bibinfo {author} {\bibfnamefont {H.}~\bibnamefont {Chen}}, \bibinfo {author} {\bibfnamefont {Q.}~\bibnamefont {Zhao}}, \ and\ \bibinfo {author} {\bibfnamefont {Y.-X.}\ \bibnamefont {Weng}},\ }\href {\doibase 10.1063/1.5115467} {\bibfield  {journal} {\bibinfo  {journal} {The Journal of Chemical Physics}\ }\textbf {\bibinfo {volume} {151}},\ \bibinfo {pages} {114704} (\bibinfo {year} {2019})}\BibitemShut {NoStop}%
\bibitem [{\citenamefont {Giustino}(2017)}]{Giustino2017SM}%
  \BibitemOpen
  \bibfield  {author} {\bibinfo {author} {\bibfnamefont {F.}~\bibnamefont {Giustino}},\ }\href {\doibase 10.1103/RevModPhys.89.015003} {\bibfield  {journal} {\bibinfo  {journal} {Rev. Mod. Phys.}\ }\textbf {\bibinfo {volume} {89}},\ \bibinfo {pages} {015003} (\bibinfo {year} {2017})}\BibitemShut {NoStop}%
\bibitem [{\citenamefont {Baroni}\ \emph {et~al.}(2001)\citenamefont {Baroni}, \citenamefont {de~Gironcoli}, \citenamefont {Dal~Corso},\ and\ \citenamefont {Giannozzi}}]{Baroni2001SM}%
  \BibitemOpen
  \bibfield  {author} {\bibinfo {author} {\bibfnamefont {S.}~\bibnamefont {Baroni}}, \bibinfo {author} {\bibfnamefont {S.}~\bibnamefont {de~Gironcoli}}, \bibinfo {author} {\bibfnamefont {A.}~\bibnamefont {Dal~Corso}}, \ and\ \bibinfo {author} {\bibfnamefont {P.}~\bibnamefont {Giannozzi}},\ }\href {\doibase 10.1103/RevModPhys.73.515} {\bibfield  {journal} {\bibinfo  {journal} {Rev. Mod. Phys.}\ }\textbf {\bibinfo {volume} {73}},\ \bibinfo {pages} {515} (\bibinfo {year} {2001})}\BibitemShut {NoStop}%
\bibitem [{\citenamefont {Giannozzi}\ \emph {et~al.}(2017)\citenamefont {Giannozzi}, \citenamefont {Andreussi}, \citenamefont {Brumme}, \citenamefont {Bunau}, \citenamefont {Nardelli}, \citenamefont {Calandra}, \citenamefont {Car}, \citenamefont {Cavazzoni}, \citenamefont {Ceresoli}, \citenamefont {Cococcioni},\ and\ \citenamefont {et~al.}}]{Giannozzi2017SM}%
  \BibitemOpen
  \bibfield  {author} {\bibinfo {author} {\bibfnamefont {P.}~\bibnamefont {Giannozzi}}, \bibinfo {author} {\bibfnamefont {O.}~\bibnamefont {Andreussi}}, \bibinfo {author} {\bibfnamefont {T.}~\bibnamefont {Brumme}}, \bibinfo {author} {\bibfnamefont {O.}~\bibnamefont {Bunau}}, \bibinfo {author} {\bibfnamefont {M.~B.}\ \bibnamefont {Nardelli}}, \bibinfo {author} {\bibfnamefont {M.}~\bibnamefont {Calandra}}, \bibinfo {author} {\bibfnamefont {R.}~\bibnamefont {Car}}, \bibinfo {author} {\bibfnamefont {C.}~\bibnamefont {Cavazzoni}}, \bibinfo {author} {\bibfnamefont {D.}~\bibnamefont {Ceresoli}}, \bibinfo {author} {\bibfnamefont {M.}~\bibnamefont {Cococcioni}}, \ and\ \bibinfo {author} {\bibnamefont {et~al.}},\ }\href {\doibase 10.1088/1361-648X/aa8f79} {\bibfield  {journal} {\bibinfo  {journal} {J. Phys. Condens. Matter}\ }\textbf {\bibinfo {volume} {29}},\ \bibinfo {pages} {465901} (\bibinfo {year} {2017})}\BibitemShut {NoStop}%
\bibitem [{\citenamefont {Mostofi}\ \emph {et~al.}(2008)\citenamefont {Mostofi}, \citenamefont {Yates}, \citenamefont {Lee}, \citenamefont {Souza}, \citenamefont {Vanderbilt},\ and\ \citenamefont {Marzari}}]{wan90SM}%
  \BibitemOpen
  \bibfield  {author} {\bibinfo {author} {\bibfnamefont {A.~A.}\ \bibnamefont {Mostofi}}, \bibinfo {author} {\bibfnamefont {J.~R.}\ \bibnamefont {Yates}}, \bibinfo {author} {\bibfnamefont {Y.-S.}\ \bibnamefont {Lee}}, \bibinfo {author} {\bibfnamefont {I.}~\bibnamefont {Souza}}, \bibinfo {author} {\bibfnamefont {D.}~\bibnamefont {Vanderbilt}}, \ and\ \bibinfo {author} {\bibfnamefont {N.}~\bibnamefont {Marzari}},\ }\href {\doibase https://doi.org/10.1016/j.cpc.2007.11.016} {\bibfield  {journal} {\bibinfo  {journal} {Computer Physics Communications}\ }\textbf {\bibinfo {volume} {178}},\ \bibinfo {pages} {685} (\bibinfo {year} {2008})}\BibitemShut {NoStop}%
\bibitem [{\citenamefont {Giustino}\ \emph {et~al.}(2007)\citenamefont {Giustino}, \citenamefont {Cohen},\ and\ \citenamefont {Louie}}]{giustino07SM}%
  \BibitemOpen
  \bibfield  {author} {\bibinfo {author} {\bibfnamefont {F.}~\bibnamefont {Giustino}}, \bibinfo {author} {\bibfnamefont {M.~L.}\ \bibnamefont {Cohen}}, \ and\ \bibinfo {author} {\bibfnamefont {S.~G.}\ \bibnamefont {Louie}},\ }\href {\doibase 10.1103/PhysRevB.76.165108} {\bibfield  {journal} {\bibinfo  {journal} {Phys. Rev. B}\ }\textbf {\bibinfo {volume} {76}},\ \bibinfo {pages} {165108} (\bibinfo {year} {2007})}\BibitemShut {NoStop}%
\bibitem [{\citenamefont {Noffsinger}\ \emph {et~al.}(2010)\citenamefont {Noffsinger}, \citenamefont {Giustino}, \citenamefont {Malone}, \citenamefont {Park}, \citenamefont {Louie},\ and\ \citenamefont {Cohen}}]{Noffsinger2010SM}%
  \BibitemOpen
  \bibfield  {author} {\bibinfo {author} {\bibfnamefont {J.}~\bibnamefont {Noffsinger}}, \bibinfo {author} {\bibfnamefont {F.}~\bibnamefont {Giustino}}, \bibinfo {author} {\bibfnamefont {B.~D.}\ \bibnamefont {Malone}}, \bibinfo {author} {\bibfnamefont {C.-H.}\ \bibnamefont {Park}}, \bibinfo {author} {\bibfnamefont {S.~G.}\ \bibnamefont {Louie}}, \ and\ \bibinfo {author} {\bibfnamefont {M.~L.}\ \bibnamefont {Cohen}},\ }\href {\doibase 10.1016/j.cpc.2010.08.027} {\bibfield  {journal} {\bibinfo  {journal} {Comput. Phys. Commun.}\ }\textbf {\bibinfo {volume} {181}},\ \bibinfo {pages} {2140} (\bibinfo {year} {2010})}\BibitemShut {NoStop}%
\bibitem [{\citenamefont {Ponc\'e}\ \emph {et~al.}(2016)\citenamefont {Ponc\'e}, \citenamefont {Margine}, \citenamefont {Verdi},\ and\ \citenamefont {Giustino}}]{Ponce2016SM}%
  \BibitemOpen
  \bibfield  {author} {\bibinfo {author} {\bibfnamefont {S.}~\bibnamefont {Ponc\'e}}, \bibinfo {author} {\bibfnamefont {E.}~\bibnamefont {Margine}}, \bibinfo {author} {\bibfnamefont {C.}~\bibnamefont {Verdi}}, \ and\ \bibinfo {author} {\bibfnamefont {F.}~\bibnamefont {Giustino}},\ }\href {\doibase 10.1016/j.cpc.2016.07.028} {\bibfield  {journal} {\bibinfo  {journal} {Comput. Phys. Commun.}\ }\textbf {\bibinfo {volume} {209}},\ \bibinfo {pages} {116} (\bibinfo {year} {2016})}\BibitemShut {NoStop}%
\bibitem [{\citenamefont {Murray}\ \emph {et~al.}(2007)\citenamefont {Murray}, \citenamefont {Fahy}, \citenamefont {Prendergast}, \citenamefont {Ogitsu}, \citenamefont {Fritz},\ and\ \citenamefont {Reis}}]{Murray2007SM}%
  \BibitemOpen
  \bibfield  {author} {\bibinfo {author} {\bibfnamefont {E.~D.}\ \bibnamefont {Murray}}, \bibinfo {author} {\bibfnamefont {S.}~\bibnamefont {Fahy}}, \bibinfo {author} {\bibfnamefont {D.}~\bibnamefont {Prendergast}}, \bibinfo {author} {\bibfnamefont {T.}~\bibnamefont {Ogitsu}}, \bibinfo {author} {\bibfnamefont {D.~M.}\ \bibnamefont {Fritz}}, \ and\ \bibinfo {author} {\bibfnamefont {D.~A.}\ \bibnamefont {Reis}},\ }\href {\doibase 10.1103/PhysRevB.75.184301} {\bibfield  {journal} {\bibinfo  {journal} {Phys. Rev. B}\ }\textbf {\bibinfo {volume} {75}},\ \bibinfo {pages} {184301} (\bibinfo {year} {2007})}\BibitemShut {NoStop}%
\bibitem [{\citenamefont {Liu}\ \emph {et~al.}(2022)\citenamefont {Liu}, \citenamefont {Mao}, \citenamefont {Zhang},\ and\ \citenamefont {Zhou}}]{liu2022SM}%
  \BibitemOpen
  \bibfield  {author} {\bibinfo {author} {\bibfnamefont {K.}~\bibnamefont {Liu}}, \bibinfo {author} {\bibfnamefont {S.}~\bibnamefont {Mao}}, \bibinfo {author} {\bibfnamefont {S.}~\bibnamefont {Zhang}}, \ and\ \bibinfo {author} {\bibfnamefont {J.}~\bibnamefont {Zhou}},\ }\href {\doibase 10.1021/acs.nanolett.2c03238} {\bibfield  {journal} {\bibinfo  {journal} {Nano Letters}\ }\textbf {\bibinfo {volume} {22}},\ \bibinfo {pages} {9006} (\bibinfo {year} {2022})}\BibitemShut {NoStop}%
\bibitem [{\citenamefont {Girotto}\ and\ \citenamefont {Novko}(2023)}]{Girotto2023SM}%
  \BibitemOpen
  \bibfield  {author} {\bibinfo {author} {\bibfnamefont {N.}~\bibnamefont {Girotto}}\ and\ \bibinfo {author} {\bibfnamefont {D.}~\bibnamefont {Novko}},\ }\href {\doibase 10.1021/acs.jpclett.3c01905} {\bibfield  {journal} {\bibinfo  {journal} {The Journal of Physical Chemistry Letters}\ }\textbf {\bibinfo {volume} {14}},\ \bibinfo {pages} {8709} (\bibinfo {year} {2023})}\BibitemShut {NoStop}%
\bibitem [{\citenamefont {van Setten}\ \emph {et~al.}(2018)\citenamefont {van Setten}, \citenamefont {Giantomassi}, \citenamefont {Bousquet}, \citenamefont {Verstraete}, \citenamefont {Hamann}, \citenamefont {Gonze},\ and\ \citenamefont {Rignanese}}]{vanSetten2018SM}%
  \BibitemOpen
  \bibfield  {author} {\bibinfo {author} {\bibfnamefont {M.~J.}\ \bibnamefont {van Setten}}, \bibinfo {author} {\bibfnamefont {M.}~\bibnamefont {Giantomassi}}, \bibinfo {author} {\bibfnamefont {E.}~\bibnamefont {Bousquet}}, \bibinfo {author} {\bibfnamefont {M.~J.}\ \bibnamefont {Verstraete}}, \bibinfo {author} {\bibfnamefont {D.~R.}\ \bibnamefont {Hamann}}, \bibinfo {author} {\bibfnamefont {X.}~\bibnamefont {Gonze}}, \ and\ \bibinfo {author} {\bibfnamefont {G.~M.}\ \bibnamefont {Rignanese}},\ }\href {\doibase 10.1016/j.cpc.2018.01.012} {\bibfield  {journal} {\bibinfo  {journal} {Comput. Phys. Commun.}\ }\textbf {\bibinfo {volume} {226}},\ \bibinfo {pages} {39} (\bibinfo {year} {2018})}\BibitemShut {NoStop}%
\bibitem [{\citenamefont {Marzari}\ \emph {et~al.}(2012)\citenamefont {Marzari}, \citenamefont {Mostofi}, \citenamefont {Yates}, \citenamefont {Souza},\ and\ \citenamefont {Vanderbilt}}]{Marzari2012SM}%
  \BibitemOpen
  \bibfield  {author} {\bibinfo {author} {\bibfnamefont {N.}~\bibnamefont {Marzari}}, \bibinfo {author} {\bibfnamefont {A.~A.}\ \bibnamefont {Mostofi}}, \bibinfo {author} {\bibfnamefont {J.~R.}\ \bibnamefont {Yates}}, \bibinfo {author} {\bibfnamefont {I.}~\bibnamefont {Souza}}, \ and\ \bibinfo {author} {\bibfnamefont {D.}~\bibnamefont {Vanderbilt}},\ }\href {\doibase 10.1103/RevModPhys.84.1419} {\bibfield  {journal} {\bibinfo  {journal} {Rev. Mod. Phys.}\ }\textbf {\bibinfo {volume} {84}},\ \bibinfo {pages} {1419} (\bibinfo {year} {2012})}\BibitemShut {NoStop}%
\end{thebibliography}
\end{document}